\documentclass{article}
\usepackage[utf8]{inputenc}
\usepackage[margin=1in]{geometry}

\usepackage[pdftex]{graphicx}

\usepackage{blindtext}
\usepackage{titlesec}

\AtBeginDocument{%
  }

\usepackage[american]{babel}

\usepackage{pifont}

\usepackage{mathtools} %
\usepackage{booktabs} %
\usepackage{tikz} %
\usepackage{wrapfig}

\usepackage{color,xcolor}
\usepackage{nicefrac}
\usepackage{hyperref}
\hypersetup{
    colorlinks=true,
    linkcolor=blue,
    filecolor=magenta,      
    urlcolor=cyan,
    pdfauthor={Name}
}
\usepackage{mathtools}
\usepackage{amsmath,amsfonts}

\usepackage{amssymb}
\usepackage[ruled,vlined]{algorithm2e}
\usepackage{algorithmic}
\usepackage{makecell}
\usepackage{diagbox}
\usepackage{multirow}
\usepackage{textgreek}
\usepackage{comment}

\usepackage{caption}

\usepackage{epsfig,graphicx,subfigure}

\usepackage{amsthm}
\usepackage{bbm}

\usepackage[nameinlink,capitalize]{cleveref}
\usepackage{etoc}

\AtBeginEnvironment{appendices}{\crefalias{section}{appendix}}
\AtBeginEnvironment{appendices}{\crefalias{subsection}{appendix}}
\AtBeginEnvironment{appendices}{\crefalias{subsubsection}{appendix}}

\crefformat{section}{§#2#1#3}

\usepackage{appendix}

\crefname{section}{\S}{\S}
\Crefname{section}{\S}{\S}
\crefname{appendix}{App.}{Apps.}
\Crefname{appendix}{App.}{Apps.}
\crefname{theorem}{Thm.}{Thms.}
\Crefname{theorem}{Thm.}{Thms.}
\crefname{proposition}{Prop.}{Props.}
\Crefname{proposition}{Prop.}{Props.}
\crefname{algorithm}{Alg.}{Algs.}
\Crefname{algorithm}{Alg.}{Algs.}
\crefname{assumption}{Asm.}{Asms.}
\Crefname{assumption}{Asm.}{Asms.}
\crefname{mechanism}{Mech.}{Mechs.}
\Crefname{mechanism}{Mech.}{Mechs.}

\usepackage{mfirstuc}
\usepackage[T1]{fontenc}

\theoremstyle{plain}
\newtheorem{theorem}{Theorem}[section]
\newtheorem{proposition}[theorem]{Proposition}
\newtheorem{lemma}[theorem]{Lemma}

\theoremstyle{definition}

\theoremstyle{remark}

\newcommand\numberthis{\addtocounter{equation}{1}\tag{\theequation}}

\newcommand{\myparatightestn}[1]{\noindent\textbf{{#1}}~}

\newcounter{packednmbr}

\newcommand{\mg}[1]{\textcolor{blue}{MG: #1}}

\definecolor{ForestGreen}{RGB}{34,139,34}
\newcommand{\EK}[1]{\textcolor{ForestGreen}{EK: #1}}

\newcommand{\calM}{\mathcal{M}}

\newcommand{\bx}{{\boldsymbol{x}}}
\newcommand{\by}{\boldsymbol{y}}

\newcommand{\bra}[1]{\left( #1 \right)}
\newcommand{\brb}[1]{\left[ #1 \right]}

\newcommand{\brba}[1]{\left[ #1 \right)}
\newcommand{\brc}[1]{\left\{ #1 \right\}}

\DeclarePairedDelimiter\abs{\lvert}{\rvert}
\newcommand{\babs}[1]{\left| #1 \right|}

\newcommand{\floor}[1]{\lfloor {#1} \rfloor}
\newcommand{\ceil}[1]{\lceil {#1} \rceil}

\newcommand{\name}{summary statistic privacy}

\newcommand{\datamechanism}{data release mechanism}

\newcommand{\distortion}{distortion}
\newcommand{\Distortion}{Distortion}

\newcommand{\privacy}{privacy}
\newcommand{\Privacy}{Privacy}

\newcommand{\secret}{secret}

\newcommand{\secrets}{secrets}
\newcommand{\Secrets}{Secrets}

\newcommand{\privacythreshold}{\epsilon}
\newcommand{\privacythresholdi}[1]{\epsilon_{#1}}
\newcommand{\privacythresholdihat}[1]{\tilde{\epsilon}_{#1}}
\newcommand{\privacybound}{T}
\newcommand{\privacynotation}{\Pi_{\privacythreshold,\paramdistribution}}

\newcommand{\privacynotationuni}{\privacynotation^{\text{union}}}
\newcommand{\privacynotationgroup}{\privacynotation^{\text{group}}}
\newcommand{\privacynotationlp}{\privacynotation^{l_p}}

\newcommand{\distortionnotation}{\Delta_{\mechanismnotation}}

\newcommand{\dimension}{d}

\newcommand{\rvprivatenotation}{X}
\newcommand{\rvreleasenotation}{X}

\newcommand{\rvprivatewithparam}[1]{\rvprivatenotation_{#1}}
\newcommand{\rvreleasewithparam}[1]{\rvreleasenotation_{#1}}
\newcommand{\rvparamnotation}{\theta}
\newcommand{\RVparamnotation}{\Theta}
\newcommand{\releaservparamnotation}{\theta'}

\newcommand{\rvprivate}{\rvprivatewithparam{\rvparamnotation}}

\newcommand{\seclen}{s}

\newcommand{\distributionnotation}{\omega}
\newcommand{\distributionof}[1]{\distributionnotation_{#1}}
\newcommand{\privatedistribution}{\distributionof{\rvprivatewithparam{\rvparamnotation}}}
\newcommand{\releasedistribution}{\distributionof{\rvreleasewithparam{\releaservparamnotation}}}

\newcommand{\paramdistribution}{\distributionof{\RVparamnotation}}

\newcommand{\privatedataset}{\mathcal{X}}

\newcommand{\secretnotation}{g}
\newcommand{\secretsnotation}{\boldsymbol{g}}
\newcommand{\secreti}[1]{\secretnotation_{#1}}
\newcommand{\secretof}[1]{\secretnotation\bra{#1}}
\newcommand{\secretvof}[1]{\secretsnotation\bra{#1}}
\newcommand{\secretiof}[2]{\secretnotation_{#1}\bra{#2}}

\newcommand{\secretofparam}{\secretof{\rvparamnotation}}
\newcommand{\secretiofparam}[1]{\secretiof{#1}{\rvparamnotation}}

\newcommand{\secretestimatenotation}{\hat{\secretnotation}}
\newcommand{\secretestimatenotationvector}{\hat{\boldsymbol{\secretnotation}}}
\newcommand{\secretestimatenotationv}{\hat{\secretsnotation}^{\bra{\attackerv}}}

\newcommand{\secretestimateof}[1]{\secretestimatenotation\bra{#1}}
\newcommand{\secretiestimateof}[2]{\secretestimatenotation_{#1}\bra{#2}}
\newcommand{\secretiestimateofv}[2]{\secretestimatenotation_{#1}^{\brb{\boldsymbol{v}}}\bra{#2}}

\newcommand{\secrethatestimateof}[1]{\tilde{\secretnotation}\bra{#1}}
\newcommand{\secrethatiestimateof}[2]{\tilde{\secretnotation}_{#1}\bra{#2}}

\newcommand{\mechanismnotation}{\calM_{\secretnotation}}

\newcommand{\mechanismof}[1]{\mechanismnotation\bra{#1}}

\newcommand{\secretindex}{i}
\newcommand{\attackerv}{\boldsymbol{v}}
\newcommand{\attackerindex}[1]{v_{#1}}
\newcommand{\attackernum}{\mathcal{N}}

\newcommand{\distanceof}[2]{\mathfrak{D}\bra{#1\|#2}}

\newcommand{\ratio}{\gamma}

\newcommand{\probnotation}{\mathbb{P}}
\newcommand{\probof}[1]{\probnotation\bra{#1}}

\newcommand{\expectationnotation}{\mathbb{E}}
\newcommand{\expectationof}[1]{\expectationnotation\bra{#1}}

\newcommand{\auxdistance}[2]{D\bra{#1, #2}}
\newcommand{\auxrange}[2]{R\bra{#1, #2}}

\newcommand{\support}[1]{\text{Supp}\bra{#1}}

\newcommand{\distanceformula}[2]{\frac{1}{2} \mathfrak{D}\bra{\distributionof{#1}\|\distributionof{#2}}}

\newcommand{\rangeiformula}[3]{\abs{\secretnotation_{#1}{({#2})-\secretnotation_{#1}{({#3})}}}}
\newcommand{\rangeiformulasquare}[3]{\bra{\secretnotation_{#1}{({#2})-\secretnotation_{#1}{({#3})}}}^2}

\newcommand{\lefti}[1]{L^{\bra{#1}}_{\releaservparamnotation}}
\newcommand{\righti}[1]{H^{\bra{#1}}_{\releaservparamnotation}}
\newcommand{\half}{\frac{1}{2}}

\newcommand{\mulower}{\underline{\mu}}

\newcommand{\sigmalower}{\underline{\sigma}}

\newcommand{\dimgaussian}{k}

\newcommand{\mui}{\mu_1}
\newcommand{\muj}{\mu_2}
\newcommand{\sigmai}{\sigma_1}
\newcommand{\sigmaj}{\sigma_2}

\newcommand{\lami}{\lambda_1}
\newcommand{\lamj}{\lambda_2}
\newcommand{\alp}{\alpha}

\newcommand{\muir}{\mu'_1}
\newcommand{\mujr}{\mu'_2}
\newcommand{\sigmair}{\sigma'_1}
\newcommand{\sigmajr}{\sigma'_2}
\newcommand{\lamir}{\lambda'_1}
\newcommand{\lamjr}{\lambda'_2}
\newcommand{\alpr}{\alpha'}
\newcommand{\aar}{\tilde{a}}
\newcommand{\bbr}{\tilde{b}}

\newcommand{\sigmaih}{\hat{\sigma}_1}

\newcommand{\aaa}{{a}}
\newcommand{\bbb}{{b}}

\newcommand{\aag}{\hat{a}}
\newcommand{\bbg}{\hat{b}}

\newcommand{\aaalower}{\underline{a}}
\newcommand{\bbblower}{\underline{b}}

\newcommand{\ta}{\delta_{\sigmai, \alpha}}
\newcommand{\tb}{\delta_{\sigmai, \beta}}
\newcommand{\tar}{\delta'_{\sigmai, \alpha}}
\newcommand{\tbr}{\delta'_{\sigmai, \beta}}
\newcommand{\tc}{\delta_{\sigmaj, \alpha}}
\newcommand{\td}{\delta_{\sigmaj, \beta}}
\newcommand{\tcr}{\delta'_{\sigmaj, \alpha}}
\newcommand{\tdr}{\delta'_{\sigmaj, \beta}}

\newcommand{\lija}{L_{\sigmai,\sigmaj}^{\bra{\alpha}}}
\newcommand{\lijb}{L_{\sigmai,\sigmaj}^{\bra{\beta}}}
\newcommand{\lia}{L_{\sigmai}^{\bra{\alpha}}}
\newcommand{\lib}{L_{\sigmai}^{\bra{\beta}}}
\newcommand{\lja}{L_{\sigmaj}^{\bra{\alpha}}}
\newcommand{\ljb}{L_{\sigmaj}^{\bra{\beta}}}

\newcommand{\psigmai}{P_{\sigmai}}
\newcommand{\psigmaj}{P_{\sigmaj}}
\newcommand{\psigmaij}{P_{\sigmai,\sigmaj}}
\newcommand{\pmu}{P_{\mu}}
\newcommand{\psigma}{P_{\sigma}}

\newcommand{\group}{b}
\newcommand{\groupofindex}{\mathcal{I}_{\group}}
\newcommand{\groupofindexi}[1]{\mathcal{I}_{\group_{#1}}}
\newcommand{\groupset}{\mathcal{B}}
\newcommand{\groupsize}{\beta}

\newcommand{\norm}[2]{\lVert #2 \rVert_{#1}}
\newcommand{\normnum}{p}

\newcommand{\odataset}{\mathcal{X}}
\newcommand{\rdataset}{\mathcal{X'}}

\newcommand{\sdistortionnotation}{\tilde{\Delta}_{\mechanismnotation}}

\newcommand{\sprivacynotationunion}{\tilde{\Pi}^{\text{union}}_{\privacythreshold,\paramdistribution}}
\newcommand{\sprivacynotationinter}{\tilde{\Pi}^{\text{inter}}_{\privacythreshold,\paramdistribution}}
\newcommand{\sprivacynotationuoi}{\tilde{\Pi}^{\text{group}}_{\privacythreshold,\paramdistribution}}
\newcommand{\sprivacynotationlp}{\tilde{\Pi}^{l_p}_{\privacythreshold,\paramdistribution}}
\newcommand{\sprivacynotationlpi}[1]{\tilde{\Pi}^{l_{#1}}_{\privacythreshold,\paramdistribution}}

\newcommand{\secretset}{\mathcal{G}}

\newcommand{\unionlower}{\Delta^{\text{union}}}
\newcommand{\interlower}{\Delta^{\text{inter}}}

\newcommand{\lplower}{\Delta^{l_\normnum}}

\newcommand{\secretsestimateof}[1]{\secretestimatenotationvector\bra{#1}}
\newcommand{\secretsof}[1]{\secretsnotation\bra{#1}}
\newcommand{\secretsofparam}{\secretsof{\rvparamnotation}}
\newcommand{\privacythresholdl}{\varepsilon_{\normnum}}
\newcommand{\privacythresholdlp}[1]{\varepsilon_{#1}}

\newcommand{\xmudim}[1]{\hat{\mu}_x^{\bra{#1}}}
\newcommand{\ymudim}[1]{\hat{\mu}_y^{\bra{#1}}}

\newcommand{\uoi}{group secrets privacy}

\newcommand{\UoI}{Group Secrets Privacy}

\newcommand{\cmax}{c_{2}}
\newcommand{\cmin}{c_{1}}
\newcommand{\ceps}{c_{\epsilon}}
\newcommand{\opt}{\Delta_{opt}}

\newcommand{\constant}{c_{\epsilon,\seclen}}

\newcommand{\cmnt}[1]{\ignorespaces}  %

\newcommand{\privacynotationinter}{\Pi^{\text{inter}}_{\privacythreshold,\paramdistribution}}
\newcommand{\privacynotationunion}{\Pi^{\text{union}}_{\privacythreshold,\paramdistribution}}
\newcommand{\privacynotationnorm}{\Pi^{l_{\normnum}}_{\privacythreshold,\paramdistribution}}
\newcommand{\privacynotationalp}{\Pi^{l_{\alpha}}_{\privacythreshold,\paramdistribution}}
\newcommand{\privacynotationtau}{\Pi^{l_{\tau}}_{\privacythreshold,\paramdistribution}}

\newcommand{\supdist}{\sup_{\substack{\rvparamnotation\in\support{\paramdistribution}, \\
    \releaservparamnotation\in\support{\mechanismof{\rvparamnotation}}}}}

\newcommand{\tmpprivacynotation}{\Pi^{\text{uni}}_{\privacythreshold,\paramdistribution}}

\usepackage[normalem]{ulem} 
\newcommand\redout{\bgroup\markoverwith{\textcolor{red}{\rule[.5ex]{2pt}{0.4pt}}}\ULon}
\newcommand\oliveout{\bgroup\markoverwith{\textcolor{olive}{\rule[.5ex]{2pt}{0.4pt}}}\ULon}

\newcommand{\neuripsdelete}[1]{}

\newcommand{\cmark}{\ding{51}}
\newcommand{\xmark}{\ding{55}}

\title{Guarding Multiple Secrets: Enhanced Summary Statistic Privacy for Data Sharing}

\def\addr{\small\it}%
\def\email{\hfill\small\it}%
\def\name{\normalsize\bf}%
\def\name{\normalsize\bf}%

\author{
      \name {Shuaiqi Wang}$^{\ast}$ \qquad
      \name Rongzhe Wei$^{\dagger}$ \\ \\
      \name Mohsen Ghassemi$^{\diamond}$ \quad
      \name Eleonora Krea\v{c}i\'{c}$^{\diamond}$ 
      \quad
      \name Vamsi K. Potluru$^{\diamond}$ \\ \\
      \addr $^{\ast}$ECE, Carnegie Mellon University
      \quad \email shuaiqiw@andrew.cmu.edu \\
      \addr $^{\dagger}$ECE, Georgia Institute of Technology \quad
      \email rongzhe.wei@gatech.edu \\
      \addr $^{\diamond}$J.P. Morgan AI Research  \quad
      \email \{mohsen.ghassemi, eleonora.kreacic, vamsi.k.potluru\}@jpmchase.com 
      }
\date{}

\begin{document}
\maketitle

\etocdepthtag.toc{mtchapter}
\etocsettagdepth{mtchapter}{subsection}
\etocsettagdepth{mtappendix}{none}

\begin{abstract}
    Data sharing enables critical advances in many research areas and business applications, but it may lead to inadvertent disclosure of sensitive summary statistics (e.g., means or quantiles). %
    Existing {literature only} focuses on protecting a single confidential quantity, %
    while in practice, {data sharing involves multiple sensitive statistics}. %
    We propose a novel 
    framework to define, analyze, and protect \textit{multi-secret summary statistics privacy} in data sharing.
    Specifically, we measure the privacy risk of any data release mechanism by the worst-case probability of an attacker successfully inferring summary statistic secrets. %
    Given an attacker's objective spanning from inferring a subset to the entirety of summary statistic secrets, we systematically design and analyze tailored privacy metrics.
    Defining the distortion as the worst-case distance between the original and released data distribution, we analyze the tradeoff between privacy and distortion. 
    Our contribution also includes designing and analyzing data release mechanisms tailored for different data distributions and secret types.
    Evaluations on real-world data demonstrate the effectiveness of our mechanisms in practical applications.
\end{abstract}

\section{Introduction}

Data sharing has become integral to collaborative research and industrial applications~\cite{lee2000information,zheng2020privacy,potluru2024synthetic}.  However, in many domains such as finance and healthcare, as well as in industrial settings,  the sensitive nature of data is cause for great concern regarding data privacy and leakage of sensitive information. If appropriate measures are not adopted, data sharing may reveal sensitive information not only about particular individuals \cite{abay2019privacy}, but also aggregate information at dataset level \cite{kawamoto2019local,tillman2023privacy,lin2023summary}. 
Summary statistics of the data (e.g., mean, standard deviation, histograms, etc.) may contain sensitive or proprietary information that the data holder needs to protect due to legal or competitive risks. 
\cmnt{While summary statistics of the data (e.g., mean, standard deviation, histograms, etc.) may not be inferred from any particular record, %
it may contain sensitive or proprietary information that the data holder needs to protect due to legal or competitive risks. \EK{For the first part of the sentence do we want to say instead that while summary may not identify any particular record? So we are already hinting this is different than DP.} %
}

In recent years, protecting the privacy of individuals has garnered significant attention, %
leading to the continued development of prominent frameworks such as differential privacy and its many variants \cite{dwork2006calibrating,kifer2014pufferfish}. In contrast, there remains a noticeable lack of research on quantifying the summary statistics leakage of shared datasets. Protecting sensitive summary statistics is especially crucial in industrial settings due to the proprietary nature of such information.
Companies, when publishing data on their performance, customer base, or equipment expenditure, are often concerned with leaking sensitive summary statistics that may reveal the firm's typical client characteristics, market segments the firm is most active in, or future business plans. Despite these risks, in many scenarios it is in the data owner's interest, as well as the public interest, to ``liberate'' data to be shared safely with other entities.
One prominent example is when a financial company shares data publicly (e.g. for academic collaborations and furthering potential technological advances in areas such as financial fraud detection and anti-money laundering).  %
Similar scenarios exist for companies in many other industries that can benefit from publishing their data for research purposes or sharing their data with consulting firms. %
In many cases, the data holder may only deem certain (and not all) summary statistics as sensitive. Moreover, as with notions of individual privacy, data holders are often comfortable with tolerating a ``sufficiently low'' leakage risk. 
This highlights the need for a privacy framework to quantify the risks of revealing summary statistics and allow for designing data release mechanisms that meet the data holder's requirements.

Recently, \cite{lin2023summary} proposed a novel privacy framework
designed to identify and analyze concerns related to summary statistics privacy. They focus {solely} on the scenario where only one summary statistic is deemed confidential by the data owner.  
In real-world scenarios, however, one often deals with high-dimensional data where potentially multiple %
summary statistics %
may contain sensitive information or be regarded as proprietary information. 
Hence, there is a compelling need to expand the current framework, ensuring it encompasses cases with multiple confidential statistics. 

The privacy metric and analysis in \cite{lin2023summary} cannot be trivially extended to address the multi-secret case. 
Given the varied data sharing contexts and the intricacies of summary statistics secrets, data holders' privacy requirements can differ. Some scenarios require \textit{none of the secrets} being disclosed, while others may only demand \textit{not all secrets} being revealed simultaneously. %
To this end, we propose a framework to quantify, analyze and protect \textit{multi-secret summary statistics privacy} under different protection scenarios.

Our proposed framework includes \textit{interpretable} privacy metrics that have clear operational meaning: {the probability that an attacker successfully (i.e. within certain error tolerance range)} guesses the secrets. 
It also includes a utility metric to evaluate the usefulness of the released data. Based on these metrics, we provide privacy-utility trade-off analyses to obtain the fundamental limits of what is achievable for any release mechanism in terms of utility for a given privacy budget, or vice versa. Understanding these limits informs mechanism design as it provides a benchmark to evaluate the ``optimality'' of any release mechanism. %

\subsection{Main contributions}\label{sec:contribution}
\myparatightestn{Metric Design (\cref{sec:formulations}, \cref{sec:alternatives})} 
We introduce and analyze several novel interpretable metrics to measure the privacy risks of data release mechanisms for multi-secret protection. %
In \cref{sec:metrics}, we tackle the strictest scenario by defining the privacy metric as the worst-case probability of an attacker correctly guessing any single secret within a specified tolerance range. %
Subsequently, in \cref{sec:alternatives}, we explore %
relaxed scenarios %
wherein the data holder aims to thwart the attacker from correctly guessing %
either a group of or the entirety of secrets simultaneously, %
\neuripsdelete{or aims to ensure a satisfactory distance (in $l_p$ norm sense) \mg{edit?} between the original and the attacker guessed secret vector,}establishing the appropriate privacy metric for each.  %

\noindent\textbf{Privacy-Distortion Tradeoff Analysis (\cref{sec:tradeoffs}, \cref{sec:alternatives})} We provide the general lower bounds on the distortion %
given a certain constraint on the privacy corresponding to various privacy metrics. Those lower bounds are non-trivial extensions of the tradeoff analysis in \cite{lin2023summary}. %
The analysis of the lower bound is general, but its value depends on data distributions and secret types. The bound also reveals how the number of secrets affect the privacy-distortion tradeoffs under different privacy metrics. We then specify the \cmnt{value of the distortion} lower bound under several specific data distributions and secret types. %

\myparatightestn{Mechanism Design and Empirical Evaluation (\cref{sec:case_study}, \cref{sec:experiments})} %
Our privacy framework is applicable to any data release mechanism, and can be used to guide designing mechanisms that effectively protect against summary statistics attacks. To this end, we develop mechanisms tailored to various data distributions and secret types and subsequently assess their performance. At a high level, these mechanisms function by quantizing the distribution parameters into bins and then randomly releasing points within the respective bins into which the parameters fit. Such mechanisms are straightforward to implement and have near-optimal privacy-distortion performance. 
We employ real-world datasets to demonstrate the effectiveness of our designed data release mechanisms and to illustrate the variations in privacy-distortion tradeoffs across different privacy metrics. %

\section{Related Work} 
\label{sec:related_work}
In this section, we provide a brief overview of related work and highlight the main distinctions between our work and prior works in \cref{table:compare}. A more detailed discussion can be found in \cref{app:related_work}.  %

\myparatightestn{Heuristics}
Although commonly adopted in many industries for data sharing~\cite{hundepool2010handbook}, heuristic methods often lack rigorous privacy guarantees and can be vulnerable in real-world scenarios \cite{elliot1999scenarios}. Particularly, methods such as subsetting~\cite{reiss2012obfuscatory}, culling~\cite{reiss2012obfuscatory}, and de-identification~\cite{garfinkel2015identification} are susceptible to re-identification or unintentional data property leakage \cite{narayanan2006break,sweeney2013matching}. %

\myparatightestn{Differential Privacy (DP) \cite{dwork2006calibrating}} This notion of privacy ensures indistinguishability between neighboring datasets. %
{Several works have applied DP to release summary statistics \cite{barak2007privacy,qardaji2014priview}. However, as a metric, DP is concerned with privacy guarantees in terms of} protecting individual record contributions rather than overarching distributional statistical properties. In other words, it provides no metric to quantify risks of leaking aggregate dataset-level information. Moreover, 
certain DP mechanisms like the Laplacian \cite{dwork2006calibrating} 
introduce zero-mean noise to samples but often leave certain integral statistics (e.g. means) less affected, rendering them unsuitable for directly protecting summary statistic privacy. %

\myparatightestn{Other Indistinguishability Approaches}  {Another relevant line of work is}  \textit{attribute privacy}~\cite{zhang2022attribute} which aims to safeguard specific dataset properties. %
However, their applications do not always align with data sharing scenarios as those methods only output certain statistical queries.
Moreover, paradigms like \textit{distribution privacy} \cite{kawamoto2019local} and \textit{distribution inference} \cite{suri2021formalizing,suri2023dissecting} prioritize preserving the confidentiality of statistical secrets (e.g., mean). Their robust nature, ensuring indistinguishability {across a broad range of distributions (e.g., a Dirac delta distribution and a Gaussian distribution with the same mean),}
{may result in adding excessive amount of noise,} leading to reduced data utility.

\myparatightestn{Leakage-Based Methods}
Many works utilize information-theoretic methods to delineate and safeguard statistical privacy, typically balancing between limiting the disclosure of private data and promoting the release of non-sensitive data. They often characterize the exposure of confidential information through the concept of leakage \cite{alvim2014additive}.%
Various measures, including Shannon entropy \cite{rassouli2021perfect}, %
min-entropy \cite{asoodeh2017privacy}, %
and gain function \cite{m2012measuring} have been employed to define leakage. A notable advancement in this area is the concept of \textit{maximal leakage} \cite{issa2019operational}, which quantifies the increased likelihood of correctly inferring secrets post-data release. %
{In contrast to our setting,} maximal leakage and its generalizations, such as \cite{gilani2023alpha},  %
{assume that} the secrets are not known {to the data holder} in advance.

\myparatightestn{Summary Statistic Privacy}
The recently introduced summary statistic privacy \cite{lin2023summary} aligns closely with our objective, emphasizing the protection of the dataset's summary statistics during data sharing. In their approach, privacy is measured by the worst-case likelihood of an adversary accurately discerning the secret within a defined range. However, their framework %
focuses on safeguarding a singular secret under one-shot attack and confines the data analysis to one dimension, which may not cater comprehensively to many practical applications. %
{Our work is a non-trivial extension of the framework in \cite{lin2023summary} to guard multiple secrets (see \cref{sec:contribution} for details).} %

\begin{table}[!t]
    \centering
    \caption{Comparison with prior privacy frameworks  %
    }
    \resizebox{\linewidth}{!}{
    \begin{tabular}{c|c|c|c}
    \toprule
         &  secret to protect & privacy-distortion analysis & suitable to data sharing\\
    \midrule
        Differential Privacy \cite{dwork2006calibrating}  & individual/group information  &\textcolor{blue}{\cmark} &  \textcolor{blue}{\cmark}\\
        Distribution Privacy \cite{kawamoto2019local} &   overall distribution   & \xmark & \textcolor{blue}{\cmark} \\
        Attribute Privacy \cite{zhang2022attribute}  & single summary statistic  & \xmark & \xmark \\
         Summary Statistic Privacy \cite{lin2023summary} & single summary statistic   & \textcolor{blue}{\cmark} & \textcolor{blue}{\cmark} \\
     \midrule
        Our Framework & \textcolor{blue}{multiple summary statistics}   & \textcolor{blue}{\cmark} & \textcolor{blue}{\cmark} \\
    \bottomrule
    \end{tabular}
    }
    \label{table:compare}
\end{table}
\section{Problem Formulation}
\label{sec:formulations}
\myparatightestn{Notation}
Let $\rvprivatenotation$ denote any random variable. Its distribution measure is represented by $\distributionof{\rvprivatenotation}$. %
When $\rvprivatenotation$ is part of a parametric family, represented by a parameter $\theta$ that lies in $\mathbb{R}^q$ ($q\geq 1$), %
our notation becomes more specific: $\rvprivate$ for the random variable and $\distributionof{\rvprivate}$ for its distribution.
Additionally, when considering $\theta$ not as a fixed value but as a realization of another random variable, denoted by $\Theta$, its distribution is captured by $\distributionof{\Theta}$. 
Note that $\distributionof{\Theta}$ acts as the prior distribution of parameter $\theta$.

\myparatightestn{Original Data and Summary Statistic Secrets to Protect}
Consider a data holder in possession of a dataset, denoted as $\privatedataset = \{x_1, x_2, \ldots, x_m\}$, comprising $m$ i.i.d. samples drawn from a certain distribution. Given the ability to represent diverse datasets using parametric generative models, we posit that this distribution belongs to a parametric family characterized by a parameter $\theta \in \mathbb{R}^q$. %
The data holder aims to hide $\dimension$ summary statistic secrets from the original data distribution $\distributionof{\rvprivate}$. We can express the secrets as $\dimension$ functions $\secretsnotation = \brb{\secreti{1}, \secreti{2}, \cdots, \secreti{\dimension}}$, where $\secretiofparam{\secretindex}: \mathbb{R}^q \rightarrow \mathbb{R}$ for each function $\secreti{\secretindex}$.

\myparatightestn{Data Release Mechanism}
To release data, the data holder passes the original distribution parameter $\theta$ through the data release mechanism $\mathcal{M}_\secretnotation$. The released data distribution parameter $\theta'$ satisfies $\theta' \sim \mathcal{M}_\secretnotation\bra{\theta}$. %

\myparatightestn{Threat Model}
We assume the attacker knows the parametric family to which the data generating distribution belongs, but does not know the distribution parameter $\theta$. The attacker also knows the released parameter $\theta'$ and the mechanism $\mathcal{M}_{\secretnotation}$, but does not know the realization of the internal randomness of the mechanism. Based on the released parameter $\theta'$, the attacker guesses the secrets $\secretsnotation(\theta)=\brb{\secretiofparam{1}, \secretiofparam{2}, \cdots, \secretiofparam{\dimension}}$ by strategies $\secretestimatenotationvector(\theta') = \brb{\secretestimatenotation_{1}(\theta'), \secretestimatenotation_{2}(\theta'), \cdots, \secretestimatenotation_{\dimension}(\theta')}$ %

\subsection{Metrics for Preserving Multiple Secrets}
\label{sec:metrics}

We define the privacy and distortion metrics and formulate the data holder's objective as follows.

\myparatightestn{Privacy Metric} Consider the case where the data holder aims to prevent attackers from guessing \textit{any} secret correctly. %
We define the \emph{union privacy} metric $\privacynotationuni$ as the probability of the attacker guessing any secret within a \textit{tolerance range}, $\privacythresholdi{\secretindex}$ for secret $g_i$, employing the best attack strategy:
\begin{align}
    \privacynotationuni  \triangleq ~\sup_{\secretestimatenotationvector} ~\mathbb P\Big(\bigcup_{\secretindex\in [\dimension]}~ \abs{\secretiestimateof{\secretindex}{ \releaservparamnotation}- \secretiofparam{\secretindex} } \leq \privacythresholdi{\secretindex}\Big) ~,
    \label{eq:privacy}
\end{align}
\cmnt{
\begin{align}
    \Pi^{\text{inter}}_{\epsilon}  \triangleq ~\sup_{\hat s} ~\probof{\bigcap_{\secretindex\in [\dimension]}~ \abs{\hat s_i- s_i}  \leq \privacythresholdi{\secretindex} }
\end{align}

\begin{align}
    \Pi_{\epsilon}  \triangleq ~\sup_{\hat s} ~\probof{ \abs{\hat s- s}  \leq \epsilon}
\end{align}
}
where the probability is taken over the randomness of the original data distribution parameter $\theta$, the data release mechanism $\mathcal{M}_\secretnotation$, and the attacker strategy $\secretestimatenotationvector$. %

Union privacy is the \textit{strictest} privacy metric for the data holder, as the attacker guessing any secret successfully will result in protection failure. %
In \cref{sec:alternatives}, we introduce alternative privacy metrics. %

Union privacy also accommodates \textit{multi-shot attack} scenario where the attacker guesses the secret multiple times, and the data holder aims to prevent success in any guess. In this scenario, $g_1 = g_2=\cdots=g_d$ and $\secretestimatenotationvector$ represents sequential strategies guessing the secret. %

\myparatightestn{Distortion Metric} 
Since the goal of data sharing is to maintain high utility of the disseminated data, it is important to discern the extent to which the released data diverges from the original. In this context, we introduce the concept of \emph{distortion}, denoted as $\distortionnotation$. Specifically, the distortion of a mechanism is characterized by the worst-case discrepancy between the original distribution and the released distribution:
\begin{align}
    \distortionnotation \triangleq \sup_{\substack{\rvparamnotation\in\support{\paramdistribution}, \\
    \releaservparamnotation\in\support{\mechanismof{\rvparamnotation}}}}\distanceof{\privatedistribution} {\releasedistribution},
    \label{eq:distortion}
\end{align}

where $\support{\cdot}$ is the support of the distribution and $\mathfrak{D}$ can be any general distance metric defined over distributions. In this paper, we specify the distance metric as Wasserstein-2 distance, as it is widely adopted in data quality estimation (e.g., \cite{korotin2021neural}).

\myparatightestn{Objective}
The data holder's objective is to choose a data release mechanism that minimizes \distortion{} metric $\distortionnotation{}$  subject to a privacy budget constraint on $\privacynotationuni $:
\begin{align}
    \begin{split}
    \min_{\mechanismnotation} & \quad \distortionnotation
    \quad\quad\quad
    \text{subject to} \quad \privacynotationuni  \leq \privacybound.
    \end{split}
    \label{eq:opt}
\end{align}

\section{General Lower Bound on \Privacy{}-\Distortion{} Tradeoffs}
\label{sec:tradeoffs}

Given the metrics defined in \cref{sec:metrics} and a \privacy{} budget $\privacybound$, we present a lower bound on \distortion{}  that applies to any data distribution and any secret $\secretsnotation = \brb{\secretnotation_1, \secretnotation_2, \cdots, \secretnotation_\dimension}$. %

\begin{theorem}[Lower bound of \privacy{}-\distortion{} tradeoff]
\label{thm:trade_off_general}
Let $\auxdistance{\rvprivatewithparam{\rvparamnotation_1}}{\rvprivatewithparam{\rvparamnotation_2}} = \distanceformula{\rvprivatewithparam{\rvparamnotation_1}}{\rvprivatewithparam{\rvparamnotation_2}}$.
Further, let
$R^{\text{union}}({\rvprivatewithparam{\rvparamnotation_1}},{\rvprivatewithparam{\rvparamnotation_2}}) = \prod_{\secretindex\in [\dimension]}\rangeiformula{\secretindex}{{\rvparamnotation_1}}{{\rvparamnotation_2}}^{1/\dimension}$ and 
\begin{align} 
\ratio^{\text{union}} = \inf_{\rvparamnotation_1, \rvparamnotation_2 \in\support{\paramdistribution}}
\frac{\auxdistance{\rvprivatewithparam{\rvparamnotation_1}}{\rvprivatewithparam{\rvparamnotation_2}}}{R^{\text{union}}({\rvprivatewithparam{\rvparamnotation_1}},{\rvprivatewithparam{\rvparamnotation_2}})}.
\label{eq:gamma}
\end{align}
{Then, for tolerance ranges $\{\epsilon_i\}$ as defined in \cref{eq:privacy} and for any mechanism $\mathcal M_g$ subject to privacy budget $\privacynotation^{\text{union}} \leq \privacybound<1$, we have}
\begin{align}
\label{eqn:distortion_lower}
\distortionnotation  > 
2 
\ratio^{\text{union}}  \Big\lceil{\frac{1}{{1-\bra{1-\privacybound}^{{1}/{\dimension}}}}-1}\Big\rceil  \Big(\prod\nolimits_{\secretindex\in [\dimension]} \privacythresholdi{\secretindex}\Big)^{{1}/{\dimension}}.
\end{align}
\end{theorem}

The proof is shown in \cref{proof:trade_off_general}, and we provide the proof sketch as below.
\cref{thm:trade_off_general} shows that the lower bound is proportional to the geometric mean of the tolerance ranges $\privacythresholdi{1}, \cdots,  \privacythresholdi{\dimension}$, and is \textit{inversely} proportional to the privacy budget $\privacybound$. 
The ratio $\ratio^{\text{union}}$ acts as a conversion parameter that bridges the difficulty of guessing the secrets and the distributional disparity, and its value depends on secret types and data distribution.
The impact of the secret number $\dimension$ on the lower bound depends on the characteristics of secrets (i.e., $\ratio^{\text{union}}$) and their tolerance ranges (i.e., the geometric mean of tolerance ranges). However, as the ceiling term in \cref{eqn:distortion_lower} increases with the growth of secret number, achieving a lower value for the lower bound becomes much more challenging with a larger number of secrets. %
This aligns with intuition: with more secrets, the attacker can more easily succeed in guessing at least one secret. For the multi-shot attack scenario, where $g_1=\cdots=g_d$ and $\epsilon_1=\cdots =\epsilon_d$, the lower bound increases as the attacker's trial count $\dimension$ grows.

\begin{wrapfigure}{l}{0.6\textwidth}
    \centering
    \includegraphics[width=\linewidth]{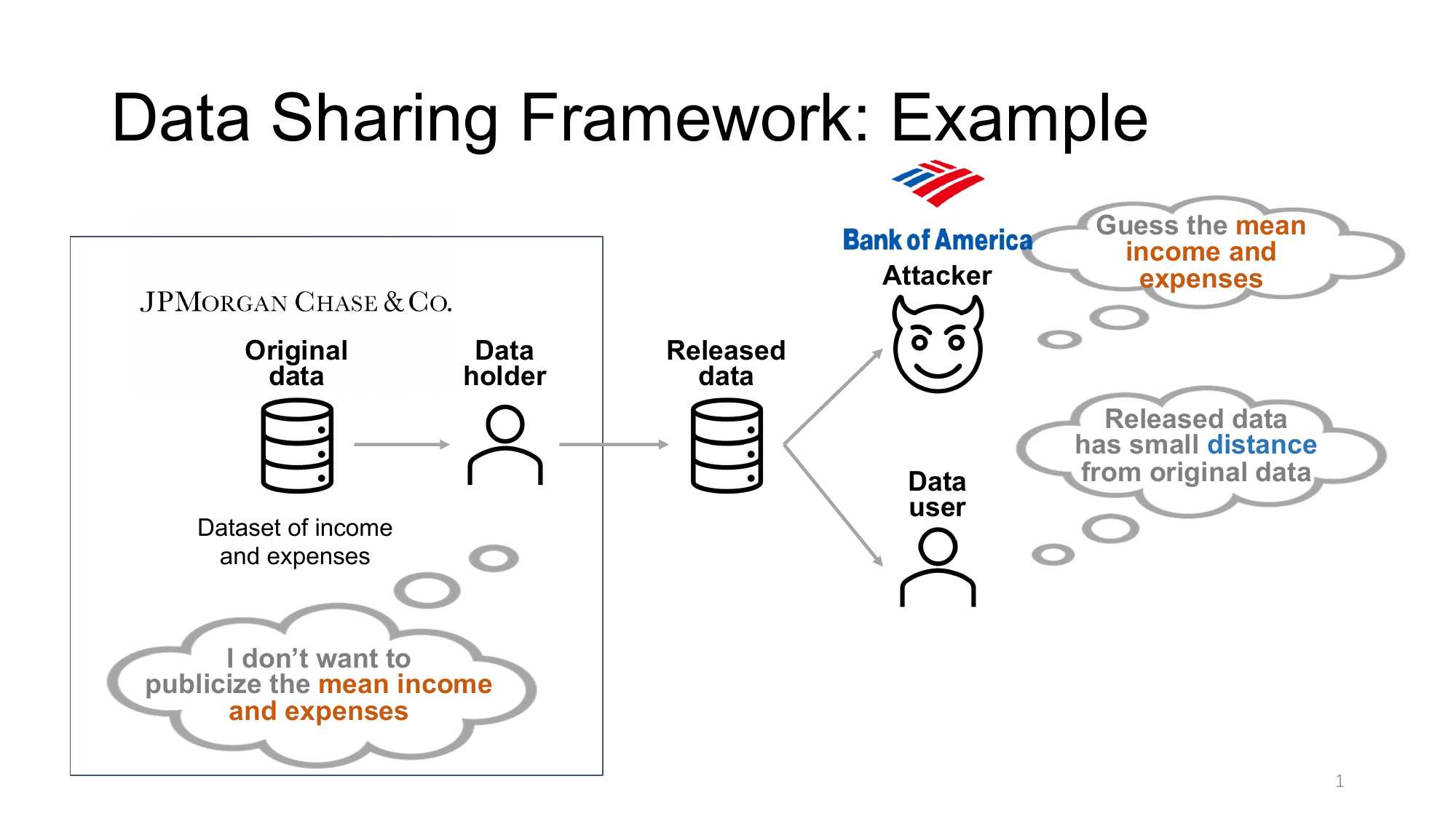}
    \caption{%
    Attacker construction for proof of \cref{thm:trade_off_general} under the $2$-secret case. The true value of secret vector lies in the intersection of the highlighted bands.}
\label{fig:proof_sketch}
\end{wrapfigure}

\myparatightestn{Proof Sketch}
We prove the tradeoff lower bound \cref{eqn:distortion_lower} %
by constructing a sequence of attackers {(agnostic of data distribution and release mechanism)}, such that some of them can successfully guess at least one secret. We take the $2$-secret case (i.e., $d=2$) as an example, as illustrated in \cref{fig:proof_sketch}. For each secret $g_i(\theta)$, we partition the range of possible secret values into $N_i$ segments of length $2\privacythresholdi{i}$ and design $N_i$ individual-secret attack strategies $\secretestimatenotation_{i}^{\bra{j}}$ ($j\in\brb{N_i}$), each guessing the midpoint of a segment. We subsequently formulate multi-secret attack strategies $\secretestimatenotationvector^{(j,k)}$ ($j\in\brb{N_1}, k\in\brb{N_2}$) by combining individual-secret strategy $\secretestimatenotation_{1}^{\bra{j}}$ for secret $g_1(\theta)$ and $\secretestimatenotation_{2}^{\bra{k}}$ for secret $g_2(\theta)$. {Note that none of these attacks depends on a released parameter $\theta'$: the sequence is solely determined by the range of the secret.} The yellow $\secretsnotation(\theta)$ region in \cref{fig:proof_sketch} represents where the attacker $\secretestimatenotationvector^{(2,2)}$ correctly guesses at least one secret within the tolerance range. %
We then establish the distortion lower bound based on the privacy constraint that the attack success rate is at most $\privacybound$ and by utilizing the conversion parameter $\ratio^{\text{union}}$, which serves as a linkage between the distributional distance and possible ranges of secrets.

\section{Alternative \Privacy{} Metrics and Analysis} %
\label{sec:alternatives}

In \cref{sec:metrics}, we define the privacy metric for the worst case: %
attacker guessing any of the secrets within the tolerance range will result in the failure of secret protection. In practice, sensitive information is sometimes significantly compromised only when the attacker successfully guesses all or a group of secrets. %
Consider, e.g., a dataset containing the ingredients purchase logs belonging to a pharmaceutical (or food product) company. %
The company may be concerned about its trade secrets (e.g. drug formula or recipe) being leaked based on the average monthly purchase amounts for the ingredients. However, in this case the company may consider its trade secret revealed only if all  (or most) of the ingredient ratios are disclosed, rather than the absolute values for one or a few ingredients.
In this section, we propose alternative privacy metrics that apply to such scenarios. We discuss a further variant metric ($l_p$ privacy metric) in \cref{app:lp_norm}.%

\subsection{Intersection Summary Statistic Privacy}
We first consider the scenario where secrets are %
{considered} compromised %
{only} when the attacker successfully guesses \textit{all} of them simultaneously. %
We define the \emph{intersection privacy} metric as the probability of attacker guessing all \secrets{} within their respective tolerance ranges, $\privacythresholdi{\secretindex}$ for secret $\secretnotation_i$, employing the best attacker strategy: %
\begin{align}
    \privacynotationinter  \triangleq ~\sup_{\secretestimatenotationvector} ~\probof{\bigcap\nolimits_{\secretindex\in [\dimension]}~ \abs{\secretiestimateof{\secretindex}{ \releaservparamnotation}- \secretiofparam{\secretindex} } \leq \privacythresholdi{\secretindex} }~.
    \label{eq:privacy_intersection}
\end{align}
The following theorem presents a lower bound on distortion, given intersection \privacy{} budget $\privacybound$.  %

\begin{theorem}[Lower bound of \privacy{}-\distortion{} tradeoff for intersection privacy]
\label{thm:trade_off_intersection}
Let $\auxdistance{\rvprivatewithparam{\rvparamnotation_1}}{\rvprivatewithparam{\rvparamnotation_2}} = \distanceformula{\rvprivatewithparam{\rvparamnotation_1}}{\rvprivatewithparam{\rvparamnotation_2}}$.
Further, let
$R^{\text{inter}}({\rvprivatewithparam{\rvparamnotation_1}},{\rvprivatewithparam{\rvparamnotation_2}})  = \frac{1}{\dimension} \sum_{\secretindex\in [\dimension]}\rangeiformula{\secretindex}{{\rvparamnotation_1}}{{\rvparamnotation_2}}$ and 
\begin{align} 
\ratio^{\text{inter}} = \inf_{\rvparamnotation_1, \rvparamnotation_2 \in\support{\paramdistribution}}
\frac{\auxdistance{\rvprivatewithparam{\rvparamnotation_1}}{\rvprivatewithparam{\rvparamnotation_2}}}{R^{\text{inter}}({\rvprivatewithparam{\rvparamnotation_1}},{\rvprivatewithparam{\rvparamnotation_2}}) }.
\label{eq:gamma_intersection}
\end{align}
{Then, for tolerance ranges $\{\epsilon_i\}$ as defined in \cref{eq:privacy_intersection} and for any mechanism $\mathcal M_g$ subject to privacy budget $\privacynotationinter \leq \privacybound<1$, we have}
\begin{align}\label{eqn:distortion_lower_inter}
\distortionnotation  > 
2\gamma^{\text{inter}}\bra{\Big\lceil{\frac{1}{\privacybound}}\Big\rceil^{1/\dimension} \Big(\prod\nolimits_{\secretindex\in [\dimension]} \privacythresholdi{\secretindex}\Big)^{{1}/{\dimension}} - \frac{1}{\dimension}\sum\nolimits_{\secretindex\in [\dimension]} \privacythresholdi{\secretindex}}.
\end{align}
\end{theorem}

(Proof in \cref{prooof:trade_off_intersection}.)
As the secret number $\dimension$ increases, \cref{thm:trade_off_intersection} shows that achieving a lower value for the distortion lower bound becomes easier, which aligns with intuition: with more secrets, it becomes increasingly challenging for the attacker to succeed in guessing all secrets. %
Since intersection privacy is the least strict privacy metric for the data holder, its least achievable distortion is no greater than that for union privacy, as demonstrated in \cref{prop:distortion_bound_compare} (proof in \cref{proof:distortion_bound_compare_inter}).

\begin{proposition}
\label{prop:distortion_bound_compare}
Given a privacy budget $\privacybound$ and tolerance ranges $\privacythresholdi{1}, \cdots, \privacythresholdi{\dimension}$, we have $\unionlower \geq \interlower$, where $\unionlower$ and $\interlower$ are the achievable distortion lower bounds for union privacy and intersection privacy.
\end{proposition}

\subsection{\hspace{-1mm}Group Secrets Summary Statistic Privacy}
\label{sec:UoI}

We then consider the case where the data holder divides secrets into distinct groups, aiming to thwart the attacker from successfully guessing an entire group of secrets. %
We define the \emph{\uoi{}} metric as the probability of attacker guessing any disjoint group $\group = \brc{\secreti{\secretindex}}_{\secretindex\in \groupofindex} \in \groupset$ of \secrets{} to within tolerance ranges, $\privacythresholdi{\secretindex}$ for secret $g_i$, adopting the best attack strategy: 
\begin{align}\label{eq:privacy_group}
    \privacynotationgroup  \triangleq ~\sup_{\secretestimatenotationvector} ~\probof{\bigcup\nolimits_{\group\in\groupset}\bra{\bigcap\nolimits_{\secretindex\in \groupofindex}~ \abs{\secretiestimateof{\secretindex}{ \releaservparamnotation}- \secretiofparam{\secretindex} } \leq \privacythresholdi{\secretindex} }}~,
\end{align}
where the secret index set $\groupofindex$ satisfies $\groupofindexi{1}\cap \groupofindexi{2} = \emptyset$ for any distinct group $\group_{1}, \group_{2} \in \groupset$. %
Note that when $\abs{\groupset} = 1$, \uoi{} reduces to intersection privacy, the least strict privacy metric. On the other hand, for $\abs{\groupset} = \dimension$, \uoi{} is equivalent to union privacy, the strictest metric.

Under \uoi{}, given a privacy budget $\privacybound$, we then present a general lower bound on \distortion{}.

\begin{theorem}[Lower bound of \privacy{}-\distortion{} tradeoff for \uoi{}]
\label{thm:trade_off_mix}
Let $\auxdistance{\rvprivatewithparam{\rvparamnotation_1}}{\rvprivatewithparam{\rvparamnotation_2}} = \distanceformula{\rvprivatewithparam{\rvparamnotation_1}}{\rvprivatewithparam{\rvparamnotation_2}}$.
Further, let
$R^{\text{group}}({\rvprivatewithparam{\rvparamnotation_1}},{\rvprivatewithparam{\rvparamnotation_2}}) = \frac{1}{\dimension} \sum_{\secretindex\in [\dimension]}\rangeiformula{\secretindex}{{\rvparamnotation_1}}{{\rvparamnotation_2}}$ and 
\begin{align} 
\ratio^{\text{group}} \triangleq \inf_{\rvparamnotation_1, \rvparamnotation_2 \in\support{\paramdistribution}}
\frac{\auxdistance{\rvprivatewithparam{\rvparamnotation_1}}{\rvprivatewithparam{\rvparamnotation_2}}}{R^{\text{group}}({\rvprivatewithparam{\rvparamnotation_1}},{\rvprivatewithparam{\rvparamnotation_2}}) }.
\label{eq:gamma_mix}
\end{align}
{Then, for tolerance ranges $\{\epsilon_i\}$ as defined in \cref{eq:privacy_group} and for any mechanism $\mathcal M_g$ subject to privacy budget $\privacynotationgroup \leq \privacybound<1$, we have}
\footnotesize
\begin{align*}
\distortionnotation \hspace{-0.5mm}>\hspace{-0.5mm} 
2\gamma^{\text{group}} \Bigg(\bigg\lceil{\frac{1}{\bra{1-\bra{1-\privacybound}^{{1}/{\groupsize}}}^{\groupsize/\dimension}}}\bigg\rceil\hspace{-0.5mm} \prod_{\secretindex\in[\dimension]}\privacythresholdi{\secretindex}^{1/\dimension} \hspace{-0.5mm}-\hspace{-0.5mm}\sum_{\secretindex\in [\dimension]} \frac{\privacythresholdi{\secretindex}}{\dimension}\Bigg),
\end{align*}
\normalsize
where $\groupsize$ is the number of groups, i.e., $\groupsize = \abs{\groupset}$.
\end{theorem}

(Proof in \cref{proof:trade_off_mix}.)
We can observe that the distortion lower bound in \cref{thm:trade_off_mix} is proportional to the group number $\groupsize$.

\section{A Case Study: Multivariate Gaussian}
\label{sec:case_study}

In this section, we instantiate the general privacy-distortion tradeoff result from \cref{sec:tradeoffs} on Gaussian distributions with multiple \secrets, %
devise a data release mechanism, and assess its privacy-distortion performance under union privacy. We defer the case studies and mechanism analysis under other privacy metrics in \cref{sec:alternatives} to \cref{app:mech_analysis}.

We focus on $\dimgaussian$-dimensional Gaussian distribution ($\dimgaussian\in \mathbb{Z}^+$) with independent variables (i.e., with diagonal covariance matrix), represented by distribution parameters $\rvparamnotation = \bra{\mu_1, \cdots, \mu_{\dimgaussian}, \sigma_1, \cdots, \sigma_{\dimgaussian}}$. We defer the analysis of the general case of multivariate Gaussian distributions to \cref{app:2dGaussian}. We aim to protect $\dimension$ \secrets{} ($\dimension \leq 2\dimgaussian$), where each \secret{} can represent either mean or standard deviation of any dimension within the distribution, i.e., $\secretnotation_{\secretindex}\in \brc{\mu_1, \cdots, \mu_{\dimgaussian}, \sigma_1, \cdots, \sigma_{\dimgaussian}}, \forall \secretindex\in\brb{\dimension}$. 
We first instantiate the privacy-distortion lower bound for union privacy in \cref{prop:multiGaussian}.

\begin{proposition}
\label{prop:multiGaussian}
    For $\dimgaussian$-dimensional Gaussian distribution with diagonal covariance matrix and distribution parameters $\rvparamnotation = \bra{\mu_1, \cdots, \mu_{\dimgaussian}, \sigma_1, \cdots, \sigma_{\dimgaussian}}$, consider $d$ \secrets{} ($d\leq 2\dimgaussian$), where each secret satisfies $\secretnotation_{\secretindex}(\theta)\in \brc{\mu_1, \cdots, \mu_{\dimgaussian}, \sigma_1, \cdots, \sigma_{\dimgaussian}}$, $\forall\secretindex\in \brb{\dimension}$. %
    {Then, for tolerance ranges $\{\epsilon_i\}$ as defined in \cref{eq:privacy} and for any mechanism $\mathcal M_g$ subject to privacy budget $\privacynotationunion \leq \privacybound<1$, we have}
    \vspace{-1mm}
    \begin{align*}
\distortionnotation> 
\sqrt{\dimension} \cdot \Big\lceil{\frac{1}{{1-\bra{1-\privacybound}^{{1}/{\dimension}}}}-1}\Big\rceil \cdot \Big(\prod\nolimits_{\secretindex\in [\dimension]} \privacythresholdi{\secretindex}\Big)^{{1}/{\dimension}}.
    \end{align*}
\end{proposition}
\vspace{-0.7mm}
(Proof in \cref{proof:multiGaussian}.)
Next, we design a quantization mechanism to approximate the tradeoff lower bound. 
Quantization mechanisms have been shown to achieve (near)-optimal privacy-utility tradeoffs across various privacy pipelines, as evidenced by \cite{lin2023summary,farokhi2021noiseless}.
Intuitively, we partition the ranges of possible secret values into intervals of lengths $\seclen_{g_i}$ for each secret $g_i$, $\secretindex\in\brb{\dimension}$. %
The mechanism randomly outputs %
a point within the intervals where the original secrets reside. Precisely, we provide the mechanism in \cref{mech:dGaussian_diagnol}, where $U$ stands for uniform distribution.

\begin{algorithm}[htpb]
    \LinesNumbered
	\BlankLine
	\SetKwInOut{Input}{Input}
	\caption{Data release mechanism for dimensionally independent multivariate Gaussian with $\dimension$ secrets.}
    \label{mech:dGaussian_diagnol}
	\Input{$\theta = \bra{\mu_1, \cdots, \mu_{\dimgaussian}, \sigma_1, \cdots, \sigma_{\dimgaussian}}$, lower bound $\underline{g_i}$ for secret $g_i$,  quantization interval $s_{g_i}$, $\forall \secretindex \in \brb{\dimension}$.}
	\BlankLine
	\textbf{for} each $\secretindex\in\brb{\dimension}$: $g_i'(\theta) \leftarrow \underline{g_i} + \floor{\frac{\secretiofparam{\secretindex}-\underline{g_i}}{s_{g_i}}}\cdot s_{g_i}+U\bra{0, s_{g_i}}$\;
        Output Gaussian distribution with secret parameter $g_i$ as $g_i'(\theta)$, $\forall\secretindex\in \brb{\dimension}$, and non-secret parameters as the original values.
\end{algorithm}

For the mechanism  performance analysis, we consider the union privacy and assume that the prior distribution of each secret distribution parameter $g_i, i\in\brb{\dimension}$, is uniform.

\begin{proposition}[Mechanism privacy-distortion tradeoff]
\label{prop:multiGaussian_performance}
Assume that secret distribution parameters $g_1, \cdots, g_\dimension$ follow uniform distributions. Let $_{\text{Alg.1}}\privacynotationunion$ denote the union privacy risk of \cref{mech:dGaussian_diagnol}. Then, we have
\begin{align*}
    _{\text{Alg.1}}\privacynotationunion = 1 - \prod\nolimits_{\secretindex\in\brb{\dimension}} \Big(1-\frac{2\privacythresholdi{i}}{\seclen_{g_i}}\Big),\qquad
    \Delta_{\text{Alg.1}} = \sqrt{\sum\nolimits_{\secretindex\in\brb{\dimension}}{\seclen_{g_i}}^2}< \constant\cdot\opt,
\end{align*}
where $\constant$ is a constant depending on the tolerance ranges and interval lengths of the mechanism, and $\opt$ %
{is the achievable distortion lower bound under the privacy budget $T=$ $  _{\text{Alg.1}}\privacynotationunion$}. %
\end{proposition}
(Proof in \cref{proof:multiGaussian_performance}.) From \cref{prop:multiGaussian_performance} we know that \cref{mech:dGaussian_diagnol} achieves order-optimal privacy-distortion performance with a constant multiplication factor.

\myparatightestn{The case of unknown distribution parameters}
In the sections above, we introduced release mechanisms for the case where the data holder knows the distribution parameters of the dataset. 
In practice, the data holder may only possess a dataset without knowing its underlying distribution's parameters.
Our data release mechanisms can be easily extended to this case where the input and the output of the mechanism are datasets. Briefly, the data holder estimates the parameters $\theta$ from the data and maps them to corresponding intervals. Once the released parameters $\theta'$ are determined, each sample is adjusted to conform to the distribution characterized by $\theta'$. %
We provide an example in \cref{sec:app_unkown_params_example} to demonstrate the extension process.
 
\section{Experiments}
\label{sec:experiments}

In this section, we employ real-world dataset to illustrate the variations in privacy-distortion tradeoffs
across different privacy metrics and present the privacy-distortion performance of our mechanism. {We include more results on alternative metrics and scalability of our mechanism in \cref{sec:app_empirical}.}

\myparatightestn{Metrics}
Our \privacy{} and \distortion{} metrics require the knowledge of prior distribution $\paramdistribution$ of the distribution parameters, which cannot be obtained in practice since the data holder can only have one dataset. Consequently, we need to design surrogate \privacy{} and \distortion{} metrics to bound the real \privacy{} and \distortion{} values in experiments.

\textit{Surrogate \privacy{} metric.}
For the original and the released datasets $\odataset$  and $\rdataset$, we define the surrogate metric for union privacy as
\begin{align*}
    \sprivacynotationunion = \max_{\secretindex \in \brb{\dimension}}\brc{-\abs{g_i(\odataset) - \hat g_i(\rdataset)} / \privacythresholdi{\secretindex}},
\end{align*}
where $\secretiof{\secretindex}{\cdot}$ is the $\secretindex$-th secret of the dataset, and $\privacythresholdi{\secretindex}$ is the tolerance range of the $\secretindex$-th \secret{}. %
We measure privacy by the difference between the secrets of the original and released datasets, using the negative sign to ensure that a smaller value signifies stronger privacy. Given varying tolerance ranges for different secrets, we use $\privacythresholdi{\secretindex}$ in the denominator to normalize the difference. 
For union privacy, the data holder aims to prevent the estimation of any secret, and therefore, we take the maximal privacy value over all secrets (i.e., worst case privacy). %
Note that when the prior distribution $\paramdistribution$ is uniform, guessing the $\secretindex$-th \secret{} as $\hat g_i(\rdataset)$ %
is the optimal attack strategy under \cref{mech:dGaussian_diagnol}. %
{Therefore, there is a mapping between $\privacynotationunion$ and $\sprivacynotationunion$, making the surrogate a suitable approximation of $\privacynotationunion$.}

Similarly, we define surrogate metrics for intersection privacy and \uoi \neuripsdelete{, and $l_{\normnum}$ norm privacy} as 
\begin{align*}
    \sprivacynotationinter &= \min_{\secretindex \in \brb{\dimension}}\brc{-\abs{\secretiof{\secretindex}{\odataset} - \secretiof{\secretindex}{\rdataset}} / \privacythresholdi{\secretindex}}, \quad
    \sprivacynotationuoi &= \max_{\group\in\groupset}\brc{\min_{\secretindex\in \groupofindex} \brc{-\abs{\secretiof{\secretindex}{\odataset} - \secretiof{\secretindex}{\rdataset}} / \privacythresholdi{\secretindex} }},\neuripsdelete{\\
    \sprivacynotationlp &= -\norm{\normnum}{\secretvof{\odataset} - \secretvof{\rdataset}} / \privacythresholdl.}
\end{align*}
where $\groupset$ represents the set of groups for \uoi{}, and $\groupofindex$ is the index set for group $b$.

\textit{Surrogate \distortion{} metric.}
We define the surrogate \distortion{} metric as the Wasserstein-2 distance between the original and the released dataset:
\begin{align*}
    \sdistortionnotation = \mathfrak{D}_{\text{Wasserstein-2}}\bra{{\omega_{\odataset}} \|{\omega_{\rdataset}}}, 
\end{align*}
where $\omega_{\mathcal{D}}$ is the empirical distribution of dataset $\mathcal{D}$. %

\myparatightestn{Dataset}%
We adopt Wikipedia Web Traffic Dataset (WWT) \cite{kaggle2019data} (license: CC0: Public Domain) to simulate the motivating scenarios.
WWT contains daily page views of 117,277 Wikipedia web pages in 2015-2016. Our goal is to release the page views (i.e., a 365$\times$117,277-dimensional dataset) 
while protecting the average daily page views of three websites (i.e., secrets are the means of three dimensions within data), %
which are $15, 68, 54$ respectively.

\myparatightestn{Baselines}
We compare our quantization mechanism \cref{mech:dGaussian_diagnol} with three widely-adopted privacy-preserving mechanisms: differentially-private density estimation \cite{wasserman2010statistical} (abbreviated as DP), attribute-private Gaussian mechanism \cite{zhang2022attribute} (abbreviated as AP), and distribution-private Laplacian mechanism \cite{kawamoto2019local} (abbreviated as DistP). DP works by partitioning the space into bins to construct a histogram of the dataset, adding zero-mean Laplacian noise to the histogram, and releasing samples drawing from the obfuscated histogram. AP works by adding zero-mean Gaussian noise to each sample. DistP works by adding zero-mean Laplacian noise to each sample.

\myparatightestn{Empirical Results}
For union, intersection, and \uoi{}, we set the tolerance ranges of three secrets as  $\privacythresholdi{1} = 1, \privacythresholdi{2} = 4, \privacythresholdi{3} = 3$, according to their respective average page views. 
For \uoi{}, we cluster the first two means as a group and consider the last mean separately.

Under different privacy metrics, we first calculate the surrogate distortion and privacy values of \cref{mech:dGaussian_diagnol}. The results are shown in \cref{fig:wiki}, where each point represents a realization of the mechanism with different interval lengths. %
Different levels of distortion are obtained by varying the bin size.
Each solid line in \cref{fig:wiki} represents a theoretical privacy-distortion lower bound with secrets as three means for a specific surrogate privacy metric. Such theoretical analysis does not require any assumptions on the data distribution (details in \cref{proof:surrogate_lower}). From \cref{fig:wiki}, we can observe that:

\begin{wrapfigure}{l}{0.6\textwidth}
    \centering
    \includegraphics[width=\linewidth]{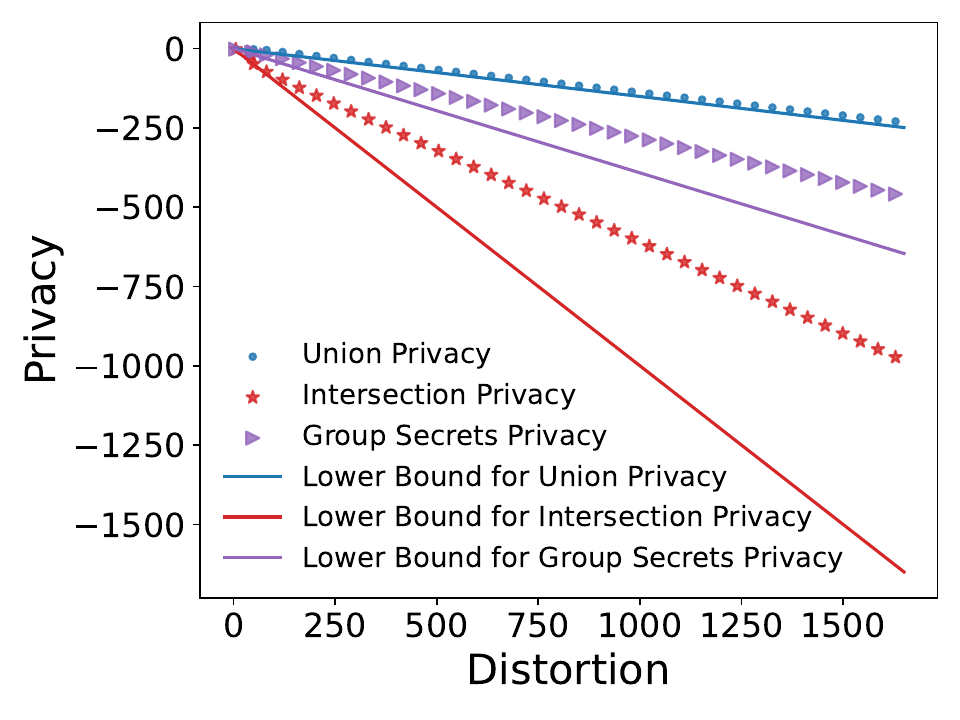}
    \caption{Privacy (lower is better) 
    and distortion of \cref{mech:dGaussian_diagnol} under WWT with different privacy metrics. For each metric, the solid line with the same color represents the theoretical lower bound.}
\label{fig:wiki}
\end{wrapfigure}

\textit{(1) The privacy-distortion tradeoffs of \cref{mech:dGaussian_diagnol} match the theoretical lower bounds (or with a mild gap) even the distribution assumption does not hold.}
For union privacy \neuripsdelete{and $l_1$ norm privacy}(color in blue\neuripsdelete{ and green}), the privacy-distortion tradeoffs of \cref{mech:dGaussian_diagnol} match the theoretical lower bounds. For the other privacy metrics, regardless of the privacy budget, the distortion of \cref{mech:dGaussian_diagnol} approximates the theoretical lower bound with a multiplication factor smaller than $2$. The results indicate that without assumptions on distribution type, our mechanism still achieves order-optimal privacy-distortion performance on real-world dataset. 

\textit{(2) The discrepancies in privacy-distortion tradeoffs
across different privacy metrics align with our theoretical results.} Under a certain realization of \cref{mech:dGaussian_diagnol} (i.e., with a fixed distortion), union privacy achieves the highest value, while intersection privacy is the lowest, reflecting the strictest and mildest metrics, respectively. %
\neuripsdelete{For both \uoi{} and $l_\normnum$ norm privacy, their privacy values are intermediate between union and intersection privacy.} %

\begin{figure*}[!t]
    \centering
\subfigure[Union Privacy]{\includegraphics[width=0.32\linewidth]{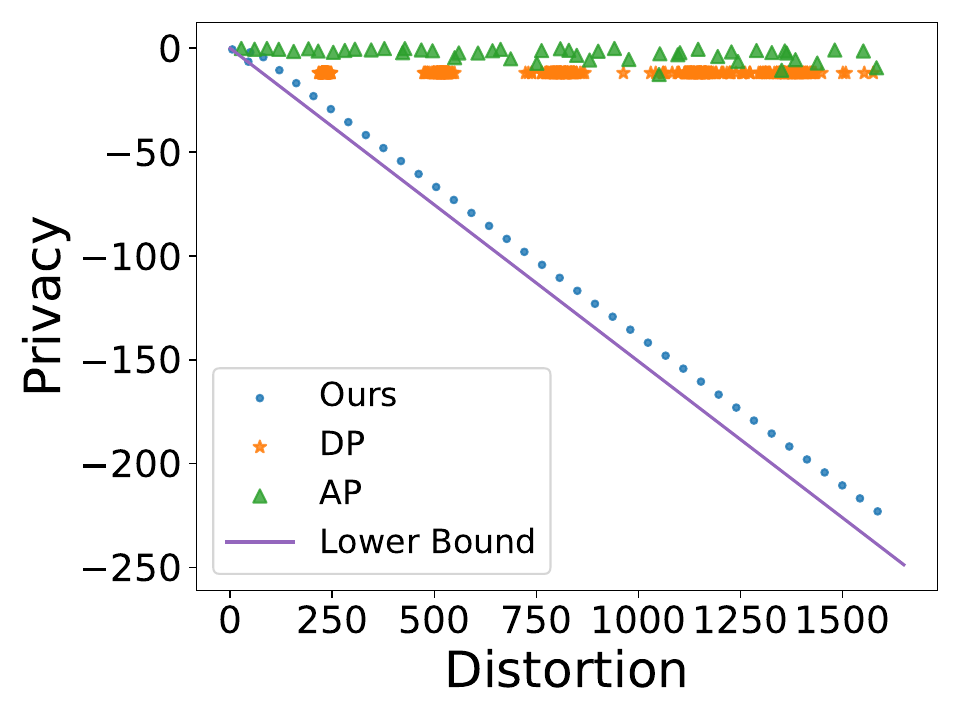}}
\subfigure[Intersection Privacy]{\includegraphics[width=0.32\linewidth]{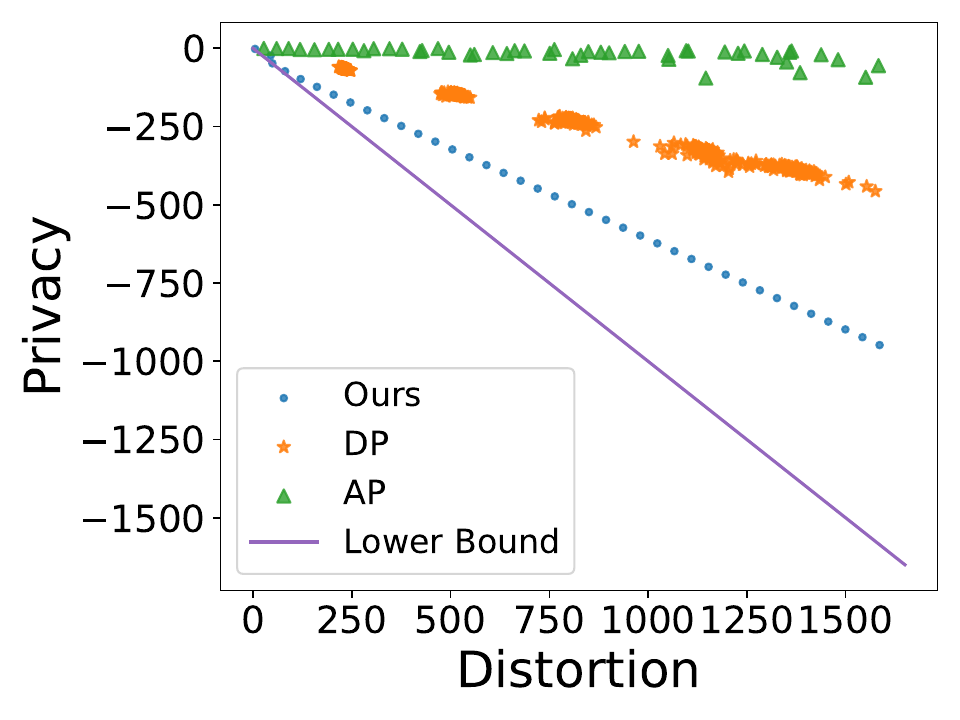}}
\subfigure[\UoI]{\includegraphics[width=0.32\linewidth]{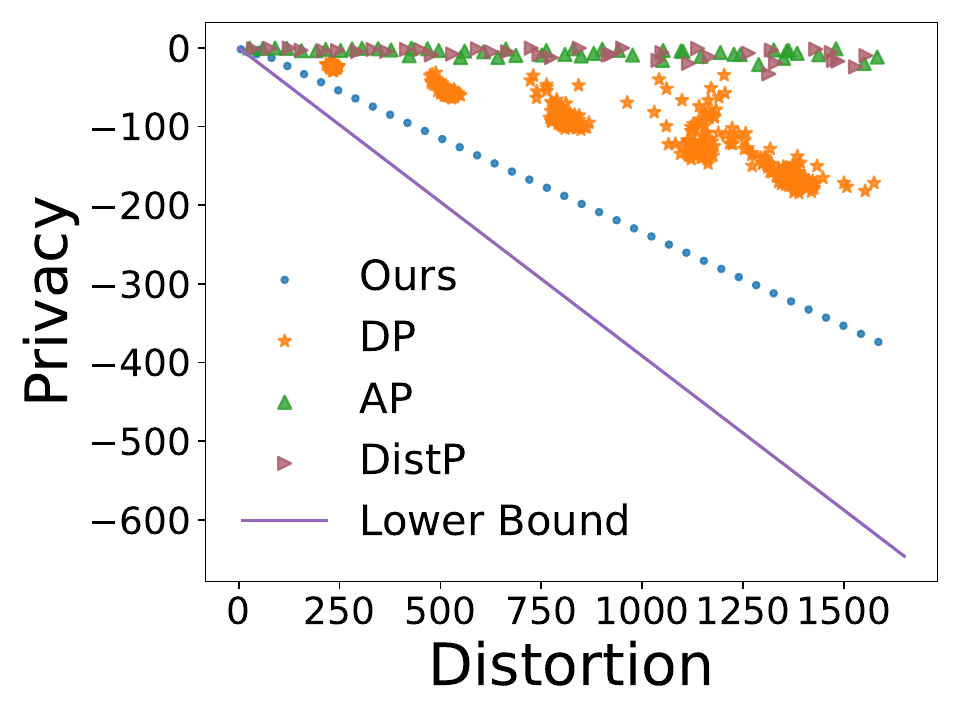}}
\caption{Privacy and distortion (lower values are better) of DP, AP, DistP, and ours under different privacy metrics. The solid line represents the theoretical lower bound of
achievable region.
    }
    \label{fig:compare}
\end{figure*}

We then compare the privacy-distortion tradeoffs between DP, AP, DistP, and our mechanism under different privacy metrics in \cref{fig:compare}. Each point represents a realization of data release mechanism with one hyper-parameter (bin size and noise level for DP, noise level for AP and DistP, interval length for ours). We can observe that \emph{our quantization mechanism achieves better privacy-distortion tradeoffs under different privacy metrics}. AP and DistP add noise to individual samples, which has less effect on means, and therefore achieves the worst privacy-distortion tradeoffs. DP instead adds noise to the histograms after the space quantization, which could protect the means more efficiently. However, since the added noise is unbounded, the privacy-distortion tradeoffs are still worse than our quantization mechanism.%

\section{Conclusion and Future Work}
\label{sec:conclusion}

We introduce a framework to quantify, analyze, and protect multi-secret summary statistic privacy in data sharing.
To accommodate varying data sharing scenarios, we propose and compare several privacy metrics, analyzing the tradeoff between privacy and distortion under each metric.
For data release mechanism design, we propose a quantization-based mechanism and demonstrate its order-optimal privacy-distortion performance in certain settings, both theoretically and empirically. 

Our proposed privacy metrics rely on the prior distribution of parameters $\distributionof{\Theta}$, which may not be known in practice. Motivated by \textit{maximal leakage} \cite{issa2019operational}, an avenue for future work is to extend the privacy metrics by considering the worst-case leakage over all possible priors.
Moreover, while we provide analysis of the adaptive composition property of summary statistic privacy under a specific setting in \cref{sec:composition}, %
we leave the comprehensive composition analysis under the general setting to future work. Extending our quantization-based release mechanism to scenarios beyond those studied in this work is another promising direction we hope to investigate in the future.

\paragraph{Disclaimer} 
This paper was prepared for informational purposes in part by the CDAO group of JPMorgan Chase \& Co and its affiliates (“J.P. Morgan”) and is not a product of the Research Department of J.P. Morgan.  J.P. Morgan makes no representation and warranty whatsoever and disclaims all liability, for the completeness, accuracy or reliability of the information contained herein.  This document is not intended as investment research or investment advice, or a recommendation, offer or solicitation for the purchase or sale of any security, financial instrument, financial product or service, or to be used in any way for evaluating the merits of participating in any transaction, and shall not constitute a solicitation under any jurisdiction or to any person, if such solicitation under such jurisdiction or to such person would be unlawful.

\newpage
\appendix
\onecolumn
\begin{appendix}

\title{\huge\bfseries\noindent Appendix}
\maketitle
\etocdepthtag.toc{mtappendix}
\etocsettagdepth{mtchapter}{none}
\etocsettagdepth{mtappendix}{subsection}
\thispagestyle{empty}
\tableofcontents
\thispagestyle{empty}
\newpage
\setcounter{page}{1}

\begin{appendix}
\section{Related Works}
\label{app:related_work}

\myparatightestn{Heuristics} 
While heuristics are widely adopted in industries for data sharing \cite{hundepool2010handbook}, many lack theoretical privacy guarantees and can be vulnerable in real-world scenarios \cite{elliot1999scenarios}. For example, \textit{subsetting} aims to protect sensitive information by only selecting a part of available data to release \cite{reiss2012obfuscatory}. However, sub-sampling data will not change the data distribution, and thus the statistical properties are still preserved. 
\textit{Culling} and \textit{de-identification} remove certain attributes of the dataset \cite{reiss2012obfuscatory}, but they may excise too much information and are risked from re-identification attacks based on side information or cross-attribute correlations \cite{narayanan2006break,sweeney2013matching,el2008protecting}.

\myparatightestn{Indistinguishability Methods}
\textit{Differential privacy} (DP) \cite{dwork2006calibrating} is one of the most widely adopted privacy metric, and it provides privacy by ensuring that any two input neighboring datasets are indistinguishable. However, DP cannot be directly applied to protect summary statistic secrets as it aims to protect whether an individual record (or group) contribute to the released data, rather than to hide statistical properties of a distribution.
For example, typical DP or Local DP approaches like Laplacian mechanism \cite{dwork2006calibrating,wasserman2010statistical} would add zero mean noise to each sample. Obfuscating data in such way will not change statistical properties like the mean of the distribution on expectation. 
While several works have applied DP to release summary statistics\cite{barak2007privacy,qardaji2014priview}, the problems they address do not align with typical data sharing scenarios, as those methods only provide a synopsis of the dataset.

Motivating by differential privacy, several works proposed approaches aiming to make pairs of datasets or distributions with similar summary statistic secrets indistinguishable \cite{zhang2022attribute,ghosh2016inferential,chen2023protecting,suri2021formalizing}. \textit{Attribute privacy} \cite{zhang2022attribute} also aims to protect properties of the dataset and parameters of the underlying distribution from
which dataset is sampled, but not under the data sharing scenarios. Adopting the Pufferfish privacy framework \cite{kifer2014pufferfish}, the paper designed mechanisms ensuring indistinguishablility. However, the attribute privacy framework only outputs certain statistical query, rather than releases the dataset, making it not suitable for data sharing.
\textit{Distribution privacy} \cite{kawamoto2019local} and \textit{distribution inference} \cite{suri2021formalizing,suri2023dissecting} also share the similar goal of protecting statistical secrets of the data. Roughly, they design indistinguishably mechanisms ensuring that the output distribution are similar for any input distribution with similar statistical secrets. Since we only aim to hide certain statistics rather than the whole distribution, this framework is far stronger than what we need and will cause worse utility. Totally different distributions may have similar statistics (e,g, a Dirac delta distribution and a Gaussian distribution may have the same mean), and prohibitive noise should be added to make the whole distributions indistinguishable.

\myparatightestn{Leakage-Based Methods}
Another category of works adopt information theoretic approaches to define and protect statistical privacy. Typically, works in this category aim to limit the disclosure of private information while maximizing disclosure of others. Specifically, those works quantify disclosure of sensitive information by the notion of leakage \cite{alvim2014additive,smith2015recent}. Leakage can be defined by various of measures, such as Shannon entropy \cite{makhdoumi2014information,rassouli2021perfect,zamani2022bounds,calmon2015fundamental}, min-entropy \cite{asoodeh2017privacy,asoodeh2018estimation,smith2009foundations}, and gain function \cite{m2012measuring,alvim2014additive,liao2019tunable,saeidian2023pointwise}. One important theme is the development of leakage measures with operational significance \cite{alvim2014additive}. \textit{Maximal leakage} \cite{issa2019operational}, an operationally-interpretable and robust measure, has been proposed recently. Maximal leakage is defined as the increase of the probability of correctly guessing the secrets after observing the released dataset. However, maximal leakage and its generalizations (e.g., \cite{gilani2023alpha,kurri2022operational}) are unsuitable to our scenario since they assume the secrets to protect are not known  in advance and therefore take the worst-case leakage over all possible secrets.

\section{Proofs}
\subsection{Proof of \NoCaseChange\cref{thm:trade_off_general}}
\label{proof:trade_off_general}
\begin{proof}

We prove the tradeoff lower bound by constructing a sequence of attackers guessing different possible secrets values, such that there exists attackers guessing at least one secret successfully. Specifically, for each secret, we divide the range of possible secret values into segments, and design a series of individual-secret attack strategies, guessing the midpoint of each segment. We subsequently formulate multi-secret attack strategies by choosing one individual-secret strategy for each secret. We then establish the distortion lower bound based on the privacy constraint that the attack success rate is at most $\privacybound$ and by utilizing the conversion parameter $\ratio^{\text{union}}$ that serves as a linkage between the distributional distance and the distance between secrets.

\begin{align*}
    \privacybound
    &\geq \privacynotationunion\\
    &= \sup_{\secretestimatenotationvector} ~\probof{\bigcup_{\secretindex\in [\dimension]}~ \abs{\secretiestimateof{\secretindex}{ \releaservparamnotation}- \secretiofparam{\secretindex} } \leq \privacythresholdi{\secretindex} }\\
    &=\sup_{\secretestimatenotationvector} \expectationof{\probof{\bigcup_{\secretindex\in [\dimension]}~ \abs{\secretiestimateof{\secretindex}{ \releaservparamnotation}- \secretiofparam{\secretindex} } \leq \privacythresholdi{\secretindex}\bigg| \releaservparamnotation } }\\
    &= \expectationof{\sup_{\secretestimatenotationvector}\probof{\bigcup_{\secretindex\in [\dimension]}~ \abs{\secretiestimateof{\secretindex}{ \releaservparamnotation}- \secretiofparam{\secretindex} } \leq \privacythresholdi{\secretindex}\bigg| \releaservparamnotation } }~~,\numberthis\label{eq:esup_supe}
\end{align*}
where \cref{eq:esup_supe} is due to the fact that $\secretestimatenotation$ only depends on $\releaservparamnotation$ and therefore one can devise an attacker that for each $\releaservparamnotation$, performs the optimal attack. 
It follows from \cref{eq:esup_supe} there exists $\releaservparamnotation$ s.t.  $\sup_{\secretestimatenotationvector}\probof{\bigcup_{\secretindex\in [\dimension]}~ \abs{\secretiestimateof{\secretindex}{ \releaservparamnotation}- \secretiofparam{\secretindex} } \leq \privacythresholdi{\secretindex}\bigg| \releaservparamnotation } \leq \privacybound$. 
For any $\secretindex\in[\dimension]$, let $\lefti{\secretindex}$ and $\righti{\secretindex}$ be the smallest and the largest possible value of secret $g_i$ given the released distribution parameter $\theta'$: 
\begin{align*}
\lefti{\secretindex} \triangleq \inf_{\rvparamnotation \in
\support{\paramdistribution}:
\mathcal{M}_g(\theta)= \releaservparamnotation} \secretiofparam{\secretindex}~,\\
\vspace{15mm}
\righti{\secretindex} \triangleq \sup_{\rvparamnotation \in
\support{\paramdistribution}:
\mathcal{M}_g(\theta)= \releaservparamnotation}\secretiofparam{\secretindex}~.
\end{align*}

For each secret $g_i$, where $i\in[d]$, we partition the range of possible secret values, i.e., $[\lefti{i}, \righti{i}]$, into segments with length $2\privacythresholdi{i}$. Subsequently, we develop a set of individual-secret attack strategies by guessing the midpoint of each segment. As a result, the number of individual-secret attack strategies, denoted as $N_i$, satisfies $\lefti{\secretindex}+2N_\secretindex\privacythresholdi{\secretindex}\geq \righti{\secretindex}>\lefti{\secretindex}+2(N_\secretindex-1)\privacythresholdi{\secretindex}$.

We then construct multi-secret attack strategies by selecting one individual-secret strategy for each secret. For the multi-secret attack strategy $\secretestimatenotationv$, where $\attackerv = \brb{\attackerindex{1}, \attackerindex{2}, \cdots, \attackerindex{\dimension}}$ and $\attackerindex{\secretindex} \in \brb{N_\secretindex}$ for all $\secretindex\in \brb{\dimension}$, it guesses the secret $g_i$ as the midpoint of the $v_i$-th segment, i.e., $\secretiestimateofv{\secretindex}{\releaservparamnotation} = \lefti{\secretindex} + 
\bra{\attackerindex{\secretindex}-0.5}\cdot 2\privacythresholdi{\secretindex}$. The number of multi-secret attack strategies, denoted as $\attackernum$, is $\attackernum = \prod_{\secretindex \in \brb{\dimension}} N_\secretindex$. We can get that
\begin{align*}
\label{eqn:privacybound}
    \privacybound\cdot \attackernum &\geq \sum_{\attackerv} \probof{\bigcup_{\secretindex\in [\dimension]}~ \abs{\secretiestimateofv{\secretindex}{ \releaservparamnotation}- \secretiofparam{\secretindex} } \leq \privacythresholdi{\secretindex}\bigg| \releaservparamnotation }\\
    &\overset{1}{=} \sum_{\attackerv} \sum_{j\in [\dimension]} \bra{-1}^{j-1} \sum_{\substack{\secretindex_1 < \secretindex_2<\cdots <\secretindex_j\\\secretindex_1, \secretindex_2, \cdots, \secretindex_j\in [\dimension]}} \probof{\bigcap_{\secretindex \in \brc{\secretindex_1, \secretindex_2,\cdots, \secretindex_j}}\abs{\secretiestimateofv{\secretindex}{ \releaservparamnotation}- \secretiofparam{\secretindex} } \leq \privacythresholdi{\secretindex}\bigg| \releaservparamnotation}\\
    &\overset{2}{\geq} \sum_{j\in [\dimension]} \bra{-1}^{j-1} \sum_{\substack{\secretindex_1 < \secretindex_2<\cdots <\secretindex_j\\\secretindex_1, \secretindex_2, \cdots, \secretindex_j\in [\dimension]}}\frac{\attackernum}{\prod_{k \in \brc{\secretindex_1, \secretindex_2, \cdots, \secretindex_j}}{N_k}}. 
    \numberthis
\end{align*}
where $\overset{1}{=}$ is because for any attack strategy $\secretestimatenotationv$, we have
\begin{align*}
&\probof{\bigcup_{\secretindex\in [\dimension]}~ \abs{\secretiestimateofv{\secretindex}{ \releaservparamnotation}- \secretiofparam{\secretindex} } \leq \privacythresholdi{\secretindex}\bigg| \releaservparamnotation } \\
&\quad= 
\sum_{\secretindex \in [\dimension]} \probof{\abs{\secretiestimateofv{\secretindex}{ \releaservparamnotation}- \secretiofparam{\secretindex} } \leq \privacythresholdi{\secretindex}\bigg| \releaservparamnotation} - \sum_{\substack{\secretindex_1 < \secretindex_2\\\secretindex_1, \secretindex_2\in [\dimension]}} \probof{\bigcap_{\secretindex \in \brc{\secretindex_1, \secretindex_2}}\abs{\secretiestimateofv{\secretindex}{ \releaservparamnotation}- \secretiofparam{\secretindex} } \leq \privacythresholdi{\secretindex}\bigg| \releaservparamnotation}\\
&\quad \ \ \ +\cdots + \bra{-1}^{j-1} \sum_{\substack{\secretindex_1 < \secretindex_2<\cdots <\secretindex_j\\\secretindex_1, \secretindex_2, \cdots, \secretindex_j\in [\dimension]}}\probof{\bigcap_{\secretindex \in \brc{\secretindex_1, \secretindex_2,\cdots, \secretindex_j}}\abs{\secretiestimateofv{\secretindex}{ \releaservparamnotation}- \secretiofparam{\secretindex} } \leq \privacythresholdi{\secretindex}\bigg| \releaservparamnotation}\\
&\quad\ \ \  +\cdots + \bra{-1}^{\dimension-1}  \probof{\bigcap_{i\in [\dimension]}\abs{\secretiestimateofv{\secretindex}{ \releaservparamnotation}- \secretiofparam{\secretindex} } \leq \privacythresholdi{\secretindex}\bigg| \releaservparamnotation}\\
&\quad= \sum_{j\in [\dimension]} \bra{-1}^{j-1} \sum_{\substack{\secretindex_1 < \secretindex_2<\cdots <\secretindex_j\\\secretindex_1, \secretindex_2, \cdots, \secretindex_j\in [\dimension]}} \probof{\bigcap_{\secretindex \in \brc{\secretindex_1, \secretindex_2,\cdots, \secretindex_j}}\abs{\secretiestimateofv{\secretindex}{ \releaservparamnotation}- \secretiofparam{\secretindex} } \leq \privacythresholdi{\secretindex}\bigg| \releaservparamnotation}.
\end{align*}
$\overset{2}{\geq}$ is because 
\begin{align*}
\sum_{\attackerv} &\sum_{j\in [\dimension]} \bra{-1}^{j-1} \sum_{\substack{\secretindex_1 < \secretindex_2<\cdots <\secretindex_j\\\secretindex_1, \secretindex_2, \cdots, \secretindex_j\in [\dimension]}} \probof{\bigcap_{\secretindex \in \brc{\secretindex_1, \secretindex_2,\cdots, \secretindex_j}}\abs{\secretiestimateofv{\secretindex}{ \releaservparamnotation}- \secretiofparam{\secretindex} } \leq \privacythresholdi{\secretindex}\bigg| \releaservparamnotation} \\
&=\sum_{j\in [\dimension]} \bra{-1}^{j-1} \sum_{\attackerv} \sum_{\substack{\secretindex_1 < \secretindex_2<\cdots <\secretindex_j\\\secretindex_1, \secretindex_2, \cdots, \secretindex_j\in [\dimension]}} \probof{\bigcap_{\secretindex \in \brc{\secretindex_1, \secretindex_2,\cdots, \secretindex_j}}\abs{\secretiestimateofv{\secretindex}{ \releaservparamnotation}- \secretiofparam{\secretindex} } \leq \privacythresholdi{\secretindex}\bigg| \releaservparamnotation}\\
& = \sum_{j\in [\dimension]} \bra{-1}^{j-1} \sum_{\substack{\secretindex_1 < \secretindex_2<\cdots <\secretindex_j\\\secretindex_1, \secretindex_2, \cdots, \secretindex_j\in [\dimension]}} \sum_{\attackerv} \probof{\bigcap_{\secretindex \in \brc{\secretindex_1, \secretindex_2,\cdots, \secretindex_j}}\abs{\secretiestimateofv{\secretindex}{ \releaservparamnotation}- \secretiofparam{\secretindex} } \leq \privacythresholdi{\secretindex}\bigg| \releaservparamnotation}\\
&\geq \sum_{j\in [\dimension]} \bra{-1}^{j-1} \sum_{\substack{\secretindex_1 < \secretindex_2<\cdots <\secretindex_j\\\secretindex_1, \secretindex_2, \cdots, \secretindex_j\in [\dimension]}}  \prod_{k_1 \in [\dimension]\setminus\brc{\secretindex_1, \secretindex_2, \cdots, \secretindex_j}} N_{k_1}\\
&= \sum_{j\in [\dimension]} \bra{-1}^{j-1} \sum_{\substack{\secretindex_1 < \secretindex_2<\cdots <\secretindex_j\\\secretindex_1, \secretindex_2, \cdots, \secretindex_j\in [\dimension]}}  \frac{\attackernum}{\prod_{k \in \brc{\secretindex_1, \secretindex_2, \cdots, \secretindex_j}}{N_k}}.
\end{align*}

Since $\righti{\secretindex}>\lefti{\secretindex}+2(N_\secretindex-1)\privacythresholdi{\secretindex}$, $\forall i\in [\dimension]$,  we can get that
\begin{align*}
\prod_{\secretindex\in [\dimension]} \bra{\righti{\secretindex}-\lefti{\secretindex}} & > \prod_{\secretindex\in [\dimension]} 2\privacythresholdi{\secretindex} \bra{N_\secretindex-1}\\
&= \prod_{\secretindex\in [\dimension]} 2\privacythresholdi{\secretindex} \cdot \prod_{\secretindex\in [\dimension]} \bra{N_\secretindex-1}\\
&= \bra{\attackernum - \sum_{j\in [\dimension]} \bra{-1}^{j-1} \sum_{\substack{\secretindex_1 < \secretindex_2<\cdots <\secretindex_j\\\secretindex_1, \secretindex_2, \cdots, \secretindex_j\in [\dimension]}}  \frac{\attackernum}{\prod_{k \in \brc{\secretindex_1, \secretindex_2, \cdots, \secretindex_j}}{N_k}}}\cdot \prod_{\secretindex\in [\dimension]} 2\privacythresholdi{\secretindex}\\
& \overset{1}{\geq} \left\lceil{\bra{1-\privacybound}\attackernum}\right\rceil \cdot \prod_{\secretindex\in [\dimension]} 2\privacythresholdi{\secretindex},
\numberthis
\label{eqn:R-L}
\end{align*}
where $\overset{1}{\geq}$ is because based on \cref{eqn:privacybound} and the fact that $N_\secretindex$ is an integer for any $\secretindex\in [\dimension]$, we have
\begin{align*}
    \sum_{j\in [\dimension]} \bra{-1}^{j-1} \sum_{\substack{\secretindex_1 < \secretindex_2<\cdots <\secretindex_j\\\secretindex_1, \secretindex_2, \cdots, \secretindex_j\in [\dimension]}}  \frac{\attackernum}{\prod_{k \in \brc{\secretindex_1, \secretindex_2, \cdots, \secretindex_j}}{N_k}} \leq \floor{\privacybound \attackernum}.
\end{align*}

We analyze the value of $\attackernum$ as follows. Based on \cref{eqn:privacybound}, we have
\begin{align*}
\privacybound\cdot \attackernum &\geq \sum_{j\in [\dimension]} \bra{-1}^{j-1} \sum_{\substack{\secretindex_1 < \secretindex_2<\cdots <\secretindex_j\\\secretindex_1, \secretindex_2, \cdots, \secretindex_j\in [\dimension]}}\frac{\attackernum}{\prod_{k \in \brc{\secretindex_1, \secretindex_2, \cdots, \secretindex_j}}{N_k}}= \attackernum - \prod_{\secretindex\in [\dimension]} \bra{N_\secretindex-1}.
\numberthis
\label{eqn:N_lowerbound}
\end{align*}
Define $a_\secretindex = N_\secretindex - 1$. Then we have $a_\secretindex \geq 0$ and $\attackernum = \prod_{\secretindex \in [d] } \bra{a_\secretindex+1}$. We can get that
\begin{align*}
\attackernum &= \prod_{\secretindex \in [d] } \bra{a_\secretindex+1}\\
&= 1 + \sum_{j\in [\dimension]} \sum_{\substack{\secretindex_1 < \secretindex_2<\cdots <\secretindex_j\\\secretindex_1, \secretindex_2, \cdots, \secretindex_j\in [\dimension]}} \prod_{k\in \brc{\secretindex_1, \secretindex_2, \cdots, \secretindex_j}} a_{k}\\
&\overset{1}{\geq} \sum_{j\in [\dimension]\cup\brc{0}}\binom{\dimension}{j} \cdot \bra{\prod_{\secretindex\in [\dimension]}a_\secretindex}^{\frac{j}{\dimension}}\\
&= \bra{\bra{\prod_{\secretindex\in [\dimension]}a_\secretindex}^\frac{1}{\dimension}+1}^\dimension\\
&= \bra{\bra{\prod_{\secretindex\in [\dimension]}\bra{N_\secretindex-1}}^\frac{1}{\dimension}+1}^\dimension,
\end{align*}
where $\overset{1}{\geq}$ is because when $j=0$, $\binom{\dimension}{j} \cdot \bra{\prod_{\secretindex\in [\dimension]}a_\secretindex}^{\frac{j}{\dimension}}=1$, and for any $j \in [\dimension]$, we have
\begin{align*}
\sum_{\substack{\secretindex_1 < \secretindex_2<\cdots <\secretindex_j\\\secretindex_1, \secretindex_2, \cdots, \secretindex_j\in [\dimension]}} \prod_{k\in \brc{\secretindex_1, \secretindex_2, \cdots, \secretindex_j}} a_{k} 
&\geq \binom{\dimension}{j} \cdot \bra{\prod_{\substack{\secretindex_1 < \secretindex_2<\cdots <\secretindex_j\\\secretindex_1, \secretindex_2, \cdots, \secretindex_j\in [\dimension]}} \prod_{k\in \brc{\secretindex_1, \secretindex_2, \cdots, \secretindex_j}} a_{k}}^{\frac{1}{\binom{\dimension}{j}}}\\
&= \binom{\dimension}{j} \cdot {\bra{\prod_{\secretindex\in [\dimension]}a_\secretindex}}^{\frac{\binom{\dimension-1}{j-1}}{\binom{\dimension}{j}}}\\
&= \binom{\dimension}{j} \cdot \bra{\prod_{\secretindex\in [\dimension]}a_\secretindex}^{\frac{j}{\dimension}}~.
\end{align*}

Therefore, we can get that
\begin{align*}
    \prod_{\secretindex\in [\dimension]}\bra{N_\secretindex-1} \leq \bra{\attackernum^{\frac{1}{\dimension}}-1}^\dimension.
\end{align*}

Combined with \cref{eqn:N_lowerbound} and the fact that $\attackernum$ is an integer, we have
\begin{align*}
    \privacybound \cdot \attackernum &\geq \attackernum - \bra{\attackernum^{\frac{1}{\dimension}}-1}^\dimension\\
    \bra{\attackernum^{\frac{1}{\dimension}}-1}^\dimension &\geq \bra{1-\privacybound} \attackernum\\
    \attackernum^{\frac{1}{\dimension}} &\geq \frac{1}{1-\bra{1-\privacybound}^\frac{1}{\dimension}}\\
    \attackernum &\geq \left\lceil{\frac{1}{\bra{1-\bra{1-\privacybound}^{{1}/{\dimension}}}^\dimension}}\right\rceil.
\end{align*}

Combined with \cref{eqn:R-L}, we can get that
\begin{align*}
\prod_{\secretindex\in [\dimension]} \bra{\righti{\secretindex}-\lefti{\secretindex}} 
& > \ceil{\bra{1-\privacybound}\attackernum} \cdot \prod_{\secretindex\in [\dimension]} 2\privacythresholdi{\secretindex}
\geq  \left\lceil{\bra{1-\privacybound}\left\lceil{\frac{1}{\bra{1-\bra{1-\privacybound}^{{1}/{\dimension}}}^\dimension}}\right\rceil}\right\rceil \cdot \prod_{\secretindex\in [\dimension]} 2\privacythresholdi{\secretindex}.
\end{align*}
Hence, we have
\begin{align*}
\prod_{\secretindex\in [\dimension]} \bra{\righti{\secretindex}-\lefti{\secretindex}}^{1/\dimension}
>  2\left\lceil{\bra{1-\privacybound}^{{1}/{\dimension}}\left\lceil{\frac{1}{{1-\bra{1-\privacybound}^{{1}/{\dimension}}}}}\right\rceil}\right\rceil \cdot \bra{\prod_{\secretindex\in [\dimension]} \privacythresholdi{\secretindex}}^{{1}/{\dimension}}.
\numberthis
\label{eqn:R-L_1/d}
\end{align*}

Then we have
\begin{align*}
\distortionnotation &=\supdist\distanceof{\privatedistribution} {\releasedistribution}\\
&\geq \sup_{\rvparamnotation_i \in
\support{\paramdistribution}, i\in\brc{1,2}:
\mathcal{M}_g(\theta_i)= \releaservparamnotation}
\auxdistance{\rvprivatewithparam{\rvparamnotation_1}}{\rvprivatewithparam{\rvparamnotation_2}}\numberthis \label{eq:lowerbound_tri}\\
&> 2 
\ratio^{\text{union}} \cdot \left\lceil{\bra{1-\privacybound}^{{1}/{\dimension}}\left\lceil{\frac{1}{{1-\bra{1-\privacybound}^{{1}/{\dimension}}}}}\right\rceil}\right\rceil \cdot \bra{\prod_{\secretindex\in [\dimension]} \privacythresholdi{\secretindex}}^{{1}/{\dimension}}\numberthis\label{eq:lowerbound_ratio_def_union}\\
&> 2 
\ratio^{\text{union}} \cdot \left\lceil{\frac{1}{{1-\bra{1-\privacybound}^{{1}/{\dimension}}}}-1}\right\rceil \cdot \bra{\prod_{\secretindex\in [\dimension]} \privacythresholdi{\secretindex}}^{{1}/{\dimension}},
\end{align*}
where in
\cref{eq:lowerbound_tri}, 
$\rvparamnotation_i$ for $\secretindex\in\{1,2\}$  denotes two arbitrary parameter vectors in the support space. \cref{eq:lowerbound_tri}
comes from the triangle inequality, and \cref{eq:lowerbound_ratio_def_union} utilizes \cref{eqn:R-L_1/d} and the definition of $\ratio^{\text{union}}$.
\end{proof}

\subsection{Proof of \NoCaseChange\cref{thm:trade_off_intersection}}
\label{prooof:trade_off_intersection}

\begin{proof}
Similar to the proof of \cref{thm:trade_off_general}, we construct a sequence of attackers guessing different possible secrets values, such that there exists attackers guessing all secrets successfully. Specifically, for each secret, we divide the range of possible secret values into segments, and design a series of individual-secret attack strategies, guessing the midpoint of each segment. We subsequently formulate multi-secret attack strategies by choosing one individual-secret strategy for each secret. We then establish the distortion lower bound based on the privacy constraint that the attack success rate is at most $\privacybound$ and by utilizing the conversion parameter $\ratio^{\text{inter}}$ that serves as a linkage between the distributional distance and the distance between secrets.

It follows from \cref{eq:esup_supe} that there exists $\releaservparamnotation$ s.t.  $\sup_{\secretestimatenotation}\probof{\bigcap_{\secretindex\in [\dimension]}~ \abs{\secretiestimateof{\secretindex}{ \releaservparamnotation}- \secretiofparam{\secretindex} } \leq \privacythresholdi{\secretindex}\bigg| \releaservparamnotation } \leq \privacybound$. 
For any $\secretindex\in[\dimension]$, let $\lefti{\secretindex}$ and $\righti{\secretindex}$ be the smallest and the largest possible value of secret $g_i$ given the released distribution parameter $\theta'$: 
\begin{align*}
\lefti{\secretindex} \triangleq \inf_{\rvparamnotation \in
\support{\paramdistribution}:
\mathcal{M}_g(\theta)= \releaservparamnotation} \secretiofparam{\secretindex}~,\\
\vspace{15mm}
\righti{\secretindex} \triangleq \sup_{\rvparamnotation \in
\support{\paramdistribution}:
\mathcal{M}_g(\theta)= \releaservparamnotation}\secretiofparam{\secretindex}~.
\end{align*}

For each secret $g_i$, where $i\in[d]$, we partition the range of possible secret values, i.e., $[\lefti{i}, \righti{i}]$, into segments with length $2\privacythresholdi{i}$. Subsequently, we develop a set of individual-secret attack strategies by guessing the midpoint of each segment. As a result, the number of individual-secret attack strategies, denoted as $N_i$, satisfies $\lefti{\secretindex}+2N_\secretindex\privacythresholdi{\secretindex}\geq \righti{\secretindex}>\lefti{\secretindex}+2(N_\secretindex-1)\privacythresholdi{\secretindex}$.

We then construct multi-secret attack strategies by selecting one individual-secret strategy for each secret. For the multi-secret attack strategy $\secretestimatenotationv$, where $\attackerv = \brb{\attackerindex{1}, \attackerindex{2}, \cdots, \attackerindex{\dimension}}$ and $\attackerindex{\secretindex} \in \brb{N_\secretindex}$ for all $\secretindex\in \brb{\dimension}$, it guesses the secret $g_i$ as the midpoint of the $v_i$-th segment, i.e., $\secretiestimateofv{\secretindex}{\releaservparamnotation} = \lefti{\secretindex} + 
\bra{\attackerindex{\secretindex}-0.5}\cdot 2\privacythresholdi{\secretindex}$. The number of multi-secret attack strategies, denoted as $\attackernum$, is $\attackernum = \prod_{\secretindex \in \brb{\dimension}} N_\secretindex$. We can get that
\begin{align*}
    \privacybound\cdot \attackernum &\geq \sum_{\attackerv} \probof{\bigcap_{\secretindex\in [\dimension]}~ \abs{\secretiestimateofv{\secretindex}{ \releaservparamnotation}- \secretiofparam{\secretindex} } \leq \privacythresholdi{\secretindex}\bigg| \releaservparamnotation }=1. 
\end{align*}
Since $\attackernum \in \mathbb{N}^+$, we have 
$
    \attackernum \geq \left\lceil{\frac{1}{\privacybound}}\right\rceil.
$

Therefore, we can get that
\begin{align*}
\sum_{\secretindex\in [\dimension]} \bra{\righti{\secretindex}-\lefti{\secretindex}} & > \sum_{\secretindex\in [\dimension]} 2\privacythresholdi{\secretindex} \bra{N_\secretindex-1}\\
&\geq 2\dimension\bra{\prod_{\secretindex\in [\dimension]}\privacythresholdi{\secretindex} N_\secretindex}^{1/\dimension}-2\sum_{\secretindex\in [\dimension]} \privacythresholdi{\secretindex} \\
&\geq 2\dimension\cdot\left\lceil{\frac{1}{\privacybound}}\right\rceil^{1/\dimension} \cdot \bra{\prod_{\secretindex\in [\dimension]} \privacythresholdi{\secretindex}}^{{1}/{\dimension}}-2\sum_{\secretindex\in [\dimension]} \privacythresholdi{\secretindex}.
\end{align*}

Hence, we have
\begin{align*}
\distortionnotation &=\supdist\distanceof{\privatedistribution} {\releasedistribution}\\
&\geq \sup_{\rvparamnotation_i \in
\support{\paramdistribution}, i\in\brc{1,2}:
\mathcal{M}_g(\theta_i)= \releaservparamnotation}
\auxdistance{\rvprivatewithparam{\rvparamnotation_1}}{\rvprivatewithparam{\rvparamnotation_2}}\\
&> 2\ratio^{\text{inter}}\cdot\left\lceil{\frac{1}{\privacybound}}\right\rceil^{1/\dimension} \cdot \bra{\prod_{\secretindex\in [\dimension]} \privacythresholdi{\secretindex}}^{{1}/{\dimension}}-2\ratio^{\text{inter}}\cdot\frac{1}{\dimension}\sum_{\secretindex\in [\dimension]} \privacythresholdi{\secretindex}.
\numberthis\label{eq:lowerbound_ratio_def}
\end{align*}
\end{proof}

\subsection{Proof of \NoCaseChange\cref{prop:distortion_bound_compare}}
\label{proof:distortion_bound_compare_inter}

\begin{proof}
We first prove that for any fixed distortion budget $\delta_0$, when $\Delta \leq \delta_0$, the achievable lower bound for union privacy, denoted as $\privacynotationunion$, is no smaller than that for intersection privacy, denoted as $\privacynotationinter$.

For any attack strategy $\secretestimatenotationvector$ and data release mechanism $\mathcal{M}_g$ that satisfies $\Delta \leq \delta_0$, we have
\begin{align*}
    \probof{\bigcap_{\secretindex\in [\dimension]}~ \abs{\secretiestimateof{\secretindex}{ \releaservparamnotation}- \secretiofparam{\secretindex} } \leq \privacythresholdi{\secretindex} } \leq \probof{\bigcup_{\secretindex\in [\dimension]}~ \abs{\secretiestimateof{\secretindex}{ \releaservparamnotation}- \secretiofparam{\secretindex} } \leq \privacythresholdi{\secretindex} }.
\end{align*}
Therefore, we can get that for any data release mechanism $\mathcal{M}_g$ that satisfies $\Delta \leq \delta_0$:
\begin{align*}
    \sup_{\secretestimatenotationvector}\probof{\bigcap_{\secretindex\in [\dimension]}~ \abs{\secretiestimateof{\secretindex}{ \releaservparamnotation}- \secretiofparam{\secretindex} } \leq \privacythresholdi{\secretindex} } \leq \sup_{\secretestimatenotationvector}\probof{\bigcup_{\secretindex\in [\dimension]}~ \abs{\secretiestimateof{\secretindex}{ \releaservparamnotation}- \secretiofparam{\secretindex} } \leq \privacythresholdi{\secretindex} },
\end{align*}
which indicates that with a fixed distortion budget $\delta_0$, $\privacynotationinter\leq\privacynotationunion$. Hence, we can easily get that with a privacy budget $T$, the achievable distortion lower bounds for union privacy $\unionlower$ and intersection privacy $\interlower$ satisfy $\unionlower\geq \interlower$. 
\end{proof}
\subsection{Proof of \NoCaseChange\cref{thm:trade_off_mix}}
\label{proof:trade_off_mix}

\begin{proof}
Similar to the proof of \cref{thm:trade_off_general}, we construct a sequence of attackers guessing different possible secrets values, such that there exists attackers successfully guessing secrets. Specifically, for each secret, we divide the range of possible secret values into segments, and design a series of individual-secret attack strategies, guessing the midpoint of each segment. We subsequently formulate multi-secret attack strategies by choosing one individual-secret strategy for each secret. We then establish the distortion lower bound based on the privacy constraint that the attack success rate is at most $\privacybound$ and by utilizing the conversion parameter $\ratio^{\text{group}}$ that serves as a linkage between the distributional distance and the distance between secrets.

It follows from \cref{eq:esup_supe} there exists $\releaservparamnotation$ s.t.  $\sup_{\secretestimatenotation}\probof{\bigcap_{\secretindex\in [\dimension]}~ \abs{\secretiestimateof{\secretindex}{ \releaservparamnotation}- \secretiofparam{\secretindex} } \leq \privacythresholdi{\secretindex}\bigg| \releaservparamnotation } \leq \privacybound$. 
For any $\secretindex\in[\dimension]$, let $\lefti{\secretindex}$ and $\righti{\secretindex}$ be the smallest and the largest possible value of secret $g_i$ given the released distribution parameter $\theta'$: 
\begin{align*}
\lefti{\secretindex} \triangleq \inf_{\rvparamnotation \in
\support{\paramdistribution}:
\mathcal{M}_g(\theta)= \releaservparamnotation} \secretiofparam{\secretindex}~,\\
\vspace{15mm}
\righti{\secretindex} \triangleq \sup_{\rvparamnotation \in
\support{\paramdistribution}:
\mathcal{M}_g(\theta)= \releaservparamnotation}\secretiofparam{\secretindex}~.
\end{align*}

For each secret $g_i$, where $i\in[d]$, we partition the range of possible secret values, i.e., $[\lefti{i}, \righti{i}]$, into segments with length $2\privacythresholdi{i}$. Subsequently, we develop a set of individual-secret attack strategies by guessing the midpoint of each segment. As a result, the number of individual-secret attack strategies, denoted as $N_i$, satisfies $\lefti{\secretindex}+2N_\secretindex\privacythresholdi{\secretindex}\geq \righti{\secretindex}>\lefti{\secretindex}+2(N_\secretindex-1)\privacythresholdi{\secretindex}$.

We then construct multi-secret attack strategies by selecting one individual-secret strategy for each secret. For the multi-secret attack strategy $\secretestimatenotationv$, where $\attackerv = \brb{\attackerindex{1}, \attackerindex{2}, \cdots, \attackerindex{\dimension}}$ and $\attackerindex{\secretindex} \in \brb{N_\secretindex}$ for all $\secretindex\in \brb{\dimension}$, it guesses the secret $g_i$ as the midpoint of the $v_i$-th segment, i.e., $\secretiestimateofv{\secretindex}{\releaservparamnotation} = \lefti{\secretindex} + 
\bra{\attackerindex{\secretindex}-0.5}\cdot 2\privacythresholdi{\secretindex}$. The number of multi-secret attack strategies, denoted as $\attackernum$, is $\attackernum = \prod_{\secretindex \in \brb{\dimension}} N_\secretindex$. We can get that
\begin{align*}
\label{eqn:privacybound_union_inter}
    \privacybound\cdot \attackernum &\geq \sum_{\attackerv} \sup_{\secretestimatenotation} ~\probof{\bigcup_{\group\in\groupset}\bra{\bigcap_{\secretindex\in \groupofindex}~ \abs{\secretiestimateofv{\secretindex}{ \releaservparamnotation}- \secretiofparam{\secretindex} } \leq \privacythresholdi{\secretindex} }}\\
    &= \sum_{\attackerv} \sum_{j\in \brb{\groupsize}} \bra{-1}^{j-1} \sum_{\substack{\group_1 \not= \group_2\not=\cdots \not=\group_j\\\group_1, \group_2, \cdots, \group_j\in \groupset}} \probof{\bigcap_{\substack{\secretindex\in \groupofindex\\ \group \in \brc{\group_1, \group_2,\cdots, \group_j}}}\abs{\secretiestimateofv{\secretindex}{ \releaservparamnotation}- \secretiofparam{\secretindex} } \leq \privacythresholdi{\secretindex}\bigg| \releaservparamnotation}\\
    &\overset{1}{\geq} \sum_{j\in [\groupsize]} \bra{-1}^{j-1} \sum_{\substack{\group_1 \not= \group_2\not=\cdots \not=\group_j\\\group_1, \group_2, \cdots, \group_j\in \groupset}} \frac{\attackernum}{\prod_{\group \in \brc{\group_1, \group_2,\cdots, \group_j}}~\prod_{k\in \groupofindex}{N_k}}\\
    &= \attackernum - \prod_{b\in\groupset} \bra{\prod_{\secretindex \in \groupofindex} N_\secretindex - 1}, 
    \numberthis
\end{align*}
where $\overset{1}{\geq}$ is because 
\begin{align*}
\sum_{\attackerv} &\sum_{j\in [\groupsize]} \bra{-1}^{j-1} \sum_{\substack{\group_1 \not= \group_2\not=\cdots \not=\group_j\\\group_1, \group_2, \cdots, \group_j\in \groupset}} \probof{\bigcap_{\substack{\secretindex\in \groupofindex\\ \group \in \brc{\group_1, \group_2,\cdots, \group_j}}}\abs{\secretiestimateofv{\secretindex}{ \releaservparamnotation}- \secretiofparam{\secretindex} } \leq \privacythresholdi{\secretindex}\bigg| \releaservparamnotation}\\
& = \sum_{j\in [\groupsize]} \bra{-1}^{j-1} \sum_{\substack{\group_1 \not= \group_2\not=\cdots \not=\group_j\\\group_1, \group_2, \cdots, \group_j\in \groupset}} \sum_{\attackerv} \probof{\bigcap_{\substack{\secretindex\in \groupofindex\\ \group \in \brc{\group_1, \group_2,\cdots, \group_j}}}\abs{\secretiestimateofv{\secretindex}{ \releaservparamnotation}- \secretiofparam{\secretindex} } \leq \privacythresholdi{\secretindex}\bigg| \releaservparamnotation}\\
&\geq \sum_{j\in [\groupsize]} \bra{-1}^{j-1} \sum_{\substack{\group_1 \not= \group_2\not=\cdots \not=\group_j\\\group_1, \group_2, \cdots, \group_j\in \groupset}} \prod_{k_1 \in [\dimension]\setminus\bigcup_{\group \in \brc{\group_1, \group_2,\cdots, \group_j}}\groupofindex\ } N_{k_1}\\
&= \sum_{j\in [\groupsize]} \bra{-1}^{j-1} \sum_{\substack{\group_1 \not= \group_2\not=\cdots \not=\group_j\\\group_1, \group_2, \cdots, \group_j\in \groupset}} \frac{\attackernum}{\prod_{\group \in \brc{\group_1, \group_2,\cdots, \group_j}}~\prod_{k\in \groupofindex}{N_k}}.
\end{align*}

Define $A_\group = \prod_{\secretindex \in \groupofindex} N_{\secretindex} - 1$. We have $A_\group \geq 0$ and $\attackernum = \prod_{\group \in \groupset} \bra{A_\group+1}$. We can get that
\begin{align*}
\attackernum &= \prod_{\group \in \groupset} \bra{A_\group+1}\\
&= 1 + \sum_{j\in [\groupsize]} \sum_{\substack{\group_1 \not= \group_2\not=\cdots \not=\group_j\\\group_1, \group_2, \cdots, \group_j\in \groupset}}\prod_{\group \in \brc{\group_1, \group_2,\cdots, \group_j}} A_{\group}\\
&\overset{1}{\geq} \sum_{j\in [\groupsize]\cup\brc{0}}\binom{\groupsize}{j} \cdot \bra{\prod_{\group \in \groupset}A_\group}^{\frac{j}{\groupsize}}\\
&= \bra{\bra{\prod_{\group \in \groupset}A_\group}^\frac{1}{\groupsize}+1}^\groupsize\\
&= \bra{\bra{\prod_{\group \in \groupset}\bra{\prod_{\secretindex \in \groupofindex} N_{\secretindex} - 1}}^\frac{1}{\groupsize}+1}^\groupsize.
\end{align*}

$\overset{1}{\geq}$ is because when $j=0$, $\binom{\groupsize}{j} \cdot \bra{\prod_{\group \in \groupset}A_\group}^{\frac{j}{\groupsize}}=1$, and for any $j \in [\groupset]$, we have
\begin{align*}
\sum_{\substack{\group_1 \not= \group_2\not=\cdots \not=\group_j\\\group_1, \group_2, \cdots, \group_j\in \groupset}}\prod_{\group \in \brc{\group_1, \group_2,\cdots, \group_j}} A_{\group}
&\geq \binom{\groupsize}{j} \cdot \bra{\prod_{\substack{\group_1 \not= \group_2\not=\cdots \not=\group_j\\\group_1, \group_2, \cdots, \group_j\in \groupset}}\prod_{\group \in \brc{\group_1, \group_2,\cdots, \group_j}} A_{\group}}^{\frac{1}{\binom{\groupsize}{j}}}\\
&= \binom{\groupsize}{j} \cdot {\bra{\prod_{\group\in \groupset}A_\group}}^{\frac{\binom{\groupsize-1}{j-1}}{\binom{\groupsize}{j}}}\\
&= \binom{\groupsize}{j} \cdot \bra{\prod_{\group\in \groupset}A_\group}^{\frac{j}{\groupsize}}~.
\end{align*}

Therefore, we have
\begin{align}
\label{eqn:upperbound_N_inter_union}
    \prod_{b\in\groupset} \bra{\prod_{\secretindex \in \groupofindex} N_\secretindex - 1} \leq \bra{\attackernum^{\frac{1}{\groupsize}}-1}^\groupsize.
\end{align}

Combining this result with \cref{eqn:privacybound_union_inter} and the fact that $\attackernum$ is an integer, we have
\begin{align*}
    \privacybound \cdot \attackernum &\geq \attackernum - \bra{\attackernum^{\frac{1}{\groupsize}}-1}^\groupsize\\
    \bra{\attackernum^{\frac{1}{\groupsize}}-1}^\groupsize &\geq \bra{1-\privacybound} \attackernum\\
    \attackernum^{\frac{1}{\groupsize}} &\geq \frac{1}{1-\bra{1-\privacybound}^\frac{1}{\groupsize}}\\
    \attackernum &\geq \left\lceil{\frac{1}{\bra{1-\bra{1-\privacybound}^{{1}/{\groupsize}}}^\groupsize}}\right\rceil.
\end{align*}

Hence, we can get that
\begin{align*}
\sum_{\secretindex\in [\dimension]} \bra{\righti{\secretindex}-\lefti{\secretindex}} 
&> \sum_{\secretindex\in [\dimension]} 2\privacythresholdi{\secretindex} \bra{N_\secretindex-1}\\
&\geq 2\dimension\bra{\prod_{\secretindex\in [\dimension]}\privacythresholdi{\secretindex} N_\secretindex}^{1/\dimension}-2\sum_{\secretindex\in [\dimension]} \privacythresholdi{\secretindex} \\
&\geq 2\dimension\cdot\left\lceil{\frac{1}{\bra{1-\bra{1-\privacybound}^{{1}/{\groupsize}}}^{\groupsize/\dimension}}}\right\rceil\cdot\bra{\prod_{\secretindex\in[\dimension]}\privacythresholdi{\secretindex}}^{1/\dimension}-2\sum_{\secretindex\in [\dimension]} \privacythresholdi{\secretindex}.
\end{align*}

Then we have
\begin{align*}
\distortionnotation &=\supdist\distanceof{\privatedistribution} {\releasedistribution}\\
&\geq \sup_{\rvparamnotation_i \in
\support{\paramdistribution}, i\in\brc{1,2}:
\mathcal{M}_g(\theta_i)= \releaservparamnotation}
\auxdistance{\rvprivatewithparam{\rvparamnotation_1}}{\rvprivatewithparam{\rvparamnotation_2}}\\
&> 2\ratio^{\text{group}}\cdot\left\lceil{\frac{1}{\bra{1-\bra{1-\privacybound}^{{1}/{\groupsize}}}^{\groupsize/\dimension}}}\right\rceil\cdot\bra{\prod_{\secretindex\in[\dimension]}\privacythresholdi{\secretindex}}^{1/\dimension}-2\ratio^{\text{group}}\cdot\frac{1}{\dimension}\sum_{\secretindex\in [\dimension]} \privacythresholdi{\secretindex}.
\end{align*}
\end{proof}

\subsection{Proof of \NoCaseChange\cref{prop:multiGaussian}}
\label{proof:multiGaussian}

\begin{proof}
Define $\rvparamnotation' = \bra{\mu'_1, \cdots, \mu'_{\dimgaussian}, \sigma'_1, \cdots, \sigma'_{\dimgaussian}}$. 
We first provide the lemma as follows.
\begin{lemma} 
\label{lemma:D_lowerbound_multi_Gaussian}
$\auxdistance{\rvprivatewithparam{\rvparamnotation}}{\rvprivatewithparam{\rvparamnotation'}}$ can be derived as:
\begin{align*}
\auxdistance{\rvprivatewithparam{\rvparamnotation}}{\rvprivatewithparam{\rvparamnotation'}}
=
\half \sqrt{\sum_{j\in \brb{\dimgaussian}}\bra{\mu_j - \mu'_j}^2 + \sum_{j\in \brb{\dimgaussian}}\bra{\sigma_j - \sigma'_j}^2}.
\end{align*}
\end{lemma}

The proof is in \cref{proof:D_lowerbound_multi_Gaussian}.

Based on \cref{lemma:D_lowerbound_multi_Gaussian}, we can get that
\begin{align*}
\frac{\auxdistance{\rvprivatewithparam{\rvparamnotation}}{\rvprivatewithparam{\rvparamnotation'}}}{R^{\text{union}}\bra{{\rvprivatewithparam{\rvparamnotation}},{\rvprivatewithparam{\rvparamnotation'}}}} 
&\geq
\frac{\sqrt{\sum_{j\in \brb{k}}\bra{\mu_j - \mu'_j}^2 + \sum_{j\in \brb{k}}\bra{\sigma_j - \sigma'_j}^2}}{2\prod_{\secretindex\in [\dimension]}\rangeiformula{\secretindex}{{\rvparamnotation}}{{\rvparamnotation'}}^{1/\dimension}}\\
& \geq 
\frac{\sqrt{\sum_{\secretindex\in [\dimension]}\rangeiformulasquare{\secretindex}{{\rvparamnotation}}{{\rvparamnotation'}}}}{2\prod_{\secretindex\in [\dimension]}\rangeiformula{\secretindex}{{\rvparamnotation}}{{\rvparamnotation'}}^{1/\dimension}}\\
& =
\half \sqrt{\frac{\sum_{\secretindex\in [\dimension]}\rangeiformulasquare{\secretindex}{{\rvparamnotation}}{{\rvparamnotation'}}}{\prod_{\secretindex\in [\dimension]}\rangeiformula{\secretindex}{{\rvparamnotation}}{{\rvparamnotation'}}^{2/\dimension}}}\\
& \geq
\half \sqrt{\dimension \cdot \frac{\prod_{\secretindex\in [\dimension]}\rangeiformula{\secretindex}{{\rvparamnotation}}{{\rvparamnotation'}}^{2/\dimension}}{\prod_{\secretindex\in [\dimension]}\rangeiformula{\secretindex}{{\rvparamnotation}}{{\rvparamnotation'}}^{2/\dimension}}}\\
& =
\frac{\sqrt{d}}{2}.
\end{align*}
Therefore, we have
\begin{align*}
    \ratio^{\text{union}} = \inf_{\rvparamnotation_1, \rvparamnotation_2 \in\support{\paramdistribution}}
\frac{\auxdistance{\rvprivatewithparam{\rvparamnotation_1}}{\rvprivatewithparam{\rvparamnotation_2}}}{R^{\text{union}}\bra{{\rvprivatewithparam{\rvparamnotation_1}},{\rvprivatewithparam{\rvparamnotation_2}}}} = \frac{\sqrt{\dimension}}{2}.
\end{align*}

Based on \cref{thm:trade_off_general}, we can get that
\begin{align*}
\distortionnotation> 
\sqrt{\dimension} \cdot \left\lceil{\frac{1}{{1-\bra{1-\privacybound}^{{1}/{\dimension}}}}-1}\right\rceil \cdot \bra{\prod_{\secretindex\in [\dimension]} \privacythresholdi{\secretindex}}^{{1}/{\dimension}}.
    \end{align*}
\end{proof}

\subsubsection{Proof of \NoCaseChange\cref{lemma:D_lowerbound_multi_Gaussian}}
\label{proof:D_lowerbound_multi_Gaussian}

\begin{proof}
From \cite{givens1984class}, we have
\begin{align*}
    \distanceof{\privatedistribution} {\releasedistribution}^2
    = \sum_{j\in \brb{\dimgaussian}}\bra{\mu_j - \mu'_j}^2 + \mathrm{Tr}\bra{\Sigma + \Sigma' - 2\bra{\Sigma^\frac{1}{2}\Sigma'\Sigma^\frac{1}{2}}^\frac{1}{2}},
\end{align*}
where diagonal covariance matrices $\Sigma, \Sigma'$ are
$
\Sigma = 
\begin{bmatrix}
    \sigma_1^2 & & & \\ & \sigma_2^2 & & \\
    & & \ddots & \\
    & & & \sigma_{\dimgaussian}^2
\end{bmatrix}
$
and
$
\Sigma' = 
\begin{bmatrix}
    {\sigma'_1}^2 & & & \\ & {\sigma'_2}^2 & & \\
    & & \ddots & \\
    & & & {\sigma'_{\dimgaussian}}^2
\end{bmatrix}
$.
We can get that 
\begin{align*}
\Sigma + \Sigma' - 2\bra{\Sigma^\frac{1}{2}\Sigma'\Sigma^\frac{1}{2}}^\frac{1}{2}
&= 
\begin{bmatrix}
    {\sigma_1}^2+{\sigma'_1}^2 & & & \\ & {\sigma_2}^2+{\sigma'_2}^2 & & \\
    & & \ddots & \\
    & & & {\sigma_{\dimgaussian}}^2+{\sigma'_{\dimgaussian}}^2
\end{bmatrix}
-2
\begin{bmatrix}
    {\sigma_1}^2{\sigma'_1}^2 & & & \\ & {\sigma_2}^2{\sigma'_2}^2 & & \\
    & & \ddots & \\
    & & & {\sigma_{\dimgaussian}}^2{\sigma'_{\dimgaussian}}^2
\end{bmatrix}^{\half}
\\
& = 
\begin{bmatrix}
    \bra{{\sigma_1}-{\sigma'_1}}^2 & & & \\ & \bra{{\sigma_2}-{\sigma'_2}}^2 & & \\
    & & \ddots & \\
    & & & \bra{{\sigma_{\dimgaussian}}-{\sigma'_{\dimgaussian}}}^2
\end{bmatrix}.
\end{align*}

Therefore, we can get that
\begin{align*}
    \distanceof{\privatedistribution} {\releasedistribution}^2
    = \sum_{j\in \brb{\dimgaussian}}\bra{\mu_j - \mu'_j}^2 + \mathrm{Tr}\bra{\Sigma + \Sigma' - 2\bra{\Sigma^\frac{1}{2}\Sigma'\Sigma^\frac{1}{2}}^\frac{1}{2}} = \sum_{j\in \brb{\dimgaussian}}\bra{\mu_j - \mu'_j}^2 + \sum_{j\in \brb{\dimgaussian}}\bra{\sigma_j - \sigma'_j}^2.
\end{align*}
Hence, we have
\begin{align*}
\auxdistance{\rvprivatewithparam{\rvparamnotation}}{\rvprivatewithparam{\rvparamnotation'}}
=
\half \sqrt{\sum_{j\in \brb{\dimgaussian}}\bra{\mu_j - \mu'_j}^2 + \sum_{j\in \brb{\dimgaussian}}\bra{\sigma_j - \sigma'_j}^2}.
\end{align*}
\end{proof}

\subsection{Proof of \NoCaseChange\cref{prop:multiGaussian_performance}}
\label{proof:multiGaussian_performance}

\begin{proof}

Based on \cref{lemma:D_lowerbound_multi_Gaussian}, we can easily can get that the distortion $\Delta_{\text{Alg.1}}$ of \cref{mech:dGaussian_diagnol} is
\begin{align*}
\distortionnotation = \sqrt{\sum_{\secretindex\in\brb{\dimension}}{\seclen_{g_i}}^2}.
\end{align*}

Since the secret distribution parameters are independent of each other and follow the uniform distributions, we can get that the privacy of \cref{mech:dGaussian_diagnol} is
\begin{align*}
\privacynotationunion &= ~\sup_{\secretestimatenotationvector}~\probof{\bigcup_{\secretindex\in [\dimension]}~ \abs{\secretiestimateof{\secretindex}{ \releaservparamnotation}- \secretiofparam{\secretindex} } \leq \privacythresholdi{\secretindex} }\\
&= 1 - ~\sup_{\secretestimatenotationvector}~\probof{\bigcap_{\secretindex\in [\dimension]}~ \abs{\secretiestimateof{\secretindex}{ \releaservparamnotation}- \secretiofparam{\secretindex} } > \privacythresholdi{\secretindex} }\\
&= 1 - ~\sup_{\secretestimatenotationvector}~\prod_{\secretindex\in\brb{\dimension}}\probof{ \abs{\secretiestimateof{\secretindex}{ \releaservparamnotation}- \secretiofparam{\secretindex} } > \privacythresholdi{\secretindex} }\\
&=1 - \prod_{\secretindex\in\brb{\dimension}} \bra{1-\frac{2\privacythresholdi{i}}{\seclen_{g_i}}}.
\end{align*}

From \cref{prop:multiGaussian}, we know that the optimal achievable distortion $\Delta_{opt}$ satisfy
\begin{align*}
\Delta_{opt} &> \sqrt{\dimension} \cdot \left\lceil{\frac{1}{{1-\bra{1-\privacynotation}^{{1}/{\dimension}}}}-1}\right\rceil \cdot \bra{\prod_{\secretindex\in [\dimension]} \privacythresholdi{\secretindex}}^{{1}/{\dimension}}\\
&= \sqrt{\dimension} \cdot \left\lceil{\frac{1}{{1-\prod_{\secretindex\in\brb{\dimension}} \bra{1-\frac{2\privacythresholdi{i}}{\seclen_{g_i}}}^{{1}/{\dimension}}}}-1}\right\rceil \cdot \bra{\prod_{\secretindex\in [\dimension]} \privacythresholdi{\secretindex}}^{{1}/{\dimension}}.
\end{align*}

Let $k=\frac{\Delta}{\Delta_{opt}}$, $x_i = \frac{\privacythresholdi{i}}{\seclen_{g_i}}, \forall \secretindex\in\brb{\dimension}$, $c_1 = \min_{\secretindex\in\brb{\dimension}}\brc{x_i}$, and $c_2 = \max_{\secretindex\in\brb{\dimension}}\brc{x_i}$, we have
\begin{align*}
k &< \frac{\sqrt{\sum_{\secretindex\in\brb{\dimension}}{\seclen_{g_i}}^2}}{\sqrt{\dimension} \cdot \left\lceil{\frac{1}{{1-\prod_{\secretindex\in\brb{\dimension}} \bra{1-\frac{2\privacythresholdi{i}}{\seclen_{g_i}}}^{{1}/{\dimension}}}}-1}\right\rceil \cdot \bra{\prod_{\secretindex\in [\dimension]} \privacythresholdi{\secretindex}}^{{1}/{\dimension}}}\\
& = \frac{\sqrt{\sum_{\secretindex\in\brb{\dimension}}{\bra{\frac{\privacythresholdi{i}}{x_i}}}^2}}{\sqrt{\dimension} \cdot \left\lceil{\frac{1}{{1-\prod_{\secretindex\in\brb{\dimension}} \bra{1-2x_i}^{{1}/{\dimension}}}}-1}\right\rceil \cdot \bra{\prod_{\secretindex\in [\dimension]} \privacythresholdi{\secretindex}}^{{1}/{\dimension}}}\\
&\leq \frac{2c_2 \sqrt{\sum_{i\in[d]}\epsilon_i^2}}{c_1(1-2c_2)\sqrt{d}\bra{\prod_{\secretindex\in [\dimension]} \privacythresholdi{\secretindex}}^{{1}/{\dimension}}}\\
& = \frac{ \sqrt{\frac{1}{d}\sum_{i\in[d]}\epsilon_i^2}}{\bra{\prod_{\secretindex\in [\dimension]} \privacythresholdi{\secretindex}}^{{1}/{\dimension}}}\cdot\frac{2c_2}{c_1(1-2c_2)}.
\end{align*}

Let $c_{\epsilon} = \frac{ \sqrt{\frac{1}{d}\sum_{i\in[d]}\epsilon_i^2}}{\bra{\prod_{\secretindex\in [\dimension]} \privacythresholdi{\secretindex}}^{{1}/{\dimension}}}$, and we can easily get that $\ceps \leq \max_{i,j\in[d]}\brc{\frac{\privacythresholdi{i}}{\privacythresholdi{j}}}$, a constant depending on the values of tolerance ranges. Denoting $c_{\epsilon,s} = \frac{2\ceps\cmax}{\cmin\bra{1-2\cmax}}$, we can finally get that
\begin{align*}
\distortionnotation = k\Delta_{opt}< c_{\epsilon,s}\opt,
\end{align*}
where $c_{\epsilon,s}$ is a constant depending on tolerance ranges and the interval lengths of the mechanism.

Specifically, when $\epsilon_1=\cdots=\epsilon_d$, and the designed data released mechanism satisfy $\frac{\privacythresholdi{1}}{\seclen_{g_1}}=\cdots =\frac{\privacythresholdi{d}}{\seclen_{g_d}}\leq \frac{1}{6}$, we can get that $\Delta< 3\Delta_{opt}$.
\end{proof}
\subsection{Theoretical Lower Bounds of Surrogate Metrics with Secrets = Three Means}
\label{proof:surrogate_lower}

Without loss of generality, we assume our objective is to protect the means of the first three dimensions of the data distribution. Let $x^{\bra{i}}$ be the $i$-th dimension of the data sample $\bx \in \mathbb{R}^{t}$ ($t\geq 3$).
For the original and released dataset $\odataset = \brc{\bx_1,\cdots,\bx_m}, \rdataset = \brc{\by_1, \cdots, \by_m}$, the empirical means are
\begin{align*}
   \xmudim{1} = \frac{1}{m}\sum_{i\in \brb{m}} x_i^{\bra{1}}, &\quad \xmudim{2} = \frac{1}{m}\sum_{i\in \brb{m}} x_i^{\bra{2}}, \quad \xmudim{3} = \frac{1}{m}\sum_{i\in \brb{m}} x_i^{\bra{3}},\\
    \ymudim{1} = \frac{1}{m}\sum_{i\in \brb{m}} y_i^{\bra{1}}, &\quad \ymudim{2} = \frac{1}{m}\sum_{i\in \brb{m}} y_i^{\bra{2}}, \quad \ymudim{3} = \frac{1}{m}\sum_{i\in \brb{m}} y_i^{\bra{3}}.
\end{align*}
Regardless of the distribution type, the surrogate distortion, i.e., Wasserstein-2 distance between $\odataset$ and $\rdataset$, satisfies
\begin{align*}
\sdistortionnotation = \distanceof{\mathcal{P}_{\odataset}} {\mathcal{P}_{\rdataset}} \geq \sqrt{\bra{\xmudim{1}-\ymudim{1}}^2 + \bra{\xmudim{2}-\ymudim{2}}^2 + \bra{\xmudim{3}-\ymudim{3}}^2}.
\end{align*}

We analyze the theoretical lower bounds of surrogate metrics under different privacy formulations as follows.

For union privacy, we have
\begin{align*}
\sdistortionnotation &\geq \sqrt{\bra{\xmudim{1}-\ymudim{1}}^2 + \bra{\xmudim{2}-\ymudim{2}}^2 + \bra{\xmudim{3}-\ymudim{3}}^2}\\
& = \sqrt{\privacythresholdi{1}^2\cdot\frac{\abs{\xmudim{1}-\ymudim{1}}^2}{\privacythresholdi{1}^2}+  \privacythresholdi{2}^2\cdot\frac{\abs{\xmudim{2}-\ymudim{2}}^2}{\privacythresholdi{2}^2} + \privacythresholdi{3}^2\cdot\frac{\abs{\xmudim{3}-\ymudim{3}}^2}{\privacythresholdi{3}^2}}\\
&\geq \sqrt{{\privacythresholdi{1}^2+\privacythresholdi{2}^2+\privacythresholdi{3}^2}}\cdot\min_{i\in [3]}\brc{\frac{\abs{\xmudim{i}-\ymudim{i}}}{\privacythresholdi{i}}}\\
&=-\sqrt{{\privacythresholdi{1}^2+\privacythresholdi{2}^2+\privacythresholdi{3}^2}}\cdot\max_{i\in [3]}\brc{-\frac{\abs{\xmudim{i}-\ymudim{i}}}{\privacythresholdi{i}}}\\
&= -\sqrt{{\privacythresholdi{1}^2+\privacythresholdi{2}^2+\privacythresholdi{3}^2}}\cdot\sprivacynotationunion.
\end{align*}

For intersection privacy, we have
\begin{align*}
\sdistortionnotation &\geq \sqrt{\bra{\xmudim{1}-\ymudim{1}}^2 + \bra{\xmudim{2}-\ymudim{2}}^2 + \bra{\xmudim{3}-\ymudim{3}}^2}\\
& = \sqrt{\privacythresholdi{1}^2\cdot\frac{\abs{\xmudim{1}-\ymudim{1}}^2}{\privacythresholdi{1}^2}+  \privacythresholdi{2}^2\cdot\frac{\abs{\xmudim{2}-\ymudim{2}}^2}{\privacythresholdi{2}^2} + \privacythresholdi{3}^2\cdot\frac{\abs{\xmudim{3}-\ymudim{3}}^2}{\privacythresholdi{3}^2}}\\
&\geq 
\min\brc{{\privacythresholdi{1}, \privacythresholdi{2}, \privacythresholdi{3}}}\cdot\max_{i\in [3]}\brc{\frac{\abs{\xmudim{i}-\ymudim{i}}}{\privacythresholdi{i}}}\\
&= -\min\brc{{\privacythresholdi{1}, \privacythresholdi{2}, \privacythresholdi{3}}}\cdot\min_{i\in [3]}\brc{-\frac{\abs{\xmudim{i}-\ymudim{i}}}{\privacythresholdi{i}}}\\
&=  -\min\brc{{\privacythresholdi{1}, \privacythresholdi{2}, \privacythresholdi{3}}}\sprivacynotationinter.
\end{align*}

For \uoi{}, let the means in the first two dimension be one group, and the third mean be one group. Then we have
\begin{align*}
\sdistortionnotation &\geq \sqrt{\bra{\xmudim{1}-\ymudim{1}}^2 + \bra{\xmudim{2}-\ymudim{2}}^2 + \bra{\xmudim{3}-\ymudim{3}}^2}\\
& = \sqrt{\privacythresholdi{1}^2\cdot\frac{\abs{\xmudim{1}-\ymudim{1}}^2}{\privacythresholdi{1}^2}+  \privacythresholdi{2}^2\cdot\frac{\abs{\xmudim{2}-\ymudim{2}}^2}{\privacythresholdi{2}^2} + \privacythresholdi{3}^2\cdot\frac{\abs{\xmudim{3}-\ymudim{3}}^2}{\privacythresholdi{3}^2}}\\
&\geq 
\sqrt{\min\brc{{\privacythresholdi{1}^2, \privacythresholdi{2}^2}}\cdot\max_{i\in [2]}\brc{\frac{\abs{\xmudim{i}-\ymudim{i}}^2}{\privacythresholdi{i}^2}} + \privacythresholdi{3}^2\cdot\frac{\abs{\xmudim{3}-\ymudim{3}}^2}{\privacythresholdi{3}^2}}\\
&\geq \sqrt{\min\brc{{\privacythresholdi{1}^2, \privacythresholdi{2}^2}}+\privacythresholdi{3}^2}\cdot \min\brc{\max_{i\in [2]}\brc{\frac{\abs{\xmudim{i}-\ymudim{i}}}{\privacythresholdi{i}}}, \frac{\abs{\xmudim{3}-\ymudim{3}}}{\privacythresholdi{3}}} \\
& = - \sqrt{\min\brc{{\privacythresholdi{1}^2, \privacythresholdi{2}^2}}+\privacythresholdi{3}^2}\cdot \max\brc{\min_{i\in [2]}\brc{-\frac{\abs{\xmudim{i}-\ymudim{i}}}{\privacythresholdi{i}}}, -\frac{\abs{\xmudim{3}-\ymudim{3}}}{\privacythresholdi{3}}} \\
&=  -\sqrt{\min\brc{{\privacythresholdi{1}^2, \privacythresholdi{2}^2}}+\privacythresholdi{3}^2}\cdot\sprivacynotationuoi.
\end{align*}

For $l_\normnum$ norm privacy with $\normnum=1$, we have
\begin{align*}
\sdistortionnotation &\geq \sqrt{\bra{\xmudim{1}-\ymudim{1}}^2 + \bra{\xmudim{2}-\ymudim{2}}^2 + \bra{\xmudim{3}-\ymudim{3}}^2}\\
&\geq \frac{\sqrt{3}}{3} \bra{\abs{\xmudim{1}-\ymudim{1}} + \abs{\xmudim{2}-\ymudim{2}} + \abs{\xmudim{3}-\ymudim{3}}}\\
& = \frac{\sqrt{3}}{3} \privacythresholdlp{1} \cdot \bra{\abs{\xmudim{1}-\ymudim{1}} + \abs{\xmudim{2}-\ymudim{2}} + \abs{\xmudim{3}-\ymudim{3}}} / \privacythresholdlp{1}\\
&= -\frac{\sqrt{3}}{3} \privacythresholdlp{1} \cdot\sprivacynotationlpi{1}.
\end{align*}

For $l_\normnum$ norm privacy with $\normnum=\infty$, we have
\begin{align*}
\sdistortionnotation &\geq \sqrt{\bra{\xmudim{1}-\ymudim{1}}^2 + \bra{\xmudim{2}-\ymudim{2}}^2 + \bra{\xmudim{3}-\ymudim{3}}^2}\\
&\geq \max_{i\in [3]}\brc{\abs{\xmudim{i}-\ymudim{i}}}\\
&\geq \privacythresholdlp{\infty}\cdot\max_{i\in [3]}\brc{\abs{\xmudim{i}-\ymudim{i}}}/\privacythresholdlp{\infty}\\
&= -\privacythresholdlp{\infty} \cdot\sprivacynotationlpi{\infty}.
\end{align*}

\section{Comparsion Between Union Privacy and \UoI}

When the group size $\groupsize$ is equal to $\dimension$, \uoi{} transforms into union privacy.
As shown in \cref{prop:tighter_distortion_bound}, \cref{thm:trade_off_general} provides a tighter (i.e., higher) distortion lower bound for the union privacy compared with \cref{thm:trade_off_mix} when $\groupsize=\dimension$.

\begin{proposition}
\label{prop:tighter_distortion_bound}
Given a privacy budget $\privacybound$ and tolerance ranges $\privacythresholdi{1}, \cdots, \privacythresholdi{\dimension}$, \cref{thm:trade_off_general} provides a tighter distortion lower bound for union privacy compared with \cref{thm:trade_off_mix} when $\groupsize=\dimension$.
\end{proposition}

\begin{proof}
When $\groupsize=\dimension$, the distortion lower bound for the union privacy in \cref{thm:trade_off_mix}, denoted as $\Delta_g$, is
\begin{align*}
    \Delta_g = 2\gamma_g \left\lceil{\frac{1}{{1-\bra{1-\privacybound}^{{1}/{\dimension}}}}}\right\rceil \bra{\prod_{\secretindex\in[\dimension]}\privacythresholdi{\secretindex}}^{1/\dimension}-2\gamma_g\cdot\frac{1}{\dimension}\sum_{\secretindex\in [\dimension]} \privacythresholdi{\secretindex},
\end{align*}
where 
$
\ratio_g \triangleq \inf_{\rvparamnotation_1, \rvparamnotation_2 \in\support{\paramdistribution}}
\frac{\distanceformula{\rvprivatewithparam{\rvparamnotation_1}}{\rvprivatewithparam{\rvparamnotation_2}}}{\frac{1}{\dimension} \sum_{\secretindex\in [\dimension]}\rangeiformula{\secretindex}{{\rvparamnotation_1}}{{\rvparamnotation_2}}}.
$

The distortion lower bound for the union privacy in \cref{thm:trade_off_general}, denoted as $\Delta_u$, is
\begin{align*}
    \Delta_u = 2 
\ratio_u  \left\lceil{\frac{1}{{1-\bra{1-\privacybound}^{{1}/{\dimension}}}}}\right\rceil  \bra{\prod_{\secretindex\in [\dimension]} \privacythresholdi{\secretindex}}^{{1}/{\dimension}}-2 
\ratio_u \bra{\prod_{\secretindex\in [\dimension]} \privacythresholdi{\secretindex}}^{{1}/{\dimension}},
\end{align*}
where 
$
\ratio_u \triangleq \inf_{\rvparamnotation_1, \rvparamnotation_2 \in\support{\paramdistribution}}
\frac{\distanceformula{\rvprivatewithparam{\rvparamnotation_1}}{\rvprivatewithparam{\rvparamnotation_2}}}{\prod_{\secretindex\in [\dimension]}\rangeiformula{\secretindex}{{\rvparamnotation_1}}{{\rvparamnotation_2}}^{1/\dimension}}.
$

According to the  inequality of arithmetic and geometric means, we have
\begin{align*}
\frac{1}{\dimension}\sum_{\secretindex\in [\dimension]} \privacythresholdi{\secretindex} &\geq \bra{\prod_{\secretindex\in [\dimension]} \privacythresholdi{\secretindex}}^{{1}/{\dimension}},\\
\frac{1}{\dimension} \sum_{\secretindex\in [\dimension]}\rangeiformula{\secretindex}{{\rvparamnotation_1}}{{\rvparamnotation_2}} &\geq \prod_{\secretindex\in [\dimension]}\rangeiformula{\secretindex}{{\rvparamnotation_1}}{{\rvparamnotation_2}}^{1/\dimension}.
\end{align*}
Therefore, we can get that $\ratio_u \geq \ratio_g$ as well as
\begin{align*}
\Delta_u &= 2 
\ratio_u  \left\lceil{\frac{1}{{1-\bra{1-\privacybound}^{{1}/{\dimension}}}}-1}\right\rceil  \bra{\prod_{\secretindex\in [\dimension]} \privacythresholdi{\secretindex}}^{{1}/{\dimension}}\\
&\geq 2 
\ratio_g  \left\lceil{\frac{1}{{1-\bra{1-\privacybound}^{{1}/{\dimension}}}}-1}\right\rceil  \bra{\prod_{\secretindex\in [\dimension]} \privacythresholdi{\secretindex}}^{{1}/{\dimension}}\\
&= 2 
\ratio_g  \left\lceil{\frac{1}{{1-\bra{1-\privacybound}^{{1}/{\dimension}}}}}\right\rceil  \bra{\prod_{\secretindex\in [\dimension]} \privacythresholdi{\secretindex}}^{{1}/{\dimension}}-2 
\ratio_g \bra{\prod_{\secretindex\in [\dimension]} \privacythresholdi{\secretindex}}^{{1}/{\dimension}}\\
&\geq 2 
\ratio_g  \left\lceil{\frac{1}{{1-\bra{1-\privacybound}^{{1}/{\dimension}}}}}\right\rceil  \bra{\prod_{\secretindex\in [\dimension]} \privacythresholdi{\secretindex}}^{{1}/{\dimension}}-2 
\ratio_g\cdot \frac{1}{\dimension}\sum_{\secretindex\in [\dimension]} \privacythresholdi{\secretindex}\\
&= \Delta_g.
\end{align*}

Hence, \cref{thm:trade_off_general} provides a tighter distortion lower bound for union privacy compared with \cref{thm:trade_off_mix} when $\groupsize=\dimension$.
\end{proof}

\section{Alternative Privacy Metrics: {$l_p$} Norm Summary Statistic Privacy}
\label{app:lp_norm}

In this section, we consider the scenario where the data holder aims to ensure a significant separation between the original and the attacker guessed secret vectors, rather than emphasizing whether a single or a group of secrets are disclosed. For example, consider the cluster performance traces with secrets as the proportions of different server types. The data holder may care more about whether the attacker can approximate the overall deployment of servers.

We adopt $l_\normnum$ norm ($\normnum > 0$) as the distance metric and define the \emph{$l_\normnum$ norm privacy} metric as the probability of the $l_p$ norm distance between the attacker guessed \secret{} vector $\secretestimatenotationvector$ and the original \secret{} vector $\secretsnotation$ being within a tolerance $\privacythresholdl$, taking the best attack strategy:
\begin{align}
    \privacynotationlp  \triangleq ~\sup_{\secretestimatenotationvector} ~\probof{ \norm{\normnum}{\secretsestimateof{ \releaservparamnotation}- \secretsofparam } \leq \privacythresholdl }~.
    \label{eq:privacy_lp}
\end{align}

Under $l_\normnum$ norm privacy, given a \privacy{} budget $\privacybound$, we then present a general lower bound on \distortion{}.

\begin{theorem}[Lower bound of \privacy{}-\distortion{} tradeoff for $l_\normnum$ norm privacy]
\label{thm:trade_off_norm}
Let $\auxdistance{{\rvprivatewithparam{\rvparamnotation_1}},{\rvprivatewithparam{\rvparamnotation_2}}}  = \distanceformula{\rvprivatewithparam{\rvparamnotation_1}}{\rvprivatewithparam{\rvparamnotation_2}}$.
Further, let
$R^{l_p}\bra{\rvprivatewithparam{\rvparamnotation_1}}{\rvprivatewithparam{\rvparamnotation_2}} = \frac{1}{\dimension} \sum_{\secretindex\in [\dimension]}\rangeiformula{\secretindex}{{\rvparamnotation_1}}{{\rvparamnotation_2}}$ and 
\begin{align*} 
\ratio^{l_p} = \inf_{\rvparamnotation_1, \rvparamnotation_2 \in\support{\paramdistribution}}
\frac{\auxdistance{\rvprivatewithparam{\rvparamnotation_1}}{\rvprivatewithparam{\rvparamnotation_2}}}{R^{{l_p}}({\rvprivatewithparam{\rvparamnotation_1}},{\rvprivatewithparam{\rvparamnotation_2}}) }.
\end{align*}
{Then, for tolerance ranges $\privacythresholdl$ as defined in \cref{eq:privacy_lp} and for any mechanism $\mathcal M_g$ subject to privacy budget $\privacynotationlp \leq \privacybound<1$, we have}
\begin{align*}
\distortionnotation > 
2\gamma^{l_p}\bra{\left\lceil{\frac{1}{\privacybound}}\right\rceil^{1/\dimension}-1}\privacythresholdl/\dimension^{\frac{1}{\normnum}}.
\end{align*}
\end{theorem}

(Proof in \cref{proof:trade_off_norm}.)
From \cref{thm:trade_off_norm}, we know that the distortion lower bound is positively correlated to the tolerance range $\privacythresholdl$ and negatively correlated to the privacy budget $\privacybound$. %
As shown in \cref{prop:lpnorm_compare}, $l_\normnum$ norm privacy metric is less strict than union privacy but stricter than intersection privacy.

\begin{proposition}
\label{prop:lpnorm_compare}
For union and intersection privacy, let $\privacythresholdi{1}, \cdots, \privacythresholdi{\dimension}$ be the tolerance ranges. Let the tolerance $\privacythresholdl$ for $l_\normnum$ norm privacy be $\privacythresholdl = \bra{\sum_{\secretindex\in [\dimension]} \privacythresholdi{\secretindex}^\normnum}^{{1}/{\normnum}}$. Given a privacy budget $\privacybound$, for any $\normnum>0$, we have
\begin{align*}
    \interlower \leq \lplower \leq \unionlower,
\end{align*}
where $\unionlower$, $\interlower$, and $\lplower$ represents the achievable distortion lower bounds for union privacy, intersection privacy, and $l_\normnum$ norm privacy respectively.
\end{proposition}
The proof and further analysis for $l_\normnum$ norm privacy with different norm order $\normnum$ are detailed in \cref{proof:lpnorm_compare}.

\subsection{Proof of \NoCaseChange\cref{thm:trade_off_norm}}
\label{proof:trade_off_norm}

\begin{proof}
    When $\privacynotation{} = ~\sup_{\secretestimatenotation} ~\probof{ \norm{\normnum}{\secretestimateof{ \releaservparamnotation}- \secretofparam } \leq \privacythresholdl } \leq \privacybound$, we can get that for any non-negative values $\privacythresholdi{1}, \privacythresholdi{2}, \cdots, \privacythresholdi{\dimension}$ that satisfy $\bra{\sum_{\secretindex\in\brb{\dimension}}\privacythresholdi{\secretindex}^{\normnum}}^{1/\normnum} = \privacythresholdl$:
\begin{align*}
    \sup_{\secretestimatenotation} ~\probof{\bigcap_{\secretindex\in [\dimension]}~ \abs{\secretiestimateof{\secretindex}{ \releaservparamnotation}- \secretiofparam{\secretindex} } \leq \privacythresholdi{\secretindex} } \leq \privacybound.
\end{align*}
This is because if there exists non-negative values $\privacythresholdihat{1}, \privacythresholdihat{2}, \cdots, \privacythresholdihat{\dimension}$ that satisfy $\bra{\sum_{\secretindex\in\dimension}\privacythresholdihat{\secretindex}^{\normnum}}^{1/\normnum} = \privacythresholdl$ and \begin{align*}
\sup_{\secretestimatenotation} ~\probof{\bigcap_{\secretindex\in [\dimension]}~ \abs{\secretiestimateof{\secretindex}{ \releaservparamnotation}- \secretiofparam{\secretindex} } \leq \privacythresholdihat{\secretindex} } > \privacybound,
\end{align*}
then we can get that 
\begin{align*}
    \sup_{\secretestimatenotation} ~\probof{ \norm{\normnum}{\secretestimateof{ \releaservparamnotation}- \secretofparam } \leq \privacythresholdl } 
    & \geq \probof{ \norm{\normnum}{\secrethatestimateof{ \releaservparamnotation}- \secretofparam } \leq \privacythresholdl }\\ 
    & \geq \probof{\bigcap_{\secretindex\in [\dimension]}~ \abs{\secrethatiestimateof{\secretindex}{ \releaservparamnotation}- \secretiofparam{\secretindex} } \leq \privacythresholdihat{\secretindex} }\\
    & = \sup_{\secretestimatenotation} ~\probof{\bigcap_{\secretindex\in [\dimension]}~ \abs{\secretiestimateof{\secretindex}{ \releaservparamnotation}- \secretiofparam{\secretindex} } \leq \privacythresholdihat{\secretindex} }\\
    & > \privacybound,
\end{align*}
which contradicts with the constraint that $\sup_{\secretestimatenotation} ~\probof{ \norm{\normnum}{\secretestimateof{ \releaservparamnotation}- \secretofparam } \leq \privacythresholdl } \leq \privacybound$.

Based on \cref{thm:trade_off_intersection}, we know that when $\sup_{\secretestimatenotation} ~\probof{\bigcap_{\secretindex\in [\dimension]}~ \abs{\secretiestimateof{\secretindex}{ \releaservparamnotation}- \secretiofparam{\secretindex} } \leq \privacythresholdi{\secretindex} } \leq \privacybound$, the \distortion{} satisfies
\begin{align*}
\distortionnotation > 
2\ratio^{l_p}\cdot\left\lceil{\frac{1}{\privacybound}}\right\rceil^{1/\dimension}\cdot\bra{\prod_{\secretindex\in[\dimension]}\privacythresholdi{\secretindex}}^{1/\dimension}-2\ratio^{l_p}\cdot\frac{1}{\dimension}\sum_{\secretindex\in [\dimension]} \privacythresholdi{\secretindex}.
\end{align*}
When $\privacythresholdi{\secretindex} = \privacythresholdl / \dimension^\frac{1}{\normnum}$ for all $\secretindex\in \brb{\dimension}$, we can get that
\begin{align*}
\distortionnotation &> 
2\ratio^{l_p}\cdot\bra{\left\lceil{\frac{1}{\privacybound}}\right\rceil^{1/\dimension}-1}\cdot\privacythresholdl/\dimension^{\frac{1}{\normnum}}.
\end{align*}
\end{proof}

\subsection{Proof of \NoCaseChange\cref{prop:lpnorm_compare} and More Analysis of \texorpdfstring{$l_{\normnum}$}\text{ Norm} Privacy}
\label{proof:lpnorm_compare}

\subsubsection{Proof of \cref{prop:lpnorm_compare}}
\label{proof:lpnorm_proof}
\begin{proof}
We first prove that for any fixed distortion budget $\delta_0$, when $\Delta \leq \delta_0$, the achievable lower bounds for union privacy $\privacynotationunion$, intersection privacy $\privacynotationinter$, and $l_{\normnum}$ norm privacy $\privacynotationnorm$  satisfy $\privacynotationinter\leq\privacynotationnorm\leq\privacynotationunion$.

For any attack strategy $\secretestimatenotationvector$ and data release mechanism $\mathcal{M}_g$ that satisfies $\Delta \leq \delta_0$, when $\abs{\secretiestimateof{\secretindex}{ \releaservparamnotation}- \secretiofparam{\secretindex} } \leq \privacythresholdi{\secretindex}$, $\forall \secretindex\in\brb{\dimension}$, we have $\norm{\normnum}{\secretsestimateof{ \releaservparamnotation}- \secretsofparam } \leq \bra{\sum_{\secretindex\in [\dimension]} \privacythresholdi{\secretindex}^\normnum}^{{1}/{\normnum}}= \privacythresholdl$. Besides, when $\norm{\normnum}{\secretsestimateof{ \releaservparamnotation}- \secretsofparam } \leq \privacythresholdl$, there exists $\secretindex\in\brb{\dimension}$, such that $\abs{\secretiestimateof{\secretindex}{ \releaservparamnotation}- \secretiofparam{\secretindex} } \leq \privacythresholdi{\secretindex}$. Therefore, we can get that 
\begin{align*}
    \probof{\bigcap_{\secretindex\in [\dimension]}~ \abs{\secretiestimateof{\secretindex}{ \releaservparamnotation}- \secretiofparam{\secretindex} } \leq  \privacythresholdi{\secretindex} } \leq \probof{ \norm{\normnum}{\secretsestimateof{ \releaservparamnotation}- \secretsofparam } \leq \privacythresholdl }\leq \probof{\bigcup_{\secretindex\in [\dimension]}~ \abs{\secretiestimateof{\secretindex}{ \releaservparamnotation}- \secretiofparam{\secretindex} } \leq \privacythresholdi{\secretindex} }.
\end{align*}
Hence, for any data release mechanism $\mathcal{M}_g$ that satisfies $\Delta \leq \delta_0$:
\begin{align*}
    \sup_{\secretestimatenotationvector}\probof{\bigcap_{\secretindex\in [\dimension]}~ \abs{\secretiestimateof{\secretindex}{ \releaservparamnotation}- \secretiofparam{\secretindex} } \leq  \privacythresholdi{\secretindex} } \leq \sup_{\secretestimatenotationvector}\probof{ \norm{\normnum}{\secretsestimateof{ \releaservparamnotation}- \secretsofparam } \leq \privacythresholdl }\leq \sup_{\secretestimatenotationvector}\probof{\bigcup_{\secretindex\in [\dimension]}~ \abs{\secretiestimateof{\secretindex}{ \releaservparamnotation}- \secretiofparam{\secretindex} } \leq \privacythresholdi{\secretindex} },
\end{align*}
which indicates that with a fixed distortion budget $\delta_0$, $\privacynotationinter\leq\privacynotationnorm\leq\privacynotationunion$. Therefore, we can easily get that with a privacy budget $T$, the achievable distortion lower bounds for union privacy $\unionlower$, intersection privacy $\interlower$, and $l_{\normnum}$ norm privacy $\lplower$ satisfy $\unionlower\geq\lplower\geq \interlower$. 
\end{proof}

\subsubsection{More Analysis of \texorpdfstring{$l_{\normnum}$}\text{ Norm} Privacy}

In this section, we compare the distortion lower bounds for $l_\normnum$ norm privacy with different norm order $\normnum$.

\begin{proposition}
Consider two $l_\normnum$ norm privacy metrics with norm orders $p=\alpha$ and $p=\tau$ respectively. If $\alpha\geq \tau>0$, and their tolerance ranges satisfy $\frac{\privacythresholdlp{\tau}}{\privacythresholdlp{\alpha}}\geq \dimension^{\frac{1}{\tau}-\frac{1}{\alpha}}$, given a privacy budget $T$, we have
\begin{align*}
    \Delta^{l_{\alpha}}\leq \Delta^{l_{\tau}},
\end{align*}
where $\Delta^{l_{\alpha}}$ and $\Delta^{l_{\tau}}$ are the achievable distortion lower bounds for $l_\normnum$ norm privacy with $p=\alpha$ and $p=\tau$ respectively.
\end{proposition}

\begin{proof}
We first prove that for any fixed distortion budget $\delta_0$, when $\Delta \leq \delta_0$, the achievable lower bound for $l_{\tau}$ privacy, denoted as $\privacynotationtau$, is no smaller than that for $l_{\alpha}$ privacy, denoted as $\privacynotationalp$.

For any attack strategy $\secretestimatenotationvector$ and data release mechanism $\mathcal{M}_g$ that satisfies $\Delta \leq \delta_0$, we have $\dimension^{\frac{1}{\tau}-\frac{1}{\alpha}}\norm{\alpha}{\secretsestimateof{ \releaservparamnotation}- \secretsofparam } \geq \norm{\tau}{\secretsestimateof{ \releaservparamnotation}- \secretsofparam}$. Therefore, when $\norm{\alpha}{\secretsestimateof{ \releaservparamnotation}- \secretsofparam }\leq \privacythresholdlp{\alpha}$, we can get that
\begin{align*}
\norm{\tau}{\secretsestimateof{ \releaservparamnotation}- \secretsofparam} \leq \dimension^{\frac{1}{\tau}-\frac{1}{\alpha}}\norm{\alpha}{\secretsestimateof{ \releaservparamnotation}- \secretsofparam } \leq \dimension^{\frac{1}{\tau}-\frac{1}{\alpha}}\privacythresholdlp{\alpha}\leq \privacythresholdlp{\tau},
\end{align*}
which indicates that
\begin{align*}
    \probof{\norm{\alpha}{\secretsestimateof{ \releaservparamnotation}- \secretsofparam }\leq \privacythresholdlp{\alpha}} \leq \probof{\norm{\tau}{\secretsestimateof{ \releaservparamnotation}- \secretsofparam }\leq \privacythresholdlp{\tau}}.
\end{align*}
Then we can get that for any data release mechanism $\mathcal{M}_g$ that satisfies $\Delta \leq \delta_0$:
\begin{align*}
    \sup_{\secretestimatenotationvector}\probof{\norm{\alpha}{\secretsestimateof{ \releaservparamnotation}- \secretsofparam }\leq \privacythresholdlp{\alpha}} \leq \sup_{\secretestimatenotationvector}\probof{\norm{\tau}{\secretsestimateof{ \releaservparamnotation}- \secretsofparam }\leq \privacythresholdlp{\tau}},
\end{align*}
which indicates that with a fixed distortion budget $\delta_0$, $\privacynotationalp\leq\privacynotationtau$. Hence, we can easily get that with a privacy budget $T$, the achievable distortion lower bounds for $l_{\alpha}$ privacy $\Delta^{l_{\alpha}}$ and $l_{\tau}$ privacy $\Delta^{l_{\tau}}$ satisfy $\Delta^{l_{\alpha}}\leq \Delta^{l_{\tau}}$. 
\end{proof}

\section{Case Study: \Secrets{} \texorpdfstring{$= \brc{\text{mean}, \text{SD}}^\dimension$}\text{, distribution} \texorpdfstring{$=$}\text{ Multivariate} Gaussian}
\label{app:2dGaussian}

In this section, we first focus on $2$-dimensional Gaussian distribution, and then generalize the data release mechanism for multivariate Gaussian distribution.

For $2$-dimensional Gaussian distribution $\mathcal{N}
\bra{\boldsymbol{\mu}, \Sigma}$, the distribution parameters can be represented as $\boldsymbol{\mu} = [\mu_1, \mu_2]$, and 
\begin{align*}
\Sigma = 
\begin{bmatrix}
    \sigma_1^2 & \sigma_{12} \\ \sigma_{21} & \sigma_{2}^2
\end{bmatrix}
= \begin{bmatrix}
    \cos{\alpha} & -\sin{\alpha} \\ \sin{\alpha} & \cos{\alpha}
\end{bmatrix}
\begin{bmatrix}
    \lambda_1 & 0 \\ 0 & \lambda_2
\end{bmatrix}
\begin{bmatrix}
    \cos{\alpha} & \sin{\alpha} \\ -\sin{\alpha} & \cos{\alpha}
\end{bmatrix},
\end{align*}
where $\alpha\in \brba{0, \pi}$. We can see that the 2-dimensional Gaussian distribution is determined by five independent parameters $\rvparamnotation = \bra{\mui, \muj, \lami, \lamj, \alp}$. 
We consider $\dimension$ \secrets{}, where $\dimension \leq 4$, and each \secret{} can be either mean or standard deviation of any dimension of the distribution, i.e., $\secretnotation_{\secretindex}\in\brc{\mu_1, \mu_2, \sigma_1, \sigma_2}, \ \forall \secretindex\in\brb{\dimension}$. Let $\secretset = \brc{g_i}_{i\in[d]}$ be the secret set. We first instantiate the privacy-distortion tradeoff lower bound for $2$-dimensional Gaussian in \cref{prop:2dGaussian}.

\begin{proposition}
\label{prop:2dGaussian}
For $2$-dimensional Gaussian distribution with distribution parameters $\rvparamnotation = \bra{\mui, \muj, \lami, \lamj, \alp}$, consider $d$ \secrets{} ($d\leq 4$), where each secret satisfies $\secretnotation_{\secretindex}(\theta)\in \brc{\mu_1, \mu_2, \sigma_1, \sigma_2}$, $\forall\secretindex\in \brb{\dimension}$. For any $\privacybound\in\bra{0,1}$, when $\privacynotation\leq \privacybound$, 
    \begin{align*}
\distortionnotation> 
\sqrt{\dimension} \cdot \left\lceil{\frac{1}{{1-\bra{1-\privacybound}^{{1}/{\dimension}}}}-1}\right\rceil \cdot \bra{\prod_{\secretindex\in [\dimension]} \privacythresholdi{\secretindex}}^{{1}/{\dimension}}.
    \end{align*}
\end{proposition}

The proof is shown in \cref{proof:2dGaussian}.
We then design a quantization \datamechanism{} to approximate the tradeoff lower bound. Intuitively, if the secret is a mean $\mu_i$ ($i\in\brc{1,2}$), we %
partition the range of possible values of it into intervals of lengths $\seclen_{\mu_i}$. Otherwise, we divide the ranges of possible values of $\sqrt{\lambda_1}, \sqrt{\lambda_2}$ into intervals of lengths $\seclen_\aaa$ and $\seclen_\bbb$.  The mechanism then outputs %
the midpoints of the respective intervals into which the original distribution parameters fit. Precisely, the designed mechanism is shown in \cref{mech:2dGaussian}. Here we use independent parameters $\bra{\mui, \muj, \lami, \lamj, \alp}$ so that the attacker cannot infer the value of a parameter based on any other parameters.

\begin{algorithm}[htpb]
    \LinesNumbered
	\BlankLine
	\SetKwInOut{Input}{Input}
 \SetKwInOut{Output}{Output}
	\caption{Data release mechanism for 2-dimensional Gaussian with $\dimension$ secrets.}
    \label{mech:2dGaussian}
	\Input{$\theta = \bra{\mui, \muj, \lami, \lamj, \alp}$, lower bounds $\underline{\mu_1}, \underline{\mu_2}, \aaalower, \bbblower$ for parameters $\mu_1, \mu_2, \aaa=\sqrt{\lambda_1}, \bbb=\sqrt{\lambda_2}$,  quantization intervals $s_{\mu_1}, s_{\mu_2}, s_{\aaa}, s_{\bbb}$, secret set $\mathcal{G}$.}
	\BlankLine
 \textbf{for} each $i\in\brc{1,2}$:
 
 \quad\textbf{if} $\mu_i\in\mathcal{G}$:\ \  $\mu_i' \leftarrow \underline{\mu_i} + \bra{\floor{\frac{\mu_i-\underline{\mu_i}}{s_{\mu_i}}}+0.5}\cdot s_{\mu_i}$\;
 \quad\textbf{else}:\ \  $\mu_i' \leftarrow \mu_i$\;
 \vspace{2mm}
 \textbf{if} $\brc{\sigma_1, \sigma_2}\cap \mathcal{G}\not= \emptyset$:

 \quad$\aaa' \leftarrow \underline{\aaa} + \bra{\floor{\frac{\aaa-\underline{\aaa}}{s_{\aaa}}}+0.5}\cdot s_{\aaa}$\;
 \quad$\bbb' \leftarrow \underline{\bbb} + \bra{\floor{\frac{\bbb-\underline{\bbb}}{s_{\bbb}}}+0.5}\cdot s_{\bbb}$\;
 \textbf{else}: $\aaa'\leftarrow\aaa$, $\bbb'\leftarrow\bbb$\;
\Output{Gaussian distribution with parameter $\theta' = \bra{\mu_1', \mu'_2, {\aaa'}^2, {\bbb'}^2, \alp}$.}
\end{algorithm}

For the mechanism  \privacy-\distortion{} analysis, we consider the case where the prior distributions of parameters $\mu_1, \mu_2, \sqrt{\lambda_1}, \sqrt{\lambda_2}$ are uniform.

\begin{proposition}[Mechanism privacy-distortion tradeoff]
\label{prop:mech_2dGaussian}
Under the assumption that the distribution parameters $\mu_1, \mu_2, \sqrt{\lambda_1}, \sqrt{\lambda_2}$ follow the uniform distribution, \cref{mech:2dGaussian} has %
\begin{align*}
    &\privacynotation \leq \pmu + \psigma - \pmu\psigma,\\
    &\distortionnotation = \half\sqrt{\sum_{i: \mu_i\in \mathcal{G}}\seclen_{\mu_i}^2+\mathbbm{1}_{\brc{\sigmai,\sigmaj}\cap\mathcal{G}\not= \emptyset}\cdot\bra{\seclen_{\aaa}^2+\seclen_{\bbb}^2}},
\end{align*}
where $\pmu = 1 - \prod_{i=\brc{1,2}}\bra{1 - \mathbbm{1}_{\mu_i\in\mathcal{G}}\cdot\frac{2\privacythresholdi{\mu_i}}{\seclen_{\mu_i}}}$, $\psigma = \mathbbm{1}_{\bra{\sigmai\in\mathcal{G}}\cap\bra{\sigmaj\not\in\mathcal{G}}}\cdot \psigmai + \mathbbm{1}_{\bra{\sigmai\not\in\mathcal{G}}\cap\bra{\sigmaj\in\mathcal{G}}}\cdot \psigmaj + \mathbbm{1}_{\bra{\sigmai\in\mathcal{G}}\cap\bra{\sigmaj\in\mathcal{G}}}\cdot \psigmaij$,
$\psigmai = 1 - \mathbbm{1}_{\bra{\lia>0}\cap\bra{\lib>0}}\cdot\half\lia\lib/ \seclen_\aaa \seclen_\bbb$,
$\psigmaj = 1 - \mathbbm{1}_{\bra{\lja>0}\cap\bra{\ljb>0}}\cdot\half\lja\ljb/ \seclen_\aaa \seclen_\bbb$,
$\psigmaij = 1 - \mathbbm{1}_{\bra{\lija>0}\cap\bra{\lijb>0}}\cdot\half\lija\lijb/ \seclen_\aaa \seclen_\bbb$, and
\begin{align*}
&\lia = \seclen_{\aaa} - 2\frac{\privacythresholdi{\sigmai}}{\cos{\alp}} - \sqrt{2}\privacythresholdi{\sigmai}\cos{\alp},\quad\quad\quad \lib = \seclen_{\bbb}-2\frac{\privacythresholdi{\sigmai}}{\sin{\alp}}-\sqrt{2}\privacythresholdi{\sigmai}\sin{\alp},\\
&\lja = \seclen_{\aaa} - 2\frac{\privacythresholdi{\sigmaj}}{\sin{\alp}} - \sqrt{2}\privacythresholdi{\sigmaj}\sin{\alp},\quad\quad\quad \ljb = \seclen_{\bbb}-2\frac{\privacythresholdi{\sigmaj}}{\cos{\alp}}-\sqrt{2}\privacythresholdi{\sigmaj}\cos{\alp},\\
&\lija = \seclen_{\aaa} - \max\brc{\frac{\privacythresholdi{\sigmai}}{\cos{\alp}}, \frac{\privacythresholdi{\sigmaj}}{\sin{\alp}}} - \max\brc{\frac{\privacythresholdi{\sigmai}}{\cos{\alp}}+\sqrt{2}\privacythresholdi{\sigmai}\cos{\alp}, \frac{\privacythresholdi{\sigmaj}}{\sin{\alp}}+\sqrt{2}\privacythresholdi{\sigmaj}\sin{\alp}},\\ &\lijb= \seclen_{\bbb} - \max\brc{\frac{\privacythresholdi{\sigmai}}{\sin{\alp}}, \frac{\privacythresholdi{\sigmaj}}{\cos{\alp}}} - \max\brc{\frac{\privacythresholdi{\sigmai}}{\sin{\alp}}+\sqrt{2}\privacythresholdi{\sigmai}\sin{\alp}, \frac{\privacythresholdi{\sigmaj}}{\cos{\alp}}+\sqrt{2}\privacythresholdi{\sigmaj}\cos{\alp}}.
\end{align*}

\end{proposition}
The proof is shown in \cref{proof:mech_2dGaussian}.

Drawing upon the similar idea used in the data release mechanism for $2$-dimensional Gaussian distribution, we proceed to design a general mechanism suitable for multivariate Gaussian distributions as follows.

For a $k$-dimensional Gaussian distribution ($k\in\mathbb{Z}^+$), it can be fully characterized by $3k-1$ independent parameters denoted as $\theta=\bra{\mu_1,\cdots, \mu_k, \lambda_1,\cdots, \lambda_k, \alp_1, \cdots,\alp_{k-1}}$. Here, $\mu_1,\cdots, \mu_k$ represent the means, while $\lambda_1,\cdots,\lambda_k$ and $\alp_1,\cdots,\alp_{k-1}$ correspond to the eigenvalues and eigenvectors of the covariance matrix, respectively. We consider $\dimension$ \secrets{}, where $\dimension \leq 2k$, and each \secret{} can be either mean or standard deviation of any dimension of the distribution, i.e., $\secretnotation_{\secretindex}\in\brc{\mu_1,\cdots, \mu_k, \sigma_1, \cdots, \sigma_k}, \ \forall \secretindex\in\brb{\dimension}$. Let $\secretset = \brc{g_i}_{i\in[d]}$ be the secret set.

Similar to \cref{mech:2dGaussian}, we design a quantization data release mechanism. If a secret is mean $\mu_i$ ($i\in[k]$), we 
partition the range of possible values of it into intervals of lengths $\seclen_{\mu_i}$. Otherwise, we divide the ranges of possible values for $\sqrt{\lambda_1},\cdots, \sqrt{\lambda_k}$ into intervals with lengths $\seclen_{\aaa_1},\cdots, \seclen_{\aaa_k}$. %
Subsequently, this mechanism outputs the midpoints of the respective intervals into which the original distribution parameters fit. Precisely, the designed mechanism is shown in \cref{mech:generalGaussian}.

\begin{algorithm}[htpb]
    \LinesNumbered
	\BlankLine
	\SetKwInOut{Input}{Input}
 \SetKwInOut{Output}{Output}
	\caption{Data release mechanism for multivariate Gaussian with $\dimension$ secrets.}
    \label{mech:generalGaussian}
	\Input{$\theta =\bra{\mu_1,\cdots, \mu_k, \lambda_1,\cdots, \lambda_k, \alp_1, \cdots,\alp_{k-1}}$, lower bounds $\underline{\mu_1}, \cdots, \underline{\mu_k}, \underline{\aaa_1}, \cdots, \underline{\aaa_k}$ for parameters $\mu_1, \cdots, \mu_k, \aaa_1=\sqrt{\lambda_1}, \cdots, \aaa_k=\sqrt{\lambda_k}$,  quantization intervals $\seclen_{\mu_1}, \cdots, \seclen_{\mu_k}, \seclen_{\aaa_1}, \cdots, \seclen_{\aaa_k}$, secret set $\mathcal{G}$.}
	\BlankLine
 \textbf{for} each $i\in[k]$:
 
 \quad\textbf{if} $\mu_i\in\mathcal{G}$:\ \  $\mu_i' \leftarrow \underline{\mu_i} + \bra{\floor{\frac{\mu_i-\underline{\mu_i}}{s_{\mu_i}}}+0.5}\cdot s_{\mu_i}$\;
 \quad\textbf{else}:\ \  $\mu_i' \leftarrow \mu_i$\;
 \vspace{2mm}
 \textbf{if} $\brc{\sigma_1,\cdots, \sigma_k}\cap \mathcal{G}\not= \emptyset$:

 \quad\textbf{for} each $i\in[k]$:
 \ $\aaa_i' \leftarrow \underline{\aaa_i} + \bra{\floor{\frac{\aaa_i-\underline{\aaa_i}}{s_{\aaa_i}}}+0.5}\cdot s_{\aaa_i}$\;
 \textbf{else}: \  \textbf{for} each $i\in[k]$:\ \ $\aaa_i'\leftarrow\aaa_i$\;
\Output{Gaussian distribution with parameter $\theta' = \bra{\mu_1', \cdots, \mu_k', {\aaa_1'}^2, \cdots, {\aaa_k'}^2,\alp_1, \cdots,\alp_{k-1}}$}
\end{algorithm}

\subsection{Proof of \NoCaseChange\cref{prop:2dGaussian}}
\label{proof:2dGaussian}

\begin{proof}
Let $\rvparamnotation' = \bra{\muir, \mujr, \lamir, \lamjr, \alpr}$. From \cite{givens1984class}, we have
\begin{align*}
    \distanceof{\privatedistribution} {\releasedistribution}^2
    = \bra{\mui-\muir}^2 + \bra{\muj-\mujr}^2 + \mathrm{Tr}\bra{\Sigma + \Sigma' - 2\bra{\Sigma^\frac{1}{2}\Sigma'\Sigma^\frac{1}{2}}^\frac{1}{2}}.
\end{align*}

We provide a lower bound on $\distanceof{\privatedistribution} {\releasedistribution}$ in \cref{lemma:w2_lower}.

\begin{lemma}
\label{lemma:w2_lower}
$\distanceof{\privatedistribution} {\releasedistribution}^2$ can be lower bounded as
\begin{align*}
\distanceof{\privatedistribution} {\releasedistribution}^2 \geq \bra{\mui-\muir}^2 + \bra{\muj-\mujr}^2 + \bra{\sigmai-\sigmair}^2 + \bra{\sigmaj-\sigmajr}^2.
\end{align*}
\end{lemma}

The proof is shown in \cref{proof:w2_lower}.

Based on \cref{lemma:w2_lower}, %
we can get that
\begin{align*}
\frac{\auxdistance{\rvprivatewithparam{\rvparamnotation}}{\rvprivatewithparam{\rvparamnotation'}}}{\auxrange{\rvprivatewithparam{\rvparamnotation}}{\rvprivatewithparam{\rvparamnotation'}}} 
&\geq
\frac{\sqrt{\sum_{j\in \brb{2}}\bra{\mu_j - \mu'_j}^2 + \sum_{j\in \brb{2}}\bra{\sigma_j - \sigma'_j}^2}}{2\prod_{\secretindex\in [\dimension]}\rangeiformula{\secretindex}{{\rvparamnotation}}{{\rvparamnotation'}}^{1/\dimension}}\\
& \geq 
\frac{\sqrt{\sum_{\secretindex\in [\dimension]}\rangeiformulasquare{\secretindex}{{\rvparamnotation}}{{\rvparamnotation'}}}}{2\prod_{\secretindex\in [\dimension]}\rangeiformula{\secretindex}{{\rvparamnotation}}{{\rvparamnotation'}}^{1/\dimension}}\\
& =
\half \sqrt{\frac{\sum_{\secretindex\in [\dimension]}\rangeiformulasquare{\secretindex}{{\rvparamnotation}}{{\rvparamnotation'}}}{\prod_{\secretindex\in [\dimension]}\rangeiformula{\secretindex}{{\rvparamnotation}}{{\rvparamnotation'}}^{2/\dimension}}}\\
& \geq
\half \sqrt{\dimension \cdot \frac{\prod_{\secretindex\in [\dimension]}\rangeiformula{\secretindex}{{\rvparamnotation}}{{\rvparamnotation'}}^{2/\dimension}}{\prod_{\secretindex\in [\dimension]}\rangeiformula{\secretindex}{{\rvparamnotation}}{{\rvparamnotation'}}^{2/\dimension}}}\\
& =
\frac{\sqrt{d}}{2}.
\end{align*}
\end{proof}

\subsection{Proof of \NoCaseChange\cref{lemma:w2_lower}}
\label{proof:w2_lower}

\begin{proof}
To proof \cref{lemma:w2_lower}, we first provide two lemmas as follows.
\begin{lemma}
\label{lemma:w2}
$\distanceof{\privatedistribution} {\releasedistribution}^2$ can be derived as:
\begin{align*}
    \distanceof{\privatedistribution} {\releasedistribution}^2 = 
    &\bra{\mui-\muir}^2 + \bra{\muj-\mujr}^2 + \lami + \lamir + \lamj + \lamjr\\
    &- 2\sqrt{\bra{\lami\lamir + \lamj\lamjr}\cos^2\bra{\alp - \alpr} + \bra{\lami\lamjr+\lamj\lamir}\sin^2\bra{\alp-\alpr} + 2\sqrt{\lami\lamir\lamj\lamjr}}.
\end{align*}
\end{lemma}  

The proof is shown in \cref{proof:w2}.

\begin{lemma}
\label{lemma:w2_lower_inproof}
\begin{align*}
    \sigmai \sigmair + \sigmaj \sigmajr \geq \sqrt{\bra{\lami\lamir + \lamj\lamjr}\cos^2\bra{\alp - \alpr} + \bra{\lami\lamjr+\lamj\lamir}\sin^2\bra{\alp-\alpr} + 2\sqrt{\lami\lamir\lamj\lamjr}}.
\end{align*}
\end{lemma}
The proof is in \cref{proof:sigma_inequ}

Based on \cref{lemma:w2} and \cref{lemma:w2_lower_inproof}, we have
\begin{align*}
\distanceof{\privatedistribution} {\releasedistribution}^2 
&= \bra{\mui-\muir}^2 + \bra{\muj-\mujr}^2 + \lami + \lamir + \lamj + \lamjr\\
&\quad - 2\sqrt{\bra{\lami\lamir + \lamj\lamjr}\cos^2\bra{\alp - \alpr} + \bra{\lami\lamjr+\lamj\lamir}\sin^2\bra{\alp-\alpr} + 2\sqrt{\lami\lamir\lamj\lamjr}}\\
&= \bra{\mui-\muir}^2 + \bra{\muj-\mujr}^2 + \sigmai^2 + \sigmaj^2 + {\sigmair}^2 + {\sigmajr}^2\\
&\quad - 2\sqrt{\bra{\lami\lamir + \lamj\lamjr}\cos^2\bra{\alp - \alpr} + \bra{\lami\lamjr+\lamj\lamir}\sin^2\bra{\alp-\alpr} + 2\sqrt{\lami\lamir\lamj\lamjr}}\\
&\geq \bra{\mui-\muir}^2 + \bra{\muj-\mujr}^2 + \sigmai^2 + \sigmaj^2 + {\sigmair}^2 + {\sigmajr}^2 - 2\sigmai \sigmair - 2\sigmaj \sigmajr\\
&= \bra{\mui-\muir}^2 + \bra{\muj-\mujr}^2 + \bra{\sigmai-\sigmair}^2 + \bra{\sigmaj-\sigmajr}^2.
\end{align*}
\end{proof}

\subsection{Proof of \NoCaseChange\cref{lemma:w2}}
\label{proof:w2}

\begin{proof}
    Define $A = \bra{\Sigma^\frac{1}{2}\Sigma'\Sigma^\frac{1}{2}}^\frac{1}{2}$, and we have
\begin{align*} 
\bra{\mathrm{Tr}A}^2 =  \mathrm{Tr}\bra{A^2} + 2 \det \bra{A}.
\end{align*}

We can get that
\begin{align*}
\mathrm{Tr}\bra{A^2} &= \mathrm{Tr}\bra{\Sigma^\frac{1}{2}\Sigma'\Sigma^\frac{1}{2}}
= \mathrm{Tr}\bra{\Sigma^\frac{1}{2}\Sigma^\frac{1}{2}\Sigma'}
= \mathrm{Tr}\bra{\Sigma\Sigma'},
\end{align*}
where
\begin{align*}
\Sigma\Sigma' 
&= \begin{bmatrix}
    \cos{\alp} & -\sin{\alp} \\ \sin{\alp} & \cos{\alp}
\end{bmatrix}
\begin{bmatrix}
    \lami & 0 \\ 0 & \lamj
\end{bmatrix}
\begin{bmatrix}
    \cos{\alp} & \sin{\alp} \\ -\sin{\alp} & \cos{\alp}
\end{bmatrix}
\begin{bmatrix}
    \cos{\alpr} & -\sin{\alpr} \\ \sin{\alpr} & \cos{\alpr}
\end{bmatrix}
\begin{bmatrix}
    \lamir & 0 \\ 0 & \lamjr
\end{bmatrix}
\begin{bmatrix}
    \cos{\alpr} & \sin{\alpr} \\ -\sin{\alpr} & \cos{\alpr}
\end{bmatrix}\\
&= \begin{bmatrix}
    \lami \cos{\alp} & -\lamj\sin{\alp} \\ \lami\sin{\alp} & \lamj\cos{\alp}
\end{bmatrix}
\begin{bmatrix}
    \cos\bra{\alpr-\alp} & -\sin\bra{\alpr-\alp} \\ \sin\bra{\alpr-\alp} & \cos\bra{\alpr-\alp}
\end{bmatrix}
\begin{bmatrix}
    \lamir\cos{\alpr} & \lamir\sin{\alpr} \\ -\lamjr\sin{\alpr} & \lamjr\cos{\alpr}
\end{bmatrix}.
\end{align*}
Therefore, we have
\begin{align*}
\mathrm{Tr}\bra{A^2} 
&= \mathrm{Tr}\bra{\Sigma\Sigma'}\\
&= \lami\lamir\cos{\alp}\cos{\alpr}\cos{\bra{\alpr-\alp}} - \lamj\lamir\sin{\alp}\cos{\alpr}\sin{\bra{\alpr-\alp}}+\lami\lamjr\cos{\alp}\sin{\alpr}\sin{\bra{\alpr-\alp}}\\
&\ \ \ +\lamj\lamjr\sin{\alp}\sin{\alpr}\cos{\bra{\alpr-\alp}} 
+\lami\lamir\sin{\alp}\sin{\alpr}\cos{\bra{\alpr-\alp}}+\lamj\lamir\cos{\alp}\sin{\alpr}\sin{\bra{\alpr-\alp}}\\
&\ \ \ -\lami\lamjr\sin{\alp}\cos{\alpr}\sin{\bra{\alpr-\alp}}+\lamj\lamjr\cos{\alp}\cos{\alpr}\cos{\bra{\alpr-\alp}}\\
&= \bra{\lami\lamir + \lamj\lamjr}\cos^2\bra{\alp - \alpr} + \bra{\lami\lamjr+\lamj\lamir}\sin^2\bra{\alp-\alpr}.
\end{align*}
As for $\det \bra{A}$, we have
\begin{align*}
    \det \bra{A} & = \sqrt{\det \bra{A^2}} = \sqrt{\det\bra{\Sigma^\frac{1}{2}\Sigma'\Sigma^\frac{1}{2}}} = \sqrt{\det\bra{\Sigma}\det\bra{\Sigma'}} = \sqrt{\lami\lamir\lamj\lamjr}.
\end{align*}

Therefore, we have
\begin{align*}
\bra{\mathrm{Tr}A}^2 &= \mathrm{Tr}\bra{A^2} + 2 \det \bra{A}\\
&= \bra{\lami\lamir + \lamj\lamjr}\cos^2\bra{\alp - \alpr} + \bra{\lami\lamjr+\lamj\lamir}\sin^2\bra{\alp-\alpr} + 2\sqrt{\lami\lamir\lamj\lamjr}.
\end{align*}
Hence,
\begin{align*}
\distanceof{\privatedistribution} {\releasedistribution}^2
&= \bra{\mui-\muir}^2 + \bra{\muj-\mujr}^2 + \mathrm{Tr}\bra{\Sigma + \Sigma' - 2A}\\
&= \bra{\mui-\muir}^2 + \bra{\muj-\mujr}^2 + \mathrm{Tr}\bra{\Sigma} + \mathrm{Tr}\bra{\Sigma'} - 2\mathrm{Tr}\bra{A}\\
&=\bra{\mui-\muir}^2 + \bra{\muj-\mujr}^2 + \lami + \lamir + \lamj + \lamjr\\
&\quad - 2\sqrt{\bra{\lami\lamir + \lamj\lamjr}\cos^2\bra{\alp - \alpr} + \bra{\lami\lamjr+\lamj\lamir}\sin^2\bra{\alp-\alpr} + 2\sqrt{\lami\lamir\lamj\lamjr}}.
\end{align*}
\end{proof}

\subsection{Proof of \NoCaseChange\cref{lemma:w2_lower_inproof}}
\label{proof:sigma_inequ}
\begin{proof}
Let $a = \sqrt{\lami}, b = \sqrt{\lamj}, \aar = \sqrt{\lamir}$ and $\bbr = \sqrt{\lamjr}$. We can get that
\begin{align*}
    A \triangleq \sigmai \sigmair + \sigmaj \sigmajr &= \sqrt{\bra{a^2\cos^2{\alp}+b^2\sin^2{\alp}} \bra{\aar^2\cos^2{\alpr}+\bbr^2\sin^2{\alpr}}} + \sqrt{\bra{a^2\sin^2{\alp}+b^2\cos^2{\alp}} \bra{\aar^2\sin^2{\alpr}+\bbr^2\cos^2{\alpr}}},
\end{align*}
and
\begin{align*}
B
&\triangleq\sqrt{\bra{\lami\lamir + \lamj\lamjr}\cos^2\bra{\alp - \alpr} + \bra{\lami\lamjr+\lamj\lamir}\sin^2\bra{\alp-\alpr} + 2\sqrt{\lami\lamir\lamj\lamjr}}\\
&=
\sqrt{\bra{a\aar+b\bbr}^2 \cos^2{\bra{\alp-\alpr}} + \bra{a\bbr+b\aar}^2 \sin^2{\bra{\alp-\alpr}}}.
\end{align*}

We then can derive that
\begin{align*}
    A^2 - B^2 &=
    2\sqrt{\bra{a^2\cos^2{\alp}+b^2\sin^2{\alp}} \bra{\aar^2\cos^2{\alpr}+\bbr^2\sin^2{\alpr}}\bra{a^2\sin^2{\alp}+b^2\cos^2{\alp}} \bra{\aar^2\sin^2{\alpr}+\bbr^2\cos^2{\alpr}}}\\
    &\quad
    - 2a\aar b\bbr - 2\cos{\alp}\cos{\alpr}\sin{\alp}\sin{\alpr}\bra{a^2\aar^2+b^2\bbr^2-a^2\bbr^2-b^2\aar^2}.
\end{align*}

Let
\begin{align*}
C&\triangleq \sqrt{\bra{a^2\cos^2{\alp}+b^2\sin^2{\alp}} \bra{\aar^2\cos^2{\alpr}+\bbr^2\sin^2{\alpr}}\bra{a^2\sin^2{\alp}+b^2\cos^2{\alp}} \bra{\aar^2\sin^2{\alpr}+\bbr^2\cos^2{\alpr}}},\\
D&\triangleq a\aar b\bbr + 
\cos{\alp}\cos{\alpr}\sin{\alp}\sin{\alpr}\bra{a^2\aar^2+b^2\bbr^2-a^2\bbr^2-b^2\aar^2}.
\end{align*}
We can get that
\begin{align*}
C^2 &= 
a^2\aar^2 b^2\bbr^2\bra{\cos^4{\alp}+\sin^4{\alp}}\bra{\cos^4{\alpr+\sin^4{\alpr}}} 
+ a^4\aar^2\bbr^2 \cos^2{\alp}\sin^2{\alp}\bra{\cos^4{\alpr}+\sin^4{\alpr}}\\
&\quad + a^2b^2\bbr^4\cos^2{\alpr}\sin^2{\alpr}\bra{\cos^4{\alp}+\sin^4{\alp}}
+ \aar^2b^4\bbr^2\cos^2{\alp}\sin^2{\alp}\bra{\cos^4{\alpr}+\sin^4{\alpr}}\\
&\quad + a^2\aar^4b^2\cos^2{\alpr}\sin^2{\alpr}\bra{\cos^4{\alp}+\sin^4{\alp}}\\
&= a^2\aar^2 b^2 \bbr^2 \bra{1-2\cos^2{\alp}\sin^2{\alp}}\bra{1-2\cos^2{\alpr}\sin^2{\alpr}} 
+ \bra{a^4+b^4}\aar^2\bbr^2\cos^2{\alp}\sin^2{\alp}\bra{1-2\cos^2{\alpr}\sin^2{\alpr}}\\
&\quad + \bra{\aar^4+\bbr^4}a^2b^2 \cos^2{\alpr}\sin^2{\alpr}\bra{1-2\cos^2{\alp}\sin^2{\alp}},
\end{align*}
and
\begin{align*}
D^2 &= a^2\aar^2 b^2\bbr^2\bra{1 + 4\cos^2{\alp}\cos^2{\alpr}\sin^2{\alp}\sin^2{\alpr}}
+ 2\cos{\alp}\cos{\alpr}\sin{\alp}\sin{\alpr}a\aar b\bbr\bra{a^2\aar^2 + b^2\bbr^2 - a^2\bbr^2 - b^2\aar^2}\\
&\quad - 2 \cos^2{\alp}\cos^2{\alpr}\sin^2{\alp}\sin^2{\alpr}\bra{\bra{a^4+b^4}\aar^2\bbr^2+\bra{\aar^4+\bbr^4}a^2b^2}.
\end{align*}
Then we have
\begin{align*}
C^2 - D^2 
&= \cos^2{\alp}\sin^2{\alp}\bra{a^4+b^4}\aar^2\bbr^2 + \cos^2{\alpr}\sin^2{\alpr}\bra{\aar^4+\bbr^4}a^2b^2 - 2a^2\aar^2 b^2\bbr^2 \bra{\cos^2{\alp}\sin^2{\alp}+\cos^2{\alpr}\sin^2{\alpr}}\\
&\quad- 2\cos{\alp}\cos{\alpr}\sin{\alp}\sin{\alpr}a\aar b\bbr\bra{a^2\aar^2 + b^2\bbr^2 - a^2\bbr^2 - b^2\aar^2}\\
&= \cos^2{\alp}\sin^2{\alp}\bra{a^2-b^2}^2\aar^2\bbr^2 + \cos^2{\alpr}\sin^2{\alpr}\bra{\aar^2-\bbr^2}^2a^2b^2 - 2\cos{\alp}\cos{\alpr}\sin{\alp}\sin{\alpr}a\aar b\bbr\bra{a^2-b^2}\bra{\aar^2-\bbr^2}\\
&\geq 2\sqrt{\cos^2{\alp}\sin^2{\alp}\cos^2{\alpr}\sin^2{\alpr}\aar^2\bbr^2a^2b^2\bra{a^2-b^2}^2\bra{\aar^2-\bbr^2}^2} - 2\cos{\alp}\cos{\alpr}\sin{\alp}\sin{\alpr}a\aar b\bbr\bra{a^2-b^2}\bra{\aar^2-\bbr^2}\\
&= 0
\end{align*}
Since $C\geq 0$, we have $C \geq D$. Therefore, we have
\begin{align*}
    A^2 - B^2 = 2\bra{C-D} \geq 0.
\end{align*}
Since $A\geq 0$, we have $A \geq B$, i.e., 
\begin{align*}
    \sigmai \sigmair + \sigmaj \sigmajr \geq \sqrt{\bra{\lami\lamir + \lamj\lamjr}\cos^2\bra{\alp - \alpr} + \bra{\lami\lamjr+\lamj\lamir}\sin^2\bra{\alp-\alpr} + 2\sqrt{\lami\lamir\lamj\lamjr}}.
\end{align*}
\end{proof}

\subsection{Proof of \NoCaseChange\cref{prop:mech_2dGaussian}}
\label{proof:mech_2dGaussian}

\begin{proof}
Regarding the notation, we define $\rvparamnotation' = \bra{\muir, \mujr, \lamir, \lamjr, \alpr}$, $\aaa=\sqrt{\lambda_1}, \bbb=\sqrt{{\lambda_2}}, \aar=\sqrt{\lambda'_1}, \bbr=\sqrt{{\lambda'_2}}$.

As for the distortion, since $\alpr=\alp$, we have 
\begin{align*}
    \distanceof{\privatedistribution} {\releasedistribution}^2 &= 
    \bra{\mui-\muir}^2 + \bra{\muj-\mujr}^2 + \lami + \lamir + \lamj + \lamjr\\
    &\quad - 2\sqrt{\bra{\lami\lamir + \lamj\lamjr}\cos^2\bra{\alp - \alpr} + \bra{\lami\lamjr+\lamj\lamir}\sin^2\bra{\alp-\alpr} + 2\sqrt{\lami\lamir\lamj\lamjr}}\\
    &= \bra{\mui-\muir}^2 + \bra{\muj-\mujr}^2 + \lami + \lamir + \lamj + \lamjr - 2\sqrt{\bra{\sqrt{\lami\lamir} + \sqrt{\lamj\lamjr}}^2}\\
    &= \bra{\mui-\muir}^2 + \bra{\muj-\mujr}^2 + \aaa^2 + \aar^2 + \bbb^2 + \bbr^2 - 2\aaa\aar - 2\bbb\bbr\\
    &= \bra{\mui-\muir}^2 + \bra{\muj-\mujr}^2 + \bra{\aaa^2 - \aar^2}^2 + \bra{\bbb^2 - \bbr^2}^2.\\
    &\leq \frac{1}{4}\bra{\sum_{i: \mu_i\in \mathcal{G}}\seclen_{\mu_i}^2+\mathbbm{1}_{\brc{\sigmai,\sigmaj}\cap\mathcal{G}\not= \emptyset}\cdot\bra{\seclen_{\aaa}^2+\seclen_{\bbb}^2}}.
\end{align*}
Therefore, we have
\begin{align*}
    \distortionnotation = \supdist\distanceof{\privatedistribution} {\releasedistribution}
    = \half\sqrt{\sum_{i: \mu_i\in \mathcal{G}}\seclen_{\mu_i}^2+\mathbbm{1}_{\brc{\sigmai,\sigmaj}\cap\mathcal{G}\not= \emptyset}\cdot\bra{\seclen_{\aaa}^2+\seclen_{\bbb}^2}}.
\end{align*}

As for the privacy, we first provide a lemma as follows.
\begin{lemma}
\label{lemma:2dGaussian_privacy_inproof}
Let $\ta = \frac{\privacythresholdi{\sigmai}}{\cos{\alp}}, \  \tb = \frac{\privacythresholdi{\sigmai}}{\sin{\alp}}, \ \tc = \frac{\privacythresholdi{\sigmaj}}{\sin{\alp}}$ and $\td = \frac{\privacythresholdi{\sigmaj}}{\cos{\alp}}$. For all $a, b \geq 0$, we have
\begin{align*}
    \sqrt{\bra{\aaa+\ta}^2\cos^2{\alp}+\bra{\bbb+\tb}^2\sin^2{\alp}} &\geq \sqrt{\aaa^2\cos^2{\alp}+\bbb^2\sin^2{\alp}} + \privacythresholdi{\sigmai},\\
    \sqrt{\bra{\aaa+\tc}^2\sin^2{\alp}+\bra{\bbb+\td}^2\cos^2{\alp}} &\geq \sqrt{\aaa^2\sin^2{\alp}+\bbb^2\cos^2{\alp}} + \privacythresholdi{\sigmaj}.
\end{align*}
Let $\tar = \frac{\privacythresholdi{\sigmai}}{\cos{\alp}} + \sqrt{2}\privacythresholdi{\sigmai}\cos{\alp}, \ \tbr = \frac{\privacythresholdi{\sigmai}}{\sin{\alp}} + \sqrt{2}\privacythresholdi{\sigmai}\sin{\alp}$. For all $a \geq \tar, b \geq \tbr$, we have
\begin{align*}
    \sqrt{\bra{\aaa-\tar}^2\cos^2{\alp}+\bra{\bbb-\tbr}^2\sin^2{\alp}} &\leq \sqrt{\aaa^2\cos^2{\alp}+\bbb^2\sin^2{\alp}} - \privacythresholdi{\sigmai}.
\end{align*}
Let $\tcr = \frac{\privacythresholdi{\sigmaj}}{\sin{\alp}} + \sqrt{2}\privacythresholdi{\sigmaj}\sin{\alp}$ and $\tdr = \frac{\privacythresholdi{\sigmaj}}{\cos{\alp}} + \sqrt{2}\privacythresholdi{\sigmaj}\cos{\alp}$. For all $a \geq \tcr, b \geq \tdr$, we have
\begin{align*}
    \sqrt{\bra{\aaa-\tcr}^2\sin^2{\alp}+\bra{\bbb-\tdr}^2\cos^2{\alp}} &\leq \sqrt{\aaa^2\sin^2{\alp}+\bbb^2\cos^2{\alp}} - \privacythresholdi{\sigmaj}.
\end{align*}
\end{lemma}

The proof is shown in \cref{proof:2dGaussian_privacy_inproof}.

Suppose the parameters \cref{mech:2dGaussian} releases satisfy $\aar = \aaalower+\bra{I_{\aaa}+\half}\cdot\seclen_{\aaa}$ and $\bbr = \bbblower+\bra{I_{\bbb}+\half}\cdot\seclen_{\bbb}$, and let the \secret{} value $\secretiofparam{\sigmai}$ that the optimal attack strategy guesses be $\sigmaih$. Then there exist $\aag$ and $\bbg$ that satisfy 
$
\aag^2\cos^2{\alp}+\bbg^2\sin^2{\alp} = \sigmaih^2
$,
where 
\begin{align*}
    &\brc{\bra{\aag < \aaalower + I_{\aaa}\seclen_{\aaa}} \cap \bra{\bbg < \bbblower + I_{\bbb}\seclen_{\bbb}}} 
    \cup \brc{\bra{I_{\aaa}\seclen_{\aaa} \leq \aag \leq \aaalower + \bra{I_{\aaa}+1}\cdot\seclen_{\aaa}} \cap \bra{\bbblower+I_{\bbb}\seclen_{\bbb} \leq \bbg \leq \bbblower+\bra{I_{\bbb}+1}\cdot\seclen_{\bbb}}} \\
    &\quad\cup \brc{\bra{\aag > \aaalower + \bra{\bra{I_{\aaa}+1}\cdot\seclen_{\aaa}}} \cap \bra{\bbg > \bbblower + \bra{I_{\bbb}+1}\cdot\seclen_{\bbb}}} = \text{True}.
\end{align*}
The probability of the attacker guessing the \secret{} $\sigmai$ within the tolerance range is:
\begin{align*}
    \probof{\abs{\secretiestimateof{\sigmai}{ \releaservparamnotation}- \secretiofparam{\sigmai} } \leq \privacythresholdi{\sigmai}}
    =
    \probof{\babs{\sqrt{\aag^2\cos^2{\alp}+\bbg^2\sin^2{\alp}}- \sqrt{\aaa^2\cos^2{\alp}+\bbb^2\sin^2{\alp}} } \leq \privacythresholdi{\sigmai}}
\end{align*}
Based on \cref{lemma:2dGaussian_privacy_inproof}, we have
\begin{align*}
    \sqrt{\bra{\aag-\tar}^2\cos^2{\alp}+\bra{\bbg-\tbr}^2\sin^2{\alp}} &\leq \sqrt{\aag^2\cos^2{\alp}+\bbg^2\sin^2{\alp}} - \privacythresholdi{\sigmai},
\end{align*}
when $\aag \geq \tar, \bbg \geq \tbr$, and 
\begin{align*}
    \sqrt{\bra{\aag+\ta}^2\cos^2{\alp}+\bra{\bbg+\tb}^2\sin^2{\alp}} &\geq \sqrt{\aag^2\cos^2{\alp}+\bbg^2\sin^2{\alp}} + \privacythresholdi{\sigmai}.
\end{align*}
Let $\aag = \aaalower+\bra{I_{\aaa}+t_{\aaa}}\cdot\seclen_{\aaa}$ and $\bbg = \bbblower+\bra{I_{\bbb}+t_\bbb}\cdot\seclen_{\bbb}$, where $t_{\aaa}, t_{\bbb} \in \bra{-\infty, 0}$ or $t_{\aaa}, t_{\bbb} \in \brb{0,1}$ or $t_{\aaa}, t_{\bbb} \in \bra{1, \infty}$. 
Besides, the original parameter $\aaa, \bbb$ satisfy $\aaa\in \brba{\aaalower+I_{\aaa}\cdot\seclen_{\aaa}, \ \aaalower+\bra{I_{\aaa}+1}\cdot\seclen_{\aaa}}$  and  $\bbb\in \brba{\bbblower+I_{\bbb}\cdot\seclen_{\bbb}, \ \bbblower+\bra{I_{\bbb}+1}\cdot\seclen_{\bbb}}$.
Therefore, we can get that
\begin{align*}
    \probof{\abs{\secretiestimateof{\sigmai}{ \releaservparamnotation}- \secretiofparam{\sigmai} } \leq \privacythresholdi{\sigmai}}
    &=\mathbb{P}\bigg[\bra{\sqrt{\aaa^2\cos^2{\alp}+\bbb^2\sin^2{\alp}}\geq 
\sqrt{\aag^2\cos^2{\alp}+\bbg^2\sin^2{\alp}} - \privacythresholdi{\sigmai}} \cup \\
&\quad\quad\ \  \bra{\sqrt{\aaa^2\cos^2{\alp}+\bbb^2\sin^2{\alp}}\leq 
\sqrt{\aag^2\cos^2{\alp}+\bbg^2\sin^2{\alp}} + \privacythresholdi{\sigmai}}\bigg]\\
& \leq \sup_{\aag,\bbg} \Big(1 - \max\brc{t_{\aaa}\seclen_{\aaa}-\tar, 0} \cdot \max\brc{t_{\bbb}\seclen_{\bbb}-\tbr, 0} / \seclen_{\aaa}\seclen_{\bbb}\\ 
&\quad\quad\quad - \max\brc{\bra{1-t_{\aaa}}\seclen_{\aaa}-\ta, 0} \cdot \max\brc{\bra{1-t_{\bbb}}\seclen_{\bbb}-\tb, 0} / \seclen_{\aaa}\seclen_{\bbb}\Big).
\end{align*}
Let $x = t_{\aaa}\seclen_{\aaa}-\tar$, $y = t_{\bbb}\seclen_{\bbb}-\tbr$, $\lia = \seclen_{\aaa} - 2\frac{\privacythresholdi{\sigmai}}{\cos{\alp}} - \sqrt{2}\privacythresholdi{\sigmai}\cos{\alp}$ and $\lib = \seclen_{\bbb}-2\frac{\privacythresholdi{\sigmai}}{\sin{\alp}}-\sqrt{2}\privacythresholdi{\sigmai}\sin{\alp}$, we have
\begin{align*}
    \probof{\abs{\secretiestimateof{\sigmai}{ \releaservparamnotation}- \secretiofparam{\sigmai} } \leq \privacythresholdi{\sigmai}} 
    &\leq \sup_{x, y} \Big(1 - \max\brc{x, 0}\cdot \max\brc{y, 0} / \seclen_{\aaa}\seclen_{\bbb} \\
    &\quad\quad\quad - \max\brc{\seclen_{\aaa}-\ta-\tar-x, 0} \cdot \max\brc{\seclen_{\bbb}-\tb-\tbr-y, 0} / \seclen_{\aaa}\seclen_{\bbb}\Big)\\
    & = 1 - \half{\max\brc{\seclen_{\aaa}-\ta-\tar, 0} \cdot \max\brc{\seclen_{\bbb}-\tb-\tbr, 0}} / \seclen_{\aaa}\seclen_{\bbb}\\
    & = 
    \begin{cases}
    1, & \lia \leq 0 \text{ or } \lib \leq 0\\
    1 - \half\lia\lib/ \seclen_\aaa \seclen_\bbb, & \text{otherwise}
    \end{cases}\\
    & = \psigmai.
\end{align*}

Similarly, let  $\lja = \seclen_{\aaa} - 2\frac{\privacythresholdi{\sigmaj}}{\sin{\alp}} - \sqrt{2}\privacythresholdi{\sigmaj}\sin{\alp}$ and $\ljb = \seclen_{\bbb}-2\frac{\privacythresholdi{\sigmaj}}{\cos{\alp}}-\sqrt{2}\privacythresholdi{\sigmaj}\cos{\alp}$, we can get that 
\begin{align*}
    \probof{\abs{\secretiestimateof{\sigmaj}{ \releaservparamnotation}- \secretiofparam{\sigmaj} } \leq \privacythresholdi{\sigmaj}} 
    & \leq \sup_{\aag,\bbg} \Big(1 - \max\brc{t_{\aaa}\seclen_{\aaa}-\tcr, 0} \cdot \max\brc{t_{\bbb}\seclen_{\bbb}-\tdr, 0} / \seclen_{\aaa}\seclen_{\bbb}\\ 
&\quad\quad\quad - \max\brc{\bra{1-t_{\aaa}}\seclen_{\aaa}-\tc, 0} \cdot \max\brc{\bra{1-t_{\bbb}}\seclen_{\bbb}-\td, 0} / \seclen_{\aaa}\seclen_{\bbb}\Big)\\
    &=
    \begin{cases}
    1, & \lja \leq 0 \text{ or } \ljb \leq 0\\
    1 - \half\lja\ljb/ \seclen_\aaa \seclen_\bbb, & \text{otherwise}
    \end{cases}\\
    & = \psigmaj.
\end{align*}
Let $\lija = \seclen_{\aaa} - \max\brc{\frac{\privacythresholdi{\sigmai}}{\cos{\alp}}, \frac{\privacythresholdi{\sigmaj}}{\sin{\alp}}} - \max\brc{\frac{\privacythresholdi{\sigmai}}{\cos{\alp}}+\sqrt{2}\privacythresholdi{\sigmai}\cos{\alp}, \frac{\privacythresholdi{\sigmaj}}{\sin{\alp}}+\sqrt{2}\privacythresholdi{\sigmaj}\sin{\alp}}$ and $\lijb= \seclen_{\bbb} - \max\brc{\frac{\privacythresholdi{\sigmai}}{\sin{\alp}}, \frac{\privacythresholdi{\sigmaj}}{\cos{\alp}}} - \max\brc{\frac{\privacythresholdi{\sigmai}}{\sin{\alp}}+\sqrt{2}\privacythresholdi{\sigmai}\sin{\alp}, \frac{\privacythresholdi{\sigmaj}}{\cos{\alp}}+\sqrt{2}\privacythresholdi{\sigmaj}\cos{\alp}}$, we can get that
\begin{align*}
\mathbb{P}&\bra{\abs{\secretiestimateof{\sigmai}{ \releaservparamnotation}- \secretiofparam{\sigmai}} \leq \privacythresholdi{\sigmai} \cup \abs{\secretiestimateof{\sigmaj}{ \releaservparamnotation}- \secretiofparam{\sigmaj} } \leq \privacythresholdi{\sigmaj}}\\
& \leq \sup_{\aag,\bbg} \Big(1 - \max\brc{t_{\aaa}\seclen_{\aaa}-\max\brc{\tar,\tcr}, 0} \cdot \max\brc{t_{\bbb}\seclen_{\bbb}-\max\bra{\tbr,\tdr}, 0} / \seclen_{\aaa}\seclen_{\bbb}\\ 
&\quad\quad\quad - \max\brc{\bra{1-t_{\aaa}}\seclen_{\aaa}-\max\bra{\ta,\tc}, 0} \cdot \max\brc{\bra{1-t_{\bbb}}\seclen_{\bbb}-\max\bra{\tb,\td}, 0} / \seclen_{\aaa}\seclen_{\bbb}\Big)\\
&=
\begin{cases}
    1, & \lija \leq 0 \text{ or } \lijb \leq 0\\
    1 - \half\lija\lijb/ \seclen_\aaa \seclen_\bbb, & \text{otherwise}
\end{cases}\\
& = \psigmaij.
\end{align*}

Besides, we have
\begin{align*}
\mathbb{P}\bra{\abs{\secretiestimateof{\mui}{ \releaservparamnotation}- \secretiofparam{\mui}} \leq \privacythresholdi{\mui}} &= \frac{2\privacythresholdi{\mui}}{\seclen_{\mui}},\\
\mathbb{P}\bra{\abs{\secretiestimateof{\muj}{ \releaservparamnotation}- \secretiofparam{\muj}} \leq \privacythresholdi{\muj}} &= \frac{2\privacythresholdi{\muj}}{\seclen_{\muj}},\\
\mathbb{P}\bra{\abs{\secretiestimateof{\mui}{ \releaservparamnotation}- \secretiofparam{\mui}} \leq \privacythresholdi{\mui}
\cup
\abs{\secretiestimateof{\muj}{ \releaservparamnotation}- \secretiofparam{\muj}}
\leq \privacythresholdi{\muj}} &= \frac{2\privacythresholdi{\mui}}{\seclen_{\mui}} + \frac{2\privacythresholdi{\muj}}{\seclen_{\muj}} - \frac{4\privacythresholdi{\mui}\privacythresholdi{\muj}}{\seclen_{\mui}\seclen_{\muj}}.
\end{align*}

Denote $\pmu = 1 - \prod_{i=\brc{1,2}}\bra{1 - \mathbbm{1}_{\mu_i\in\mathcal{G}}\cdot\frac{2\privacythresholdi{\mu_i}}{\seclen_{\mu_i}}}$ and $\psigma = \mathbbm{1}_{\bra{\sigmai\in\mathcal{G}}\cap\bra{\sigmaj\not\in\mathcal{G}}}\cdot \psigmai + \mathbbm{1}_{\bra{\sigmai\not\in\mathcal{G}}\cap\bra{\sigmaj\in\mathcal{G}}}\cdot \psigmaj + \mathbbm{1}_{\bra{\sigmai\in\mathcal{G}}\cap\bra{\sigmaj\in\mathcal{G}}}\cdot \psigmaij$.
We can get that the privacy of \cref{mech:2dGaussian} satisfies
\begin{align*}
    \privacynotation \leq \pmu + \psigma - \pmu\psigma.
\end{align*}

\end{proof}

\subsection{Proof of \NoCaseChange\cref{lemma:2dGaussian_privacy_inproof}}
\label{proof:2dGaussian_privacy_inproof}

\begin{proof}
    Let $A = \sqrt{\bra{\aaa+\ta}^2\cos^2{\alp}+\bra{\bbb+\tb}^2\sin^2{\alp}}$ and $B = \sqrt{\aaa^2\cos^2{\alp}+\bbb^2\sin^2{\alp}} + \privacythresholdi{\sigmai}$, we can get that
    \begin{align*}
        A^2 - B^2 &= 2\aaa\privacythresholdi{\sigmai} \cos{\alp} + 2\bbb\privacythresholdi{\sigmai}\sin{\alp} + \privacythresholdi{\sigmai}^2 - 2\privacythresholdi{\sigmai}\sqrt{\aaa^2\cos^2{\alp}+\bbb^2\sin^2{\alp}}\\
        &\geq 2\aaa\privacythresholdi{\sigmai} \cos{\alp} + 2\bbb\privacythresholdi{\sigmai}\sin{\alp} + \privacythresholdi{\sigmai}^2 - 2\privacythresholdi{\sigmai}\bra{\aaa\cos{\alp}+\bbb\sin{\alp}}\\
        &= \privacythresholdi{\sigmai}^2\\
        &\geq 0.
    \end{align*}
Since $A\geq 0$, we have $A \geq B$, i.e., 
\begin{align*}
    \sqrt{\bra{\aaa+\ta}^2\cos^2{\alp}+\bra{\bbb+\tb}^2\sin^2{\alp}} &\geq \sqrt{\aaa^2\cos^2{\alp}+\bbb^2\sin^2{\alp}} + \privacythresholdi{\sigmai}.
\end{align*}
Similarly, we can get that
\begin{align*}
    \sqrt{\bra{\aaa+\tc}^2\sin^2{\alp}+\bra{\bbb+\td}^2\cos^2{\alp}} &\geq \sqrt{\aaa^2\sin^2{\alp}+\bbb^2\cos^2{\alp}} + \privacythresholdi{\sigmaj}.
\end{align*}

Let $C=\sqrt{\bra{\aaa-\tar}^2\cos^2{\alp}+\bra{\bbb-\tbr}^2\sin^2{\alp}}, \ D=  \sqrt{\aaa^2\cos^2{\alp}+\bbb^2\sin^2{\alp}} - \privacythresholdi{\sigmai}$. We have 
\begin{align*}
    C^2 - D^2 &= \privacythresholdi{\sigmai}^2 - 2\aaa\privacythresholdi{\sigmai} \cos{\alp} - 2\bbb\privacythresholdi{\sigmai}\sin{\alp} + 2\privacythresholdi{\sigmai}^2 \bra{\cos^4{\alp}+\sin^4{\alp}} -2\sqrt{2}\aaa\privacythresholdi{\sigmai}\cos^3{\alp}-2\sqrt{2}\bbb\privacythresholdi{\sigmai}\sin^3{\alp} + 2\sqrt{2}\privacythresholdi{\sigmai}^2\\
    &\quad + 2\privacythresholdi{\sigmai} \sqrt{\aaa^2\cos^2{\alp}+\bbb^2\sin^2{\alp}}\\
    &\leq \privacythresholdi{\sigmai}^2 + 2\privacythresholdi{\sigmai}^2 \bra{\cos^4{\alp}+\sin^4{\alp}} -2\sqrt{2}\aaa\privacythresholdi{\sigmai}\cos^3{\alp}-2\sqrt{2}\bbb\privacythresholdi{\sigmai}\sin^3{\alp} + 2\sqrt{2}\privacythresholdi{\sigmai}^2\\
    &\leq \privacythresholdi{\sigmai}^2 + 2\privacythresholdi{\sigmai}^2 \bra{\cos^4{\alp}+\sin^4{\alp}} -2\sqrt{2}\bra{\frac{\privacythresholdi{\sigmai}}{\cos{\alp}} + \sqrt{2}\privacythresholdi{\sigmai}\cos{\alp}}\privacythresholdi{\sigmai}\cos^3{\alp}\\
    & \quad -2\sqrt{2}\bra{\frac{\privacythresholdi{\sigmai}}{\sin{\alp}} + \sqrt{2}\privacythresholdi{\sigmai}\sin{\alp}}\privacythresholdi{\sigmai}\sin^3{\alp} + 2\sqrt{2}\privacythresholdi{\sigmai}^2\\
    &= \privacythresholdi{\sigmai}^2 - 2\privacythresholdi{\sigmai}^2 \bra{\cos^4{\alp}+\sin^4{\alp}}\\
    &\leq 0.
\end{align*}
Since $D\geq 0$, we have $C\leq D$, i.e., 
\begin{align*}
    \sqrt{\bra{\aaa-\tar}^2\cos^2{\alp}+\bra{\bbb-\tbr}^2\sin^2{\alp}} &\leq \sqrt{\aaa^2\cos^2{\alp}+\bbb^2\sin^2{\alp}} - \privacythresholdi{\sigmai}.
\end{align*}
Similarly, we can get that
\begin{align*}
    \sqrt{\bra{\aaa-\tcr}^2\sin^2{\alp}+\bra{\bbb-\tdr}^2\cos^2{\alp}} &\leq \sqrt{\aaa^2\sin^2{\alp}+\bbb^2\cos^2{\alp}} - \privacythresholdi{\sigmaj}.
\end{align*}
\end{proof}

\section{Case studies under Alternative Privacy Metrics}
\label{app:mech_analysis}

Under intersection privacy, union privacy and $l_{\normnum}$ norm privacy metrics, we instantiate the privacy-distortion lower bounds for multivariate Gaussian with dimensionally independent variables, and analyze the performance of \cref{mech:dGaussian_diagnol}.

\subsection{Intersection Privacy}

Under the intersection privacy metric, we focus on the multivariate Gaussian with dimensionally independent variables, and provide the privacy-distortion lower bound in \cref{prop:multiG_inter}.

\begin{proposition}
\label{prop:multiG_inter}
    For $\dimgaussian$-dimensional Gaussian distribution with diagonal covariance matrix and distribution parameters $\rvparamnotation = \bra{\mu_1, \cdots, \mu_{\dimgaussian}, \sigma_1, \cdots, \sigma_{\dimgaussian}}$, consider $d$ \secrets{} ($d\leq 2\dimgaussian$), where each secret satisfies $\secretnotation_{\secretindex}(\theta)\in \brc{\mu_1, \cdots, \mu_{\dimgaussian}, \sigma_1, \cdots, \sigma_{\dimgaussian}}$, $\forall\secretindex\in \brb{\dimension}$. For any $\privacybound\in\bra{0,1}$, when $\privacynotation\leq \privacybound$, 
    \begin{align*} 
\distortionnotation > 
\sqrt{\dimension}\cdot\left\lceil{\frac{1}{\privacybound}}\right\rceil^{1/\dimension} \cdot \bra{\prod_{\secretindex\in [\dimension]} \privacythresholdi{\secretindex}}^{{1}/{\dimension}}-\frac{1}{\sqrt{\dimension}}\sum_{\secretindex\in [\dimension]} \privacythresholdi{\secretindex}.
    \end{align*}
\end{proposition}

The proof is shown in \cref{proof:multiG_inter}. We then analyze the performance of \cref{mech:dGaussian_diagnol} under intersection privacy as follows.

\begin{proposition}[Mechanism privacy-distortion tradeoff under intersection privacy]
\label{prop:multiG_inter_performance}
Under the assumption that secret distribution parameters $g_1, \cdots, g_\dimension$ follow the uniform distribution, \cref{mech:dGaussian_diagnol} has %
\begin{align*}
    &\privacynotationinter = \prod_{\secretindex\in\brb{\dimension}} \frac{2\privacythresholdi{i}}{\seclen_{g_i}},\\
    &\distortionnotation = \sqrt{\sum_{\secretindex\in\brb{\dimension}}{\seclen_{g_i}}^2}< \constant\cdot\opt.
\end{align*}
where $\constant$ is a constant depending on tolerance ranges and the interval lengths of the mechanism, and $\opt$ is the optimal achievable distortion under the privacy achieved by $\cref{mech:dGaussian_diagnol}$.
\end{proposition}
The proof is shown in \cref{proof:multiG_inter_performance}. From \cref{prop:multiG_inter_performance} we know that \cref{mech:dGaussian_diagnol} is order-optimal with a constant multiplication factor.

\subsubsection{Proof of \cref{prop:multiG_inter}}
\label{proof:multiG_inter}

Define $\rvparamnotation' = \bra{\mu'_1, \cdots, \mu'_{\dimgaussian}, \sigma'_1, \cdots, \sigma'_{\dimgaussian}}$. 
Based on \cref{lemma:D_lowerbound_multi_Gaussian}, we can get that
\begin{align*}
\frac{\auxdistance{\rvprivatewithparam{\rvparamnotation}}{\rvprivatewithparam{\rvparamnotation'}}}{\auxrange{\rvprivatewithparam{\rvparamnotation}}{\rvprivatewithparam{\rvparamnotation'}}} 
&\geq
\frac{\sqrt{\sum_{j\in \brb{k}}\bra{\mu_j - \mu'_j}^2 + \sum_{j\in \brb{k}}\bra{\sigma_j - \sigma'_j}^2}}{\frac{2}{\dimension}\sum_{\secretindex\in [\dimension]}\rangeiformula{\secretindex}{{\rvparamnotation}}{{\rvparamnotation'}}}\\
& \geq 
\frac{\sqrt{\sum_{\secretindex\in [\dimension]}\rangeiformulasquare{\secretindex}{{\rvparamnotation}}{{\rvparamnotation'}}}}{\frac{2}{\dimension}\sum_{\secretindex\in [\dimension]}\rangeiformula{\secretindex}{{\rvparamnotation}}{{\rvparamnotation'}}}\\
& \geq 
\frac{\frac{1}{\sqrt{\dimension}}\sum_{\secretindex\in [\dimension]}\rangeiformula{\secretindex}{{\rvparamnotation}}{{\rvparamnotation'}}}{\frac{2}{\dimension}\sum_{\secretindex\in [\dimension]}\rangeiformula{\secretindex}{{\rvparamnotation}}{{\rvparamnotation'}}}
\\
& =
\frac{\sqrt{d}}{2}.
\end{align*}

Therefore, we have
\begin{align*}
    \ratio^{\text{inter}} = \inf_{\rvparamnotation_1, \rvparamnotation_2 \in\support{\paramdistribution}}
\frac{\auxdistance{\rvprivatewithparam{\rvparamnotation_1}}{\rvprivatewithparam{\rvparamnotation_2}}}{\auxrange{\rvprivatewithparam{\rvparamnotation_1}}{\rvprivatewithparam{\rvparamnotation_2}}} = \frac{\sqrt{\dimension}}{2}.
\end{align*}

Based on \cref{thm:trade_off_intersection}, we can get that
\begin{align*}
    \distortionnotation > 
\sqrt{\dimension}\cdot\left\lceil{\frac{1}{\privacybound}}\right\rceil^{1/\dimension} \cdot \bra{\prod_{\secretindex\in [\dimension]} \privacythresholdi{\secretindex}}^{{1}/{\dimension}}-\frac{1}{\sqrt{\dimension}}\sum_{\secretindex\in [\dimension]} \privacythresholdi{\secretindex}.
\end{align*}

\subsubsection{Proof of \cref{prop:multiG_inter_performance}}
\label{proof:multiG_inter_performance}

\begin{proof}

Based on \cref{lemma:D_lowerbound_multi_Gaussian}, we can easily can get that the distortion $\Delta$ of \cref{mech:dGaussian_diagnol} is
\begin{align*}
\distortionnotation = \sqrt{\sum_{\secretindex\in\brb{\dimension}}{\seclen_{g_i}}^2}.
\end{align*}

Since the secret distribution parameters are independent of each other and follow the uniform distributions, we can get that the privacy of \cref{mech:dGaussian_diagnol} is
\begin{align*}
\privacynotationinter &= ~\sup_{\secretestimatenotationvector}~ \probof{\bigcap_{\secretindex\in [\dimension]}~ \abs{\secretiestimateof{\secretindex}{ \releaservparamnotation}- \secretiofparam{\secretindex} } \leq \privacythresholdi{\secretindex} }\\
&=~\sup_{\secretestimatenotationvector}~ \prod_{\secretindex\in\brb{\dimension}}\probof{ \abs{\secretiestimateof{\secretindex}{ \releaservparamnotation}- \secretiofparam{\secretindex} } \leq \privacythresholdi{\secretindex} }\\
&=\prod_{\secretindex\in\brb{\dimension}} \frac{2\privacythresholdi{i}}{\seclen_{g_i}}.
\end{align*}

From \cref{prop:multiG_inter}, we know that the optimal achievable distortion $\Delta_{opt}$ satisfy
\begin{align*}
\Delta_{opt} &>\sqrt{\dimension}\cdot\left\lceil{\frac{1}{\prod_{\secretindex\in\brb{\dimension}} \frac{2\privacythresholdi{i}}{\seclen_{g_i}}}}\right\rceil^{1/\dimension} \cdot \bra{\prod_{\secretindex\in [\dimension]} \privacythresholdi{\secretindex}}^{{1}/{\dimension}}-\frac{1}{\sqrt{\dimension}}\sum_{\secretindex\in [\dimension]} \privacythresholdi{\secretindex}\\
&= \frac{\sqrt{\dimension}}{2}\cdot\bra{\prod_{\secretindex\in\brb{\dimension}} \seclen_{g_i}}^{1/\dimension}-\frac{1}{\sqrt{\dimension}}\sum_{\secretindex\in [\dimension]} \privacythresholdi{\secretindex}.
\end{align*}

Let $k=\frac{\Delta}{\Delta_{opt}}$, $x_i = \frac{\privacythresholdi{i}}{\seclen_{g_i}}, \forall \secretindex\in\brb{\dimension}$, $c_1 = \min_{\secretindex\in\brb{\dimension}}\brc{x_i}$, and $c_2 = \max_{\secretindex\in\brb{\dimension}}\brc{x_i}$, we have
\begin{align*}
k &< \frac{2\sqrt{\sum_{\secretindex\in\brb{\dimension}}{\seclen_{g_i}}^2}}{\sqrt{\dimension}\cdot\bra{\prod_{\secretindex\in\brb{\dimension}} \seclen_{g_i}}^{1/\dimension}-\frac{2}{\sqrt{\dimension}}\sum_{\secretindex\in [\dimension]} \privacythresholdi{\secretindex}}\\
& = \frac{2\sqrt{\sum_{\secretindex\in\brb{\dimension}}{\bra{\frac{\privacythresholdi{i}}{x_i}}}^2}}{\sqrt{\dimension}\cdot\bra{\prod_{\secretindex\in\brb{\dimension}} \frac{\privacythresholdi{i}}{x_i}}^{1/\dimension}-\frac{2}{\sqrt{\dimension}}\sum_{\secretindex\in [\dimension]} \privacythresholdi{\secretindex}}\\
&\leq \frac{2c_2 \sqrt{\sum_{i\in[d]}\epsilon_i^2}}{c_1\sqrt{\dimension}\cdot\bra{\prod_{\secretindex\in\brb{\dimension}} \privacythresholdi{i}}^{1/\dimension}-\frac{2c_1c_2}{\sqrt{\dimension}}\sum_{\secretindex\in [\dimension]} \privacythresholdi{\secretindex}}\\
& = \frac{2c_2 \sqrt{\sum_{i\in[d]}\epsilon_i^2/\dimension}}{c_1\bra{\prod_{\secretindex\in\brb{\dimension}} \privacythresholdi{i}}^{1/\dimension}-2c_1c_2\sum_{\secretindex\in [\dimension]} \privacythresholdi{\secretindex}/\dimension}\\
&\triangleq c_{\epsilon,s}.
\end{align*}

Therefore, we can get that
\begin{align*}
\distortionnotation = k\Delta_{opt}< c_{\epsilon,s}\opt,
\end{align*}
where $c_{\epsilon,s}$ is a constant depending on tolerance ranges and the interval lengths of the mechanism.

Specifically, when $\epsilon_1=\cdots=\epsilon_d$, and the designed data released mechanism satisfy $\frac{\privacythresholdi{1}}{\seclen_{g_1}}=\cdots =\frac{\privacythresholdi{d}}{\seclen_{g_d}}\leq \frac{1}{6}$, we can get that $\Delta< 3\Delta_{opt}$.
\end{proof}

\subsection{\UoI{}}

Under the \uoi{} metric, we focus on the multivariate Gaussian with dimensionally independent variables, and provide the privacy-distortion lower bound in \cref{prop:multiG_uoi}.

\begin{proposition}
\label{prop:multiG_uoi}
    For $\dimgaussian$-dimensional Gaussian distribution with diagonal covariance matrix and distribution parameters $\rvparamnotation = \bra{\mu_1, \cdots, \mu_{\dimgaussian}, \sigma_1, \cdots, \sigma_{\dimgaussian}}$, consider $d$ \secrets{} ($d\leq 2\dimgaussian$), where each secret satisfies $\secretnotation_{\secretindex}(\theta)\in \brc{\mu_1, \cdots, \mu_{\dimgaussian}, \sigma_1, \cdots, \sigma_{\dimgaussian}}$, $\forall\secretindex\in \brb{\dimension}$. For any $\privacybound\in\bra{0,1}$, when $\privacynotation\leq \privacybound$, 
    \begin{align*} 
\distortionnotation > 
\sqrt{\dimension} \left\lceil{\frac{1}{\bra{1-\bra{1-\privacybound}^{{1}/{\groupsize}}}^{\groupsize/\dimension}}}\right\rceil \bra{\prod_{\secretindex\in[\dimension]}\privacythresholdi{\secretindex}}^{1/\dimension}-\frac{1}{\sqrt{\dimension}}\sum_{\secretindex\in [\dimension]} \privacythresholdi{\secretindex}.
    \end{align*}
\end{proposition}

The proof is shown in \cref{proof:multiG_uoi}. We then analyze the performance of \cref{mech:dGaussian_diagnol} under \uoi{} as follows.

\begin{proposition}[Mechanism privacy-distortion tradeoff under \uoi{}]
\label{prop:multiG_uoi_performance}
Denote $\groupset$ as the set of the secrets groups, and $\groupofindex$ as the secret index set of group $\group\in\groupset$.
Under the assumption that secret distribution parameters $g_1, \cdots, g_\dimension$ follow the uniform distribution, \cref{mech:dGaussian_diagnol} has %
\begin{align*}
    &\privacynotationgroup = 1-\prod_{\group\in\groupset}\bra{1 - \prod_{\secretindex\in \groupofindex}\frac{2\privacythresholdi{i}}{\seclen_{g_i}}},\\
    &\distortionnotation = \sqrt{\sum_{\secretindex\in\brb{\dimension}}{\seclen_{g_i}}^2}< c_{\privacythreshold,\seclen,\groupset}\cdot\opt.
\end{align*}
where $c_{\privacythreshold,\seclen,\groupset}$ is a constant depending on tolerance ranges, the interval lengths of the mechanism, and the set of secret groups $\groupset$. $\opt$ is the optimal achievable distortion under the privacy achieved by $\cref{mech:dGaussian_diagnol}$.
\end{proposition}
The proof is shown in \cref{proof:multiG_uoi_performance}. From \cref{prop:multiG_uoi_performance} we know that \cref{mech:dGaussian_diagnol} is order-optimal with a constant multiplication factor.

For the $1$-dimensional Gaussian scenario where $\theta=\bra{\mu, \sigma}$, when $\mu$ and $\sigma$  are treated as distinct secret groups (i.e., $\groupset=\brc{\brc{\mu}, \brc{\sigma}}$), the \uoi{} is equivalent to the union privacy.  If $\mu$ and $\sigma$ are in the same secret group (i.e., $\groupset=\brc{\brc{\mu, \sigma}}$), the \uoi{} is equivalent to the intersection privacy.

\subsubsection{Proof of \cref{prop:multiG_uoi}}
\label{proof:multiG_uoi}

Define $\rvparamnotation' = \bra{\mu'_1, \cdots, \mu'_{\dimgaussian}, \sigma'_1, \cdots, \sigma'_{\dimgaussian}}$. 
Based on \cref{lemma:D_lowerbound_multi_Gaussian}, we can get that
\begin{align*}
\frac{\auxdistance{\rvprivatewithparam{\rvparamnotation}}{\rvprivatewithparam{\rvparamnotation'}}}{\auxrange{\rvprivatewithparam{\rvparamnotation}}{\rvprivatewithparam{\rvparamnotation'}}} 
&\geq
\frac{\sqrt{\sum_{j\in \brb{k}}\bra{\mu_j - \mu'_j}^2 + \sum_{j\in \brb{k}}\bra{\sigma_j - \sigma'_j}^2}}{\frac{2}{\dimension}\sum_{\secretindex\in [\dimension]}\rangeiformula{\secretindex}{{\rvparamnotation}}{{\rvparamnotation'}}}\\
& \geq 
\frac{\sqrt{\sum_{\secretindex\in [\dimension]}\rangeiformulasquare{\secretindex}{{\rvparamnotation}}{{\rvparamnotation'}}}}{\frac{2}{\dimension}\sum_{\secretindex\in [\dimension]}\rangeiformula{\secretindex}{{\rvparamnotation}}{{\rvparamnotation'}}}\\
& \geq 
\frac{\frac{1}{\sqrt{\dimension}}\sum_{\secretindex\in [\dimension]}\rangeiformula{\secretindex}{{\rvparamnotation}}{{\rvparamnotation'}}}{\frac{2}{\dimension}\sum_{\secretindex\in [\dimension]}\rangeiformula{\secretindex}{{\rvparamnotation}}{{\rvparamnotation'}}}
\\
& =
\frac{\sqrt{d}}{2}.
\end{align*}

Therefore, we have
\begin{align*}
    \ratio^{\text{group}} = \inf_{\rvparamnotation_1, \rvparamnotation_2 \in\support{\paramdistribution}}
\frac{\auxdistance{\rvprivatewithparam{\rvparamnotation_1}}{\rvprivatewithparam{\rvparamnotation_2}}}{\auxrange{\rvprivatewithparam{\rvparamnotation_1}}{\rvprivatewithparam{\rvparamnotation_2}}} = \frac{\sqrt{\dimension}}{2}.
\end{align*}

Based on \cref{thm:trade_off_mix}, we can get that
\begin{align*}
\distortionnotation > 
\sqrt{\dimension} \left\lceil{\frac{1}{\bra{1-\bra{1-\privacybound}^{{1}/{\groupsize}}}^{\groupsize/\dimension}}}\right\rceil \bra{\prod_{\secretindex\in[\dimension]}\privacythresholdi{\secretindex}}^{1/\dimension}-\frac{1}{\sqrt{\dimension}}\sum_{\secretindex\in [\dimension]} \privacythresholdi{\secretindex}.
\end{align*}

\subsubsection{Proof of \cref{prop:multiG_uoi_performance}}
\label{proof:multiG_uoi_performance}

\begin{proof}

Based on \cref{lemma:D_lowerbound_multi_Gaussian}, we can easily can get that the distortion $\Delta$ of \cref{mech:dGaussian_diagnol} is
\begin{align*}
\distortionnotation = \sqrt{\sum_{\secretindex\in\brb{\dimension}}{\seclen_{g_i}}^2}.
\end{align*}

Since the secret distribution parameters are independent of each other and follow the uniform distributions, we can get that the privacy of \cref{mech:dGaussian_diagnol} is
\begin{align*}
\privacynotationgroup &= ~\sup_{\secretestimatenotationvector} ~\probof{\bigcup_{\group\in\groupset}\bra{\bigcap_{\secretindex\in \groupofindex}~ \abs{\secretiestimateof{\secretindex}{ \releaservparamnotation}- \secretiofparam{\secretindex} } \leq \privacythresholdi{\secretindex} }}\\
&=1-~\sup_{\secretestimatenotationvector} ~\probof{\bigcap_{\group\in\groupset}\bra{\bigcup_{\secretindex\in \groupofindex}~ \abs{\secretiestimateof{\secretindex}{ \releaservparamnotation}- \secretiofparam{\secretindex} } > \privacythresholdi{\secretindex} }}\\
&=1-~\sup_{\secretestimatenotationvector} ~\prod_{\group\in\groupset}\probof{{\bigcup_{\secretindex\in \groupofindex}~ \abs{\secretiestimateof{\secretindex}{ \releaservparamnotation}- \secretiofparam{\secretindex} } > \privacythresholdi{\secretindex} }}\\
&=1-\prod_{\group\in\groupset}\bra{1 - \prod_{\secretindex\in \groupofindex}\frac{2\privacythresholdi{i}}{\seclen_{g_i}}}.
\end{align*}

From \cref{prop:multiG_uoi}, we know that the optimal achievable distortion $\Delta_{opt}$ satisfy
\begin{align*}
\Delta_{opt} &>\sqrt{\dimension} \left\lceil{\frac{1}{\bra{1-\brb{\prod_{\group\in\groupset}\bra{1 - \prod_{\secretindex\in \groupofindex}\frac{2\privacythresholdi{i}}{\seclen_{g_i}}}}^{{1}/{\groupsize}}}^{\groupsize/\dimension}}}\right\rceil \bra{\prod_{\secretindex\in[\dimension]}\privacythresholdi{\secretindex}}^{1/\dimension}-\frac{1}{\sqrt{\dimension}}\sum_{\secretindex\in [\dimension]} \privacythresholdi{\secretindex}.
\end{align*}

Let $k=\frac{\Delta}{\Delta_{opt}}$, $x_i = \frac{\privacythresholdi{i}}{\seclen_{g_i}}, \forall \secretindex\in\brb{\dimension}$, $c_1 = \min_{\secretindex\in\brb{\dimension}}\brc{x_i}$, and $c_2 = \max_{\secretindex\in\brb{\dimension}}\brc{x_i}$, we have
\begin{align*}
k &< \frac{\sqrt{\sum_{\secretindex\in\brb{\dimension}}{\bra{\frac{\privacythresholdi{i}}{x_i}}}^2}}{\sqrt{\dimension} \left\lceil{\frac{1}{\bra{1-\brb{\prod_{\group\in\groupset}\bra{1 - \prod_{\secretindex\in \groupofindex}2x_i}}^{{1}/{\groupsize}}}^{\groupsize/\dimension}}}\right\rceil \bra{\prod_{\secretindex\in[\dimension]}\privacythresholdi{\secretindex}}^{1/\dimension}-\frac{2}{\sqrt{\dimension}}\sum_{\secretindex\in [\dimension]} \privacythresholdi{\secretindex}}\\
&\leq \frac{\sqrt{\sum_{i\in[d]}\epsilon_i^2}}{c_1\sqrt{\dimension} \cdot{\frac{1}{\bra{1-\brb{\prod_{\group\in\groupset}\bra{1 - \bra{2c_2}^{|\group|}}}^{{1}/{\groupsize}}}^{\groupsize/\dimension}}}\cdot\bra{\prod_{\secretindex\in[\dimension]}\privacythresholdi{\secretindex}}^{1/\dimension}-\frac{2c_1}{\sqrt{\dimension}}\sum_{\secretindex\in [\dimension]} \privacythresholdi{\secretindex}}\\
& = \frac{ \sqrt{\sum_{i\in[d]}\epsilon_i^2/\dimension}}{{\frac{c_1}{\bra{1-\brb{\prod_{\group\in\groupset}\bra{1 - \bra{2c_2}^{|\group|}}}^{{1}/{\groupsize}}}^{\groupsize/\dimension}}}\bra{\prod_{\secretindex\in\brb{\dimension}} \privacythresholdi{i}}^{1/\dimension}-c_1\sum_{\secretindex\in [\dimension]} \privacythresholdi{\secretindex}/\dimension}\\
&\triangleq c_{\epsilon,s,\groupset}.
\end{align*}

Therefore, we can get that
\begin{align*}
\distortionnotation = k\Delta_{opt}< c_{\epsilon,s,\groupset}\cdot\opt,
\end{align*}
where $c_{\epsilon,s,\groupset}$ is a constant depending on the tolerance ranges, the interval lengths of the mechanism, and the set of secret groups $\groupset$.

Specifically, we can get that $\Delta< 3\Delta_{opt}$ when the size of each secret group are the same (i.e., $|\group_{1}|=|\group_{2}|, \forall\group_{1}, \group_{2} \in \groupset$ ), $\epsilon_1=\cdots=\epsilon_d$, and the designed data released mechanism satisfy $\frac{\privacythresholdi{1}}{\seclen_{g_1}}=\cdots =\frac{\privacythresholdi{d}}{\seclen_{g_d}}\leq \frac{1}{6}$.
\end{proof}

\subsection{\texorpdfstring{$l_\normnum$}\text{ Norm} Privacy}

Under the $l_\normnum$ norm privacy metric, we focus on the multivariate Gaussian with dimensionally independent variables, and provide the privacy-distortion lower bound in \cref{prop:multiG_norm}.

\begin{proposition}
\label{prop:multiG_norm}
    For $\dimgaussian$-dimensional Gaussian distribution with diagonal covariance matrix and distribution parameters $\rvparamnotation = \bra{\mu_1, \cdots, \mu_{\dimgaussian}, \sigma_1, \cdots, \sigma_{\dimgaussian}}$, consider $d$ \secrets{} ($d\leq 2\dimgaussian$), where each secret satisfies $\secretnotation_{\secretindex}(\theta)\in \brc{\mu_1, \cdots, \mu_{\dimgaussian}, \sigma_1, \cdots, \sigma_{\dimgaussian}}$, $\forall\secretindex\in \brb{\dimension}$. For any $\privacybound\in\bra{0,1}$, when $\privacynotation\leq \privacybound$, 
    \begin{align*} 
\distortionnotation > 
\sqrt{\dimension}\cdot\bra{\left\lceil{\frac{1}{\privacybound}}\right\rceil^{1/\dimension}-1}\cdot\privacythresholdl/\dimension^{\frac{1}{\normnum}}.
    \end{align*}
\end{proposition}

The proof is shown in \cref{proof:multiG_norm}. We then analyze the performance of \cref{mech:dGaussian_diagnol} under $l_\normnum$ norm privacy as follows.

\begin{proposition}[Mechanism privacy-distortion tradeoff under $l_\normnum$ norm privacy]
\label{prop:multiG_norm_performance}
Under the assumption that secret distribution parameters $g_1, \cdots, g_\dimension$ follow the uniform distribution, \cref{mech:dGaussian_diagnol} has %
\begin{align*}
    &\privacynotationlp \leq 1 - \prod_{\secretindex\in\brb{\dimension}} \bra{1-\frac{2\privacythresholdl}{\dimension^{\frac{1}{\normnum}}\cdot\seclen_{g_i}}},\\
    &\distortionnotation = \sqrt{\sum_{\secretindex\in\brb{\dimension}}{\seclen_{g_i}}^2}.%
\end{align*}
\end{proposition}
The proof is shown in \cref{proof:multiG_norm_performance}. From \cref{prop:multiG_norm_performance} we know that \cref{mech:dGaussian_diagnol} is order-optimal with a constant multiplication factor.

\subsubsection{Proof of \cref{prop:multiG_norm}}
\label{proof:multiG_norm}

Define $\rvparamnotation' = \bra{\mu'_1, \cdots, \mu'_{\dimgaussian}, \sigma'_1, \cdots, \sigma'_{\dimgaussian}}$. 
Based on \cref{lemma:D_lowerbound_multi_Gaussian}, we can get that
\begin{align*}
\frac{\auxdistance{\rvprivatewithparam{\rvparamnotation}}{\rvprivatewithparam{\rvparamnotation'}}}{\auxrange{\rvprivatewithparam{\rvparamnotation}}{\rvprivatewithparam{\rvparamnotation'}}} 
&\geq
\frac{\sqrt{\sum_{j\in \brb{k}}\bra{\mu_j - \mu'_j}^2 + \sum_{j\in \brb{k}}\bra{\sigma_j - \sigma'_j}^2}}{\frac{2}{\dimension}\sum_{\secretindex\in [\dimension]}\rangeiformula{\secretindex}{{\rvparamnotation}}{{\rvparamnotation'}}}\\
& \geq 
\frac{\sqrt{\sum_{\secretindex\in [\dimension]}\rangeiformulasquare{\secretindex}{{\rvparamnotation}}{{\rvparamnotation'}}}}{\frac{2}{\dimension}\sum_{\secretindex\in [\dimension]}\rangeiformula{\secretindex}{{\rvparamnotation}}{{\rvparamnotation'}}}\\
& \geq 
\frac{\frac{1}{\sqrt{\dimension}}\sum_{\secretindex\in [\dimension]}\rangeiformula{\secretindex}{{\rvparamnotation}}{{\rvparamnotation'}}}{\frac{2}{\dimension}\sum_{\secretindex\in [\dimension]}\rangeiformula{\secretindex}{{\rvparamnotation}}{{\rvparamnotation'}}}
\\
& =
\frac{\sqrt{d}}{2}.
\end{align*}

Therefore, we have
\begin{align*}
    \ratio^{l_p} = \inf_{\rvparamnotation_1, \rvparamnotation_2 \in\support{\paramdistribution}}
\frac{\auxdistance{\rvprivatewithparam{\rvparamnotation_1}}{\rvprivatewithparam{\rvparamnotation_2}}}{\auxrange{\rvprivatewithparam{\rvparamnotation_1}}{\rvprivatewithparam{\rvparamnotation_2}}} = \frac{\sqrt{\dimension}}{2}.
\end{align*}

Based on \cref{thm:trade_off_norm}, we can get that
\begin{align*}
    \distortionnotation > 
\sqrt{\dimension}\cdot\bra{\left\lceil{\frac{1}{\privacybound}}\right\rceil^{1/\dimension}-1}\cdot\privacythresholdl/\dimension^{\frac{1}{\normnum}}.
\end{align*}

\subsubsection{Proof of \cref{prop:multiG_norm_performance}}
\label{proof:multiG_norm_performance}

\begin{proof}

Based on \cref{lemma:D_lowerbound_multi_Gaussian}, we can easily can get that the distortion $\Delta$ of \cref{mech:dGaussian_diagnol} is
\begin{align*}
\distortionnotation = \sqrt{\sum_{\secretindex\in\brb{\dimension}}{\seclen_{g_i}}^2}.
\end{align*}

We denote $\tmpprivacynotation$ as the union privacy metric with tolerance ranges $\privacythresholdi{1},\cdots,\privacythresholdi{\dimension}$.
From \cref{proof:lpnorm_proof}, we know that $\privacynotationlp\leq \tmpprivacynotation$ when $\bra{\sum_{\secretindex\in [\dimension]} \privacythresholdi{\secretindex}^\normnum}^{{1}/{\normnum}}= \privacythresholdl$. Therefore, we can get that
\begin{align*}
\privacynotationlp &\leq \min_{\substack{\privacythresholdi{1},\cdots,\privacythresholdi{\dimension}:\\\bra{\sum_{\secretindex\in [\dimension]} \privacythresholdi{\secretindex}^\normnum}^{{1}/{\normnum}}= \privacythresholdl}}\tmpprivacynotation\\
&= \min_{\substack{\privacythresholdi{1},\cdots,\privacythresholdi{\dimension}:\\\bra{\sum_{\secretindex\in [\dimension]} \privacythresholdi{\secretindex}^\normnum}^{{1}/{\normnum}}= \privacythresholdl}} 1 - \prod_{\secretindex\in\brb{\dimension}} \bra{1-\frac{2\privacythresholdi{i}}{\seclen_{g_i}}}.
\end{align*}

By setting $\privacythresholdi{\secretindex} = \privacythresholdl / \dimension^\frac{1}{\normnum}$ for all $\secretindex\in \brb{\dimension}$, we can get that
\begin{align*}
\privacynotationlp &\leq
1 - \prod_{\secretindex\in\brb{\dimension}} \bra{1-\frac{2\privacythresholdl}{\dimension^{\frac{1}{\normnum}}\cdot\seclen_{g_i}}}.
\end{align*}

\end{proof}

\section{The case of unknown distribution parameters: an example}
\label{sec:app_unkown_params_example}

We take $1$-dimensional Gaussian distribution as an example to illustrate how the extension works. 
For a dataset $\privatedataset=\brc{x_1,\ldots, x_m}$, the concrete steps of the extended mechanism are: %
\begin{enumerate}
    \item Calculate the empirical mean and standard deviation from the dataset: 
    $\hat{\mu} = \frac{1}{m}\sum_{i\in \brb{m}} x_i$, $\hat{\sigma} = \sqrt{\frac{1}{m}\sum_{i\in \brb{m}} \bra{x_i - \hat{\mu}}^2}$.
    
    \item Determine the indices $i, j$ of the intervals that $\hat{\mu}, \hat{\sigma}$ fall:
    $i = \floor{\frac{\hat{\mu} - \mulower}{\seclen_\mu}}$, 
    $j = \floor{\frac{\hat{\sigma} - \sigmalower}{\seclen_\sigma}}$.

    \item Determine the released parameters: %
    $\mu_{r} = \mulower + i\cdot \seclen_\mu + U\bra{0, \seclen_\mu}$,
    $\sigma_{r} = \sigmalower + j\cdot \seclen_\sigma + U\bra{0, \seclen_\sigma}$. %
    \item
     Release the dataset $\privatedataset'=\brc{x'_1,\ldots, x'_m}$ such that $x'_i = \frac{\sigma_r}{\hat{\sigma}}\bra{x_i-\hat{\mu}}+\mu_r$.

\end{enumerate}

This mechanism can also be integrated with generative models to alter the summary statistical properties of training samples or the generated dataset.

\section{Additional Empirical Results}
\label{sec:app_empirical}

In this section, we conduct additional experiments on different privacy metrics and analyze the scalability of our mechanism.

\subsection{Experiments on $l_p$ Norm Privacy}

We include $l_p$ norm privacy to illustrate the variations in privacy-distortion tradeoffs
across different privacy metrics and show the privacy-distortion performance of our mechanism. 
We define surrogate privacy metrics for $l_{\normnum}$ norm privacy as
\begin{align*}
\sprivacynotationlp &= -\norm{\normnum}{\secretvof{\odataset} - \secretvof{\rdataset}} / \privacythresholdl,
\end{align*}
and consider the same empirical settings and baselines as those in \cref{sec:experiments}. %
We consider $l_\normnum$ norm privacy with $\normnum = 1$ and $\normnum = \infty$, and set the tolerance as $\privacythresholdl = \bra{\privacythresholdi{1}^{\normnum}+\privacythresholdi{2}^{\normnum}+\privacythresholdi{3}^{\normnum}}^{\frac{1}{\normnum}}$ (i.e., $\privacythresholdlp{1} = 8$ for $l_1$ privacy and $\privacythresholdlp{\infty} = 4$ for $l_{\infty}$ privacy).

\begin{figure}[htbp]
\centering
\begin{minipage}[t]{0.48\textwidth}
\centering
\includegraphics[width=\linewidth]{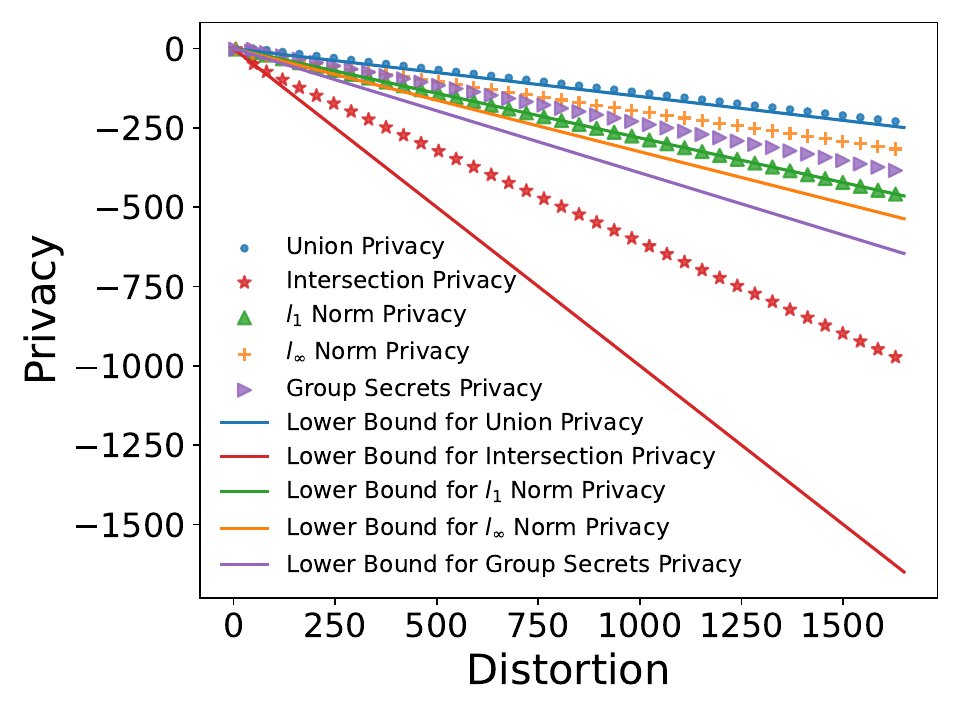}
\caption{Privacy (lower is better) and distortion of \cref{mech:dGaussian_diagnol} under Wikipedia Web Traffic Dataset with different privacy metrics. For each privacy formulation, the soild line with the same color represents the theoretical lower bound of achievable region.}
\label{fig:full_tradeoffs}
\end{minipage}
\hfill
\begin{minipage}[t]{0.48\textwidth}
\centering
\includegraphics[width=\linewidth]{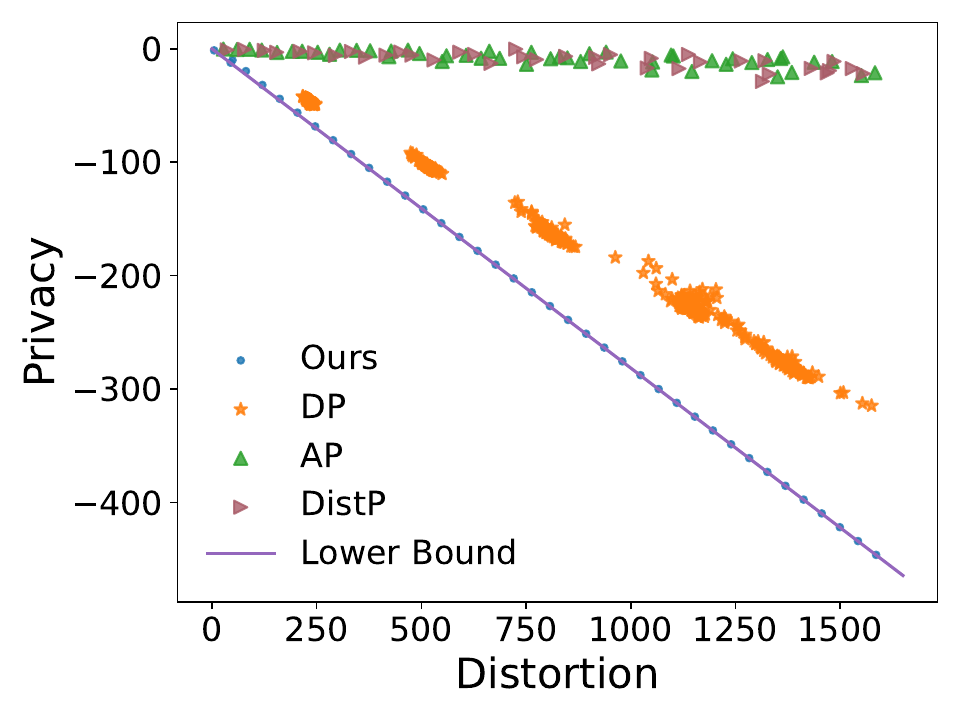}
\caption{Privacy and distortion (lower bottom is better) of DP, AP, DistP, and ours under $l_1$ norm privacy. The solid line represents the theoretical lower bound of
achievable region.}
\label{fig:l1_performance}
\end{minipage}
\end{figure}

Including $l_1$ norm and $l_\infty$ norm privacy, \cref{fig:full_tradeoffs} shows distortion and privacy values of \cref{mech:dGaussian_diagnol} as well as theoretical privacy-distortion lower bounds under different privacy metrics. Similar to \cref{fig:wiki}, we can observe from \cref{fig:full_tradeoffs} that:
(1) The privacy-distortion tradeoffs of \cref{mech:dGaussian_diagnol} match the theoretical lower bounds (or with a mild gap) even the distribution assumption does not hold; (2) The discrepancies in privacy-distortion tradeoffs across different privacy metrics align with the theoretical justifications: the values of $l_p$ norm privacy are intermediate between union and intersection privacy.

In \cref{fig:l1_performance}, we compare the privacy-distortion tradeoffs between DP, AP, DistP, and our mechanism under $l_1$ norm privacy. Each point represents a realization of data release mechanism with one hyper-parameter (bin size and noise level for DP, noise level for AP and DistP, interval length for ours). Similar to the results in \cref{fig:compare}, our quantization mechanism achieves better privacy-distortion tradeoffs.

\subsection{Scalability of Our Mechanism}

In this section, we analyze the scalability of our mechanism \cref{mech:dGaussian_diagnol} with respect to the input data size and the number of secrets.

\myparatightestn{Data Size}
If the data holder knows the underlying distribution, the computational complexity of our mechanism does not increase with data size, since the mechanism inputs are distribution parameters. Otherwise, the computational cost grows linearly with the data size (see \cref{sec:case_study}). 

We conduct experiments on a synthetic dataset whose underlying distribution is a 10-dimensional Gaussian. We vary the data size from $10^2$ to $10^8$, and record the running time of our mechanism (on an 8-core Apple M2 CPU), which is shown as follows. 

\begin{table*}[htbp]
\caption{Running time of \cref{mech:dGaussian_diagnol} under different input data size.}
\label{table:data_size}
\centering
\vspace{4mm}
\begin{tabular}{ccccc}
\toprule
Data Size & $10^2$ & $10^4$ & $10^6$ & $10^8$ \\\midrule
Runtime (s) & $<0.01$ & $<0.01$ & $0.03\pm 0.001$ & $6.90 \pm 0.01$\\
\bottomrule
\end{tabular}
\end{table*}

\cref{table:data_size} shows that even if the data size is $10^8$, our mechanism is still computationally efficient.

\myparatightestn{Secret Number}
The computational cost of our mechanism grows only linearly with the number of secrets. We conduct experiments under the web-traffic dataset and vary the secret number from $10^2$ to $10^5$, and record the running time of our mechanism (on an 8-core Apple M2 CPU), which is shown in \cref{table:secret_number}. 

\begin{table*}[htbp]
\caption{Running time of \cref{mech:dGaussian_diagnol} under different secret number.}
\label{table:secret_number}
\centering
\vspace{4mm}
\begin{tabular}{ccccc}
\toprule
Secret Number & $10^2$ & $10^3$ & $10^4$ & $10^5$ \\\midrule
Runtime (s) & $<0.01$ & $<0.01$ & $0.07\pm 0.001$ & $0.90 \pm 0.01$\\
\bottomrule
\end{tabular}
\end{table*}

\cref{table:secret_number} shows that even if the secret number is as large as $10^5$, our mechanism is still computationally efficient.

\section{Adaptive Composition Property of Summary Statistic Privacy
}
\label{sec:composition}

In this section, we take a first step towards analyzing the adaptive composition property of the summary statistic privacy. We focus on the setting where the distribution parameter $\theta$ belongs to a finite set, there is only one summary statistic secret, and the tolerance range $\epsilon$ is $0$. We also assume that $\sup_{\theta}\mathbb{P}\bra{\theta}=a<1$ and $\inf_{\theta}\mathbb{P}\bra{\theta}=b>0$. Under such setting, the summary statistic privacy of a mechanism $\mathcal{M}$ can be written as 
\begin{align*}
\Pi_{\omega_{\Theta}}^{\mathcal{M}} \triangleq \sup_{\hat{g}}\mathbb{P}\bra{\hat{g}\bra{\theta'}=g\bra{\theta}}.
\end{align*}
The following theorem shows the adaptive composition guarantee of the summary statistic privacy under such setting.
\begin{theorem}[Adaptive Composition]
Consider a data holder sequentially applies $m$ data release mechanisms to the original dataset. For the $i$-th mechanism $\mathcal{M}_i$, $\forall i\in \brb{m}$, it takes the original distribution parameter $\theta$ and all previous released parameters $\theta'_1, \ldots, \theta'_{i-1}$ as input and output $\theta'_i$. Suppose the summary statistic privacy of mechanism $\mathcal{M}_i$ is $\Pi_{\omega_{\Theta}}^{\mathcal{M}_i}$, $\forall i\in \brb{m}$. Let $\boldsymbol{\mathcal{M}}$ be the composition of these $m$ mechanisms, and suppose the adversary can get access to all released parameters $\theta'_1, \ldots, \theta'_m$. The summary statistic privacy of $\boldsymbol{\mathcal{M}}$ can be bounded as $\Pi_{\omega_{\Theta}}^{\boldsymbol{\mathcal{M}}}\leq a\cdot \prod_{i\in \brb{m}}\frac{\Pi_{\omega_{\Theta}}^{\mathcal{M}_i}}{b}$.
\end{theorem}

\begin{proof}
For the summary statistic privacy of $\boldsymbol{\mathcal{M}}$, we can get that 
\begin{align*}
\Pi_{\omega_{\Theta}}^{\boldsymbol{\mathcal{M}}}
&=\sup_{\hat{g}}\mathbb{P}\bra{\hat{g}\bra{\theta'_1, \ldots, \theta'_m}=g\bra{\theta}}\\
&= \sum_{\theta'_1, \ldots, \theta'_m}\mathbb{P}\bra{\theta'_1, \ldots, \theta'_m}\cdot\bra{\sup_{\mathrm{g}}\sum_{\theta: g\bra{\theta}=\mathrm{g}} \mathbb{P}\bra{\theta| \theta'_1, \ldots, \theta'_m}}\\
&= \sum_{\theta'_1, \ldots, \theta'_m}\sup_{\mathrm{g}}\sum_{\theta: g\bra{\theta}=\mathrm{g}} \mathbb{P}\bra{\theta, \theta'_1, \ldots, \theta'_m}\\
&= \sum_{\theta'_1, \ldots, \theta'_m}\sup_{\mathrm{g}}\sum_{\theta: g\bra{\theta}=\mathrm{g}} \mathbb{P}\bra{\theta}\cdot\mathbb{P}\bra{\theta'_1, \ldots, \theta'_m|\theta}\\
&\leq a\cdot \sum_{\theta'_1, \ldots, \theta'_m}\sup_{\mathrm{g}}\sum_{\theta: g\bra{\theta}=\mathrm{g}} \mathbb{P}\bra{\theta'_1, \ldots, \theta'_m|\theta}\\
&= a\cdot \sum_{\theta'_1, \ldots, \theta'_m}\sup_{\mathrm{g}}\sum_{\theta: g\bra{\theta}=\mathrm{g}} \prod_{i\in \brb{m}}\mathbb{P}\bra{\theta'_i|\theta, \theta'_1, \ldots, \theta'_{i-1}}\\
&\leq a\cdot \sum_{\theta'_1, \ldots, \theta'_m}\sup_{\mathrm{g}}\prod_{i\in \brb{m}}\sum_{\theta: g\bra{\theta}=\mathrm{g}} \mathbb{P}\bra{\theta'_i|\theta, \theta'_1, \ldots, \theta'_{i-1}}\\
&\leq a\cdot \sum_{\theta'_1, \ldots, \theta'_m}\prod_{i\in \brb{m}}\sup_{\mathrm{g}}\sum_{\theta: g\bra{\theta}=\mathrm{g}} \mathbb{P}\bra{\theta'_i|\theta, \theta'_1, \ldots, \theta'_{i-1}}\\
&\leq a\cdot \prod_{i\in \brb{m}}\sum_{\theta'_1, \ldots, \theta'_i}\sup_{\mathrm{g}}\sum_{\theta: g\bra{\theta}=\mathrm{g}} \mathbb{P}\bra{\theta'_i|\theta, \theta'_1, \ldots, \theta'_{i-1}}\\
&= a\cdot \prod_{i\in \brb{m}}\sum_{\theta'_1, \ldots, \theta'_i}\sup_{\mathrm{g}}\sum_{\theta: g\bra{\theta}=\mathrm{g}} \frac{\mathbb{P}\bra{\theta'_i}\mathbb{P}\bra{\theta|\theta'_1, \ldots, \theta'_{i}}}{\mathbb{P}\bra{\theta}}\\
&\leq a\cdot \prod_{i\in \brb{m}}\frac{\Pi_{\omega_{\Theta}}^{\mathcal{M}_i}}{b}.
\end{align*}
\end{proof}

\section{Broader Impact}
\label{sec:broader_impact}

This paper proposes a framework to quantify and mitigate the leakage of distributional properties of released datasets or data distributions in data-sharing scenarios, ensuring that data can be shared safely with other entities. However, improper adoption of our framework may result in the miscalculation of privacy risks, potentially leading to unsafe data sharing. To prevent such issues, one must follow our definitions closely to calculate privacy risks accurately. Moreover, extending our proposed release mechanisms to scenarios outside of those investigated in this paper requires careful attention (which is left to future work). Naïve adoption of these mechanisms outside of the investigated scenarios may expose the data holder to privacy risks and is not recommended.

\end{appendix}
\end{appendix}

\end{document}